\documentclass[11pt, reqno]{amsart}
\usepackage{amsmath, amsthm, amssymb, a4wide,xcolor,hyperref}
\numberwithin{equation}{section}
\usepackage{mathrsfs}
\usepackage{graphicx}
\usepackage{array}
\usepackage{placeins}
\usepackage{floatrow}
\usepackage{psfrag}
\usepackage{subcaption}
\usepackage{pgfplots}
\usepackage{tikz}
\usetikzlibrary{external}
\pgfplotsset{compat=1.18}
\usepackage[abspath]{currfile}
\usepackage{standalone}
\newlength\figureheight
\newlength\figurewidth

\usepackage{soul}
\usepackage{commath}
\usepackage[shortlabels]{enumitem}
\usepackage{geometry}
\usepackage{mathtools}
\usepackage{esdiff}
\newtheorem{defi}{Definition}[section]
\newtheorem{lem}[defi]{Lemma}
\newtheorem{thm}[defi]{Theorem}

\newtheorem{prop}[defi]{Proposition}

\theoremstyle{remark}
\newtheorem{rem}[defi]{Remark}

\newcommand{\N}{\mathbb{N}}

\newcommand{\E}{\mathbb{E}}

\newcommand{\R}{\mathbb{R}}
\newcommand{\eps}{\varepsilon}

\definecolor{goetheblau}{cmyk}{1.00 0.20 00 0.40}
\definecolor{hellgrau}{cmyk}{0.04 0.04 0.05 0.02}
\definecolor{sandgrau}{cmyk}{0.12 0.09 0.13 0}
\definecolor{dunkelgrau}{cmyk}{0.25 0.25 0.30 0.75}
\definecolor{purple}{cmyk}{0.08 1.00 0.30 0.36}
\definecolor{emorot}{cmyk}{0.04 1.00 0.80 0.07}
\definecolor{senfgelb}{cmyk}{0.01 0.25 1.00 0.05}
\definecolor{gruen}{cmyk}{0.62 0.40 0.87 0.09}
\definecolor{green}{cmyk}{0.92 0.30 0.67 0.03}
\definecolor{magenta}{cmyk}{0.08 0.86 0.12 0.12}
\definecolor{orange}{cmyk}{0 0.70 1.00 0.04}
\definecolor{sonnengelb}{cmyk}{0 0.12 0.95 0}
\definecolor{hellesgruen}{cmyk}{0.40 0.17 0.81 0.07}
\definecolor{lichtblau}{cmyk}{0.80 00 0.06 0.04}

\hypersetup{%
	colorlinks = true,
	linkcolor  = emorot,
	citecolor = hellesgruen
}

\mathcode`l="8000
\begingroup
\makeatletter
\lccode`\~=`\l
\DeclareMathSymbol{\lsb@l}{\mathalpha}{letters}{`l}
\lowercase{\gdef~{\ifnum\the\mathgroup=\m@ne \ell \else \lsb@l \fi}}%
\endgroup

\begin{document}
\allowdisplaybreaks
\title[Muller's ratchet in a near-critical regime]{Muller's ratchet in a near-critical regime:\\ 
tournament versus fitness proportional selection.}
\author[J.L. Igelbrink]{Jan Lukas Igelbrink}
\address{Jan Lukas Igelbrink, Institut f\"ur Mathematik, 
  Johannes Gutenberg-Universit\"at Mainz and Goethe-Universit\"at Frankfurt, Germany.}
\email{jigelbri@uni-mainz.de, igelbrin@math.uni-frankfurt.de}
\author[A. Gonz\'alez Casanova]{Adri\'an Gonz\'alez Casanova}
\address{Adri\'an Gonz\'alez Casanova, Universidad Nacional Aut\'onoma de M\' exico (UNAM), Instituto de Matem\'aticas, Circuito exterior, Ciudad Universitaria, 04510, M\'exico, 11; \newline University of California, Berkeley.
Department of Statistics, 
367 Evans Hall, University of California,
Berkeley, CA 94720-386}
\email{adriangcs@matem.unam.mx, gonzalez.casanova@berkeley.edu}

\author[C. Smadi]{Charline Smadi}
\address{Charline Smadi, Univ. Grenoble Alpes, INRAE, LESSEM, 38000 Grenoble, France
 and Univ. Grenoble Alpes, CNRS, Institut Fourier, 38000 Grenoble, France}
\email{charline.smadi@inrae.fr}

\author[A. Wakolbinger]{Anton Wakolbinger}
\address{Anton Wakolbinger, Goethe-Universit\"at, Institut f\"ur Mathematik, 60629 Frankfurt am Main, Germany}
\email{wakolbinger@math.uni-frankfurt.de}
\begin{abstract}
Muller's ratchet, in its prototype version, models a haploid, asexual population whose size~$N$ is constant over the generations. Slightly deleterious mutations are acquired along the lineages at a constant rate, and individuals carrying less mutations have a selective advantage. The classical variant  considers {\it fitness proportional} selection, but other fitness schemes are conceivable as well. Inspired by the work of Etheridge et al. \cite{EPW09} we propose a parameter scaling which fits well to the ``near-critical'' regime that was in the focus of \cite{EPW09} (and in which the mutation-selection ratio diverges logarithmically as $N\to \infty$). Using a Moran model, we investigate the``rule of thumb'' given in \cite{EPW09} for the click rate of the ``classical ratchet'' by putting it into the context of new results on the long-time evolution of the size of the best class of the ratchet with (binary) tournament selection. This variant of Muller's ratchet was introduced in \cite{GSW23}, and was analysed there in a subcritical parameter regime. Other than that of the classical ratchet, the size of the best class of the tournament ratchet follows an autonomous dynamics up to the time of its extinction. It turns out that, under a suitable correspondence of the model parameters, this dynamics coincides with the so called Poisson profile approximation of the dynamics of the best class of the classical ratchet. 

\end{abstract}
\maketitle
\newpage
\tableofcontents
\section{Introduction}\label{secIntro}
Muller's ratchet is a prototype model in population genetics. Originally it was conceived to explain the ubiquity of sexual reproduction among eukaryotes despite its many costs \cite{M64, felsenstein1974evolutionary}.  In its bare bones version, Muller's ratchet models a haploid, asexual population whose size~$N$ is constant over the generations. The neutral part of the random reproduction is given by a Wright-Fisher or a Moran dynamics. Slightly deleterious mutations are acquired along the lineages at a rate $m$, and individuals carrying less mutations have a selective advantage. The classical variant of Muller's ratchet considers {\it fitness proportional} selection, where the selective advantage of an individual carrying $\kappa$ deleterious mutations over a contemporanean that carries a larger number $\kappa'$ of deleterious mutations is 
$\frac{\mathfrak s}N(\kappa'-\kappa)$. Since the mutation mechanism is assumed to be unidirectional, every once in a while the type with the currently smallest number of mutations $\kappa$ will disappear from the population. As Herbert Muller puts it in his pioneering paper \cite{M64}, ``{\it an irreversible ratchet mechanism exists in the non-recombining species \ldots that prevents selection, even if intensified, from reducing the mutational loads below the lightest \ldots, whereas, contrariwise, 'drift', and what might be called 'selective noise' must allow occasional slips of the lightest loads in the direction of increased weight.}''
\FloatBarrier
It is these ``slips of the lightest loads''  which are called {\it clicks of the ratchet}. The question  ``How often does the ratchet click?''  was  asked by Etheridge, Pfaffelhuber and one of the present authors in \cite{EPW09}, and there it was found that 
\begin{equation}\label{WUgamma}
  \gamma := \frac m{\mathfrak s\log(Nm)}  
\end{equation}
is ``{\it an important factor in determining the rate of the ratchet}''. Specifically, under the assumption $1\ll Nm \ll N$, \cite{EPW09} states the following {\it Rule of Thumb} for the classical ratchet:
\\
{\it {\bf (RTC)} The rate of the (classical) ratchet  is of the order $N^{\gamma-1}{m}_{}^\gamma$ for $\gamma \in (\frac 12, 1)$, whereas it is exponentially slow in $(Nm)^{1-\gamma}$ for $\gamma < \frac 12$.}
\\\\
With the
{\it mutation-selection ratio}
$$\theta := \frac m{\mathfrak s},$$
(RTC) predicts the expected interclick time in the case $\gamma \in (\frac 12, 1)$ as 
$$N(Nm)^{-\gamma} = Ne^{-\theta}.$$
\begin{figure}[ht]
  \centering
  \begin{subfigure}{0.45\textwidth}
    \input{\currfiledir tikz/BReal_beta_0.8_gamma_0.4_REV.tex}
    \caption{
    }\label{Fig1a}
  \end{subfigure}
  \begin{subfigure}{0.45\textwidth}
    \input{\currfiledir tikz/BReal_beta_0.6_gamma_0.7_REV.tex}
    \caption{
    }\label{Fig1b}
  \end{subfigure}
  \caption{This is an illustration of the {\it Rule of Thumb} (RTC) predicting the order of magnitude of the interclick times of the classical ratchet. 
Each data point was obtained  by pooling the interclick times no.~50 to 150 from 100 simulations of the (classical) ratchet for the corresponding parameter configuration $(N, \beta, \delta)$ in the $(\beta, \delta)$-scaling \eqref{bdclassi}. In the exponential regime,  (RTC) predicts an order of magnitude $\exp ( c \, N^{1-\beta-\delta})$ for the interclick times. In panel (\subref{Fig1a}), we see that the constant $c$ is difficult to estimate from simulations up to $N=10^4$, but $c =2.3$ as chosen there gives a reasonable fit. For the polynomial regime, (RTC) predicts the order of magnitude $N^{1-\delta}$, which fits very well the data in the situation of panel~(\subref{Fig1b}).
  }\label{fig:1}
\end{figure}
\FloatBarrier
As observed by John  Haigh (\cite{H78}), in the deterministic limit ($N\to \infty$ and $m, \mathfrak s$ not depending on~$N$)  the type frequency profile in equilibrium becomes Poisson with parameter~$\theta$. Consequently, for $\gamma \in (\frac 12, 1)$ the rule (RTC) goes along with Haigh's prediction that the rate of the ratchet  should be proportional to the inverse of the size of the best class. 

For a polynomial mutation rate $m= N^{-\beta}$, $0<\beta < 1$, the condition that $\gamma$ remains constant (or at least bounded away from 0 and $\infty$) as $N\to \infty$ amounts to the requirement that the mutation-selection ratio $\theta$
is of the order $\log N$ as $N\to \infty$. 
 
For the purpose of illustration we will consider a family of parameter scalings for $(m,\theta)$ which we call the {\it $(\beta,\delta)$-scaling of the classical ratchet}:
\begin{equation}\label{bdclassi}
m=N^{-\beta}, \qquad \theta = \delta \log N.
\end{equation}
This amounts to {\it moderate mutation-selection}, with  the mutation-selection ratio $\theta$ diverging logarithmically in $N$. The factor $\delta$ in front of $\log N$ turns out to be critical for the click rate.   Indeed, in the $(\beta,\delta)$-scaling, \eqref{WUgamma} takes the  form
\begin{equation*}
    \gamma(\beta,\delta) =  \frac \delta{1-\beta}. 
\end{equation*}
The condition $0<\gamma < 1$ from (RTC) restricts the pair $(\beta, \delta)$ to the triangle
\begin{equation}\label{defDelta}
\Delta := \{(\beta, \delta): 0<\beta,\,   0 <\delta < 1-\beta\}.
\end{equation}
The  {\it polynomial} and the {\it exponential regime} predicted by (RTC) correspond to 
\begin{equation*}
\mathcal P := \{\tfrac 12 < \gamma(\beta,\delta) < 1\} = \{(\beta, \delta) \in \Delta: \tfrac 12 (1-\beta) < \delta < 1-\beta\},
\end{equation*}
  \begin{equation*}
  \mathcal E := \{0 < \gamma(\beta,\delta) < \tfrac 12\} = \{(\beta, \delta)\in \Delta: 0 < \delta < \tfrac 12 (1-\beta)\},
\end{equation*}
and the predictions for the orders of magnitude of the expected interclick times take the form
\begin{eqnarray}\label{RT1}
 \phantom{\exp\left({\rm const}(Nm)^{1-\gamma}\right)}N (Nm)^{-\gamma}
= \phantom{\,}N^{1-\delta}\phantom{\exp\left({\rm const}\, N^{1-\beta-\delta}\right)}  &\mbox { for }& \gamma \in (\tfrac 12, 1),\\
\label{RT2}
\phantom{N (Nm)^{-\gamma}}\exp\left({\rm const}(Nm)^{1-\gamma}\right)
= \exp\left({\rm const}\, N^{1-\beta-\delta}\right)\phantom{ N^{1-\delta}} &\mbox { for }& \gamma \in (0, \tfrac 12).
\end{eqnarray}
In view of the predicted transition from polynomial to exponential click rates we refer to $\mathcal P \cup \mathcal E$ as a {\it near-critical regime}. See Figure~\ref{fig:1} for an illustration of (RTC) via simulations.

The evidence for  (RTC) that is given in \cite{EPW09} is based on a diffusion approximation for the evolution of the relative  size $X_0$ of the {\it best class} (which consists of the individuals that carry the least amount of mutations in the current population).  Because of the fitness proportional selection, the drift coefficient in this diffusion approximation  contains the first moment $M$ of the type frequency configuration $(X_0, X_1, \ldots)$. In order to obtain an approximate autonomous dynamics for $X_0$, the empirical first moment $M$ has to be predicted based on $X_0$. A classical way to do this uses the so-called  {\it Poisson profile approximation}, which we will explain  in some detail in Section~\ref{correspond}.

In the present paper we will consider a variant of Muller's ratchet in which fitness proportional selection is replaced by {\it (binary) tournament selection}. This kind of selection has been studied in the context of evolutionary computation (\cite{BT961,EvComp1}) and has found attention also in the biological literature \cite{Pai2015}.  In the ratchet's context this means that selective advantage of an individual carrying $\kappa$ deleterious mutations over a contemporanean that carries a larger number $\kappa'$ of deleterious mutations is constant (say~$\frac s N$ for some $s = s_N > 0$), irrespective of the value of the difference $\kappa'-\kappa$. For the Moran version of the tournament ratchet, which was introduced in \cite{GSW23} and whose definition we recall in  Section \ref{mura}, this means that  ``pairwise selective fights'' are always won by the fitter individual.

Other than in the classical ratchet, the size of the $(m,s)$-tournament ratchet's best class follows an autonomous dynamics {\em up to its time of extinction}; at this time the class which was so far the second-best becomes the best one. As we will see in Section \ref {correspond},  this dynamics  is {\it equal} to that  of the Poisson profile approximation of the size of the classical $(m,\mathfrak s)$-ratchet's best class, provided that
\begin{equation} \label{rhodelta}
    \rho := \frac ms = 1-\exp(- m/\mathfrak s) = 1-e^{-\theta}.
\end{equation}

We now state a main finding of the present paper. 

{\bf Rule of thumb for the near-critical tournament ratchet (RTT):}\\{\it As $N\to \infty$, the expected time between clicks  is}
  \begin{eqnarray}\label{RT3}
\asymp \phantom{\,} \sqrt{\frac Nm} \phantom{\exp\left(Nm(1-\rho)^2\right)}&\mbox { if }& Nm(1-\rho)^2\to 0,\\
\label{RT4}
\asymp   \exp\left(Nm(1-\rho)^2\right) \phantom{\sqrt{\frac Nm}}&\mbox { if }&Nm(1-\rho)^2\to \infty.
\end{eqnarray}
Here and below, $\asymp$ stands for logarithmic equivalence, i.e.
$a_N \asymp b_N$ means $\log a_N \sim \log b_N$, or equivalently $\frac{\log a_N}{\log b_N} \to 1$.  

We will not give a complete proof of (RTT) in this work, but  will present Theorem \ref{mainth} which gives strong evidence for its validity. See Figure~\ref{fig:2} for an illustration of (RTT) in the light of Theorem \ref{mainth}. In Remark \ref{missing} we will discuss what are the ingredients missing to go from Theorem \ref{mainth} to a proof of (RTT), and we will also indicate a different route to the proof of (RTT), using the technique developed in \cite{GSW23}. 
  
We emphasise that, in view  of the correspondence \eqref{rhodelta}, Theorem \ref{mainth} also is a result on the asymptotics of the Poisson profile approximation of the classical ratchet, here in terms of Moran processes with mutation and selection. A similar asymptotics was obtained in  \cite{EPW09} heuristically by passing right away to the diffusion approximation for logistic branching processes.

In view of \eqref{rhodelta} we define, in analogy to  \eqref{bdclassi}, the $(\beta,\delta)$-scaling for the tournament ratchet as
\begin{equation*}
 m = N^{-\beta}, \qquad \rho = \frac ms = 1-N^{-\delta}  . 
\end{equation*}
With this scaling, (RTT) takes the following form: {\it As $N\to \infty$, the expected time between clicks  is}
  \begin{eqnarray}\label{RT3scal}
\asymp \phantom{\,}N^{\frac{1+\beta}2} \phantom{ \exp\left(N^{1-\beta-2\delta}\right) }&\mbox { if }& (\beta, \delta) \in \mathcal P,\\
\label{RT4scal}
\asymp \exp\left(N^{1-\beta-2\delta}\right) \phantom{N^{\frac{1+\beta}2}}&\mbox { if }& (\beta, \delta) \in \mathcal E.
\end{eqnarray}
While both (RTC) and (RTT) state the same boundary ($\gamma = \frac 12$) between the polynomial and the exponential regime,
the exponents differ between \eqref{RT1} and~\eqref{RT3scal} as well as between \eqref{RT2} and~\eqref{RT4scal}. Specifically, in the polynomial regime $\mathcal P$ the exponent $\frac {1+\beta}2$ for the tournament ratchet is larger than the exponent $1-\delta$ for the classical ratchet. 

Here is an explanation  for the polynomial regime. The centers of attraction of the equilibrium profile weights of the best and the second best class differ asymptotically by the factor $\sqrt{1-\rho}=N^{\frac \delta 2}$ for the tournament ratchet (see \eqref{centattY}), while they are given by the Poisson weights $e^{-\theta}$ and $\theta e^{-\theta} $ for the classical ratchet and hence for the latter differ only by the factor $\theta = \delta \log N$ (and thus have the same polynomial order $N^{1-\delta}$). 
This latter factor is only logarithmic in~$N$; therefore, when starting the ``new best class'' at the time of a click in its ``old'' center of attraction, the tournament ratchet has a longer way to go than the classical ratchet. 
The exponent $\frac {1+\beta}2$ in \eqref{RT3} will be obtained by a Green function analysis in the proof of Theorem~\ref{mainth}. This analysis will also explain the exponent $1-\delta$ in \eqref{RT1}, which corresponds to Haigh's prediction, saying that ``the interclick times are of the order of the size of the best class''. An intuitive explanation for the appearance of the exponent $1-\beta-2\delta$ in \eqref{RT4} will be given at the end of Section~\ref{secPPA}. The reason why this exponent is different from the one appearing in \eqref{RT2} is that  \cite{EPW09} work here not with the Poisson profile approximation, but with (a rescaling of the diffusion approximation of) the so-called {\it relaxed Poisson profile approximation}. 

Similar as \cite{EPW09},  the papers \cite{pfaffelhuber2012muller,neher2012fluctuations,audiffren2013muller, MPV20,brautigam2022diffusion} used  a diffusion approximation for the classical ratchet and modifications thereof. Metzger and Eule \cite{ME13} consider, as a proxy to the classical ratchet, a two-type Moran model with selective advantage $s$ of type $0$ over type $1$ and mutation rate $m$ from type $0$ to type $1$. Their formula~(8) corresponds to our formula \eqref{rhodelta} but their approximations for the classical ratchet concentrate on a regime in which $\theta$ remains bounded (see the discussion around \cite[(23)]{ME13}, and also \mbox{\cite[(7),(8)]{waxman2010stochastic}}), whereas we focus here on a regime in which $\theta=\theta_N$ diverges logarithmically in $N$.

In \cite{GSW23} it was discovered that the tournament ratchet has a dual which consists of a hierarchy of competing logistic processes. The main results of \cite{GSW23} (on the click rate of the tournament ratchet and its type frequency profile between clicks) were obtained for the so-called subcritical regime (see Sec.~\ref{Secsubc}) and were proved there via duality, with the help of recent results on logistic processes (see \cite{L,sagitov2013extinction,chazottes2016sharp}). This ``backward in time'' view, which comes on top of an Ancestral Selection Graph decorated with mutation events, 
opens a route for proving the above stated result (RTT) and for analysing the type frequency profile of the tournament ratchet also in the near-critical regime. This will be pursued in future work.

In \cite{GSW23} the rate of the tournament ratchet was identified in the subcritical regime (i.e. for $\rho = m/s < 1$ and not depending on $N$)  up to logarithmic equivalence. Thus our Theorem \ref{mainth} \ref{th:mainth:b}, which is  valid both for the near-critical {\em and} the subcritical regime, provides an essential step in sharpening the rate asymptotics of \cite{GSW23}  from logarithmic equivalence to asymptotic equivalence, see  Remark~\ref{missing}\ref{remark36a}.   

\section{Muller's ratchet as a Moran process with mutation and selection}\label{mura}
\subsection{Model and basic concepts}
In the Moran version of Muller's ratchet, neutral resampling within any  ordered pair of individuals happens at rate~$\frac 1{2N}$, and  mutation from $\kappa$ to $\kappa+1$ takes place at rate $m/N$ along each individual lineage. Selective reproduction for an individual $i$ of type $\kappa(i)$ happens at rate $\frac 1N \sum_{j}\Phi(\kappa(j)-\kappa(i))$, where the sum is taken over all those individuals $j$ whose type $\kappa(j)$ is larger (and therefore ``worse'') than $\kappa(i)$. Here $\Phi$ is the {\it fitness function}, with $\Phi(0)=0$ and $\Phi(-d) = -\Phi(d)$ for $d \in \N$. For the classical case of {\it proportional selection}, one has $\Phi(\kappa'-\kappa) = \mathbf s\,(\kappa'-\kappa)$, while for the case of {\it (binary)  tournament selection} one has 
$\Phi(\kappa'-\kappa) = \mathbf s\, (\mathbf 1_{\{\kappa' > \kappa\}}-\mathbf 1_{\{\kappa' < \kappa\}}) $. In the sequel we will refer to these two Moran variants of  Muller's ratchet briefly as the {\it classical ratchet} and the {\it tournament ratchet}. Both models have $(N,m,\mathbf s)$ as their parameter triple, and in both models a crucial role is played by the {\it mutation-selection ratio} $\frac {m}{\mathbf s}$. In this section we reserve the symbol $\mathbf s$ for the selection parameter. Later, this will be specified as different parameters $s$ and $\mathfrak s$ for the tournament and the classical ratchet, respectively. The following definition gives the rates for the type frequencies of the two ratchets.
\begin{defi}\label{defdyn}\
\begin{enumerate}[a)] 
\item Writing $N_\kappa$ for the current number of individuals of type $\kappa$, the jump rates are specified as follows:

- Resampling: for $\kappa \neq \kappa'$,    \\
\phantom{AAA} $(N_\kappa, N_{\kappa'})$ jumps to $(N_\kappa+1, N_{\kappa'}-1)$ at rate $\frac 1{2N} N_\kappa N_{\kappa'}$ 

\medskip
- Mutation: for $\kappa$, \\
\phantom{AAA} $(N_\kappa, N_{\kappa+1})$ jumps to $(N_\kappa-1, N_{\kappa+1}+1)$ at rate $m N_\kappa   $ 

\medskip
- Selection: for $\kappa < \kappa'$, \\
\hspace*{-0.375cm} $(N_\kappa, N_{\kappa'})$ jumps to $(N_\kappa+1, N_{\kappa'}-1)$ at rate $\begin{cases}\frac {\mathbf s}N N_\kappa N_{\kappa'}(\kappa'-\kappa) \,\, \mbox {for the classical ratchet}\\
 \frac {\mathbf s}N N_\kappa N_{\kappa'}\quad \quad \mbox {for the tournament ratchet}
 \end{cases}
 $ 
 \item The {\rm currently best type} is
$$K^\ast(t):=\min\left\{\kappa \in \mathbb N_0: N_\kappa(t)>0\right\}.$$
\item The {\rm click times of the ratchet} are the jump times of $K^\ast$, i.e. the times at which the currently best type is lost from the population. The {\rm  type frequency profile seen from the currently best type} has the (random) weights
\begin{equation}\label{typefprof}
    X_k^{(N)}(t) := \frac 1N N_{K^\ast(t)+k}(t), \qquad k= 0,1,2\ldots
\end{equation}
\end{enumerate}

\end{defi}

We say that a  (non-random) type frequency profile $(p_k)_{k\in \N_0}$ obeys the mutation-selection equilibrium conditions (for the parameters $m$ and $\mathbf s$) if
\begin{equation}\label{mutseleq}
m(p_{k}-p_{k-1}) = \mathbf s\,p_k\left(\sum_{k'\in \N_0}p_{k'}\Phi(k'-k)\right),\quad k= 0,1,2\ldots,
\end{equation}
where we put $p_{-1} := 0$.

For the classical ratchet, \eqref{mutseleq} turns into
\begin{equation}\label{mutseleqclass}
m(p_{k}-p_{k-1}) = \mathbf s\,p_k(\mu-k),\quad k= 0,1,2\ldots,
\end{equation}
where $\mu := \sum_{\ell}\ell p_\ell$ is the first moment of the profile. As already noticed by John Haigh (\cite{H78}), \eqref{mutseleqclass} is solved by the {\it Poisson weights} with first moment $\mu = \frac{m}{\mathbf s}$. Indeed, this is the unique solution of \eqref{mutseleqclass} under the  condition $p_0 > 0$.

For the tournament ratchet, \eqref{mutseleq} turns into
\begin{eqnarray}\label{systeq}  m \left(p_{k}-p_{k-1}\right) =  \mathbf s\,p_{k}\left(\sum_{k'\in \mathbb N_0} p_{k'}\left(\mathbf 1_{\{k'>k\}}- \mathbf 1_{\{k'<k\}}\right)\right),  \quad k= 0,1,2\ldots
 \end{eqnarray}
 Here the  condition $p_0 > 0$ leads to the requirement  $m<\mathbf s$ and yields $p_0 = 1-\frac {m}{\mathbf s}$. Various properties of the solution  $(p_{k'})$ of \eqref{systeq} are stated in \cite{GSW23} Theorem 2.3. The r.h.s. of \eqref{systeq}  equals
 \begin{equation}\label{systeqrhs}   \mathbf s\,p_{k}\left(1- p_k -2\sum_{k'=0}^{k-1} p_{k'}\right),  \quad k= 0,1,2\ldots
 \end{equation}
 A formal analogy between \eqref {mutseleqclass} and \eqref{systeq} results because \eqref{systeqrhs} 
is close to $2\mathbf s\,p_k (\frac 12-g(k))$, where $g$ is the cumulative distribution function of~$(p_{k'})$. In this sense the role played by the profile's first moment in \eqref {mutseleqclass} is taken by the profile's median in 
 \eqref{systeq}.
 \subsection{The subcritical regime of the tournament ratchet} \label{Secsubc}~\\We now report briefly on the main results of the recent paper \cite{GSW23}. The parameters of the tournament ratchet will be denoted   by $(m,s)$ and its mutation-selection ratio by
$\rho := \frac ms.$  In \cite{GSW23},  
 as $N\to \infty$, the mutation-selection ratio $\rho= \frac ms$ is kept constant and smaller than $1$, and it is assumed that $m\to 0$ and $mN \to \infty$.
  (For technical reasons, $mN$ is assumed to be of larger order of $\log \log N$, which keeps the regime slightly away from that of weak mutation, in which $mN$ would be of order one as $N\to \infty$.) We will refer to this regime as the {\it subcritical regime} of the tournament ratchet.
Informally stated, the  main results of \cite{GSW23} are

$\bullet$ {\it In the subcritical regime the click rate of the tournament ratchet on the \mbox{$\frac 1m$-timescale} is, as $N\to \infty$, logarithmically equivalent to 
\begin{equation}
  \label{subcras}  
e^{-2Nm\left(\frac 1\rho - 1 +\log \rho\right)}.
\end{equation}}

 $\bullet$ {\it In the subcritical regime and for $N$ large, the empirical type frequency profile  at generic time points between clicks of the tournament ratchet is with high
probability close to the mutation-selection equilibrium system~$(p_k)$ given by~\eqref{systeq}  with 
 $p_0 = 1-\rho$.}

See Theorems 2.2 and 2.3 in \cite{GSW23}, which there are proved via a hierarchical duality. As discussed in Remark~\ref{missing}.\ref{remark36a}, Theorem \ref{mainth} \ref{th:mainth:b} can be considered as a significant step in sharpening \eqref{subcras} to an asymptotic equivalence.

\section{A synopsis of the classical and the tournament ratchet}
\label{correspond}
\subsection{The dynamics of  the best classes}
~ \\ For $k=0,1,\ldots$ let $Y_k^{\rm C}(t)= N_{K^*+k}^{\rm C}(t)$ and $Y_k^{\rm T}(t)= N_{K^*+k}^{\rm T}(t)$ be the sizes of the \\$(k+1)^{\rm st}$-best class of the classical and the tournament ratchet, where $(N_\kappa^{\rm C})_{\kappa \in \N_0}$ and $(N_\kappa^{\rm T})_{\kappa \in \N_0}$ follow the dynamics specified in Definition \ref{defdyn}. Here we assume that the mutation rate $m$ is equal for both ratchets, but the selection coefficients are different: 
$$\mathbf s =\begin{cases}\frac m \theta =: \mathfrak s \quad \quad \mbox {for the classical ratchet}\\
 \frac m \rho =: s\quad \quad \mbox {for the tournament ratchet}.
 \end{cases}
  $$
  The jump rates from $n$ to $n-1$ are given for both $Y_0^{\rm C}$ and $Y_0^{\rm T}$ by
  \begin{equation}\label{bestdown}
n\left(\frac 12\left(1-\frac nN\right)+m\right),
\end{equation}
but the jump rates from $n$ to $n+1$ are different:
those of $Y_0^{\rm T}$ are
\begin{equation}\label{bestup}
n\left(\frac 12\left(1-\frac nN\right) + \frac m\rho \left(1-\frac nN\right)\right),
\end{equation}
while those of $Y_0^{\rm C}$ are
\begin{equation}\label{bestupclass}
n\left(\frac 12\left(1-\frac nN\right)+ \frac m \theta \sum_{k=1}^\infty k X_k\right)
\end{equation}
where $(X_k(t))_{k\in \N_0}$ is the type frequency profile as defined in \eqref{typefprof}, with  $(N_\kappa^{\rm C})$ in place of $(N_\kappa)$.
Writing $$M(t):= \sum_{k=1}^\infty kX_k(t)$$ for the first moment of the  type frequency profile $\left(X_k\right)$, the upward jump rate \eqref{bestupclass} takes the form
\begin{equation}\label{bestupclass1}
n\left(\frac 12\left(1-\frac nN\right)+m \frac M \theta\right).
\end{equation}
An inspection of the jump rates in Definition \ref{defdyn} reveals that for each $k \in \N$ the process $(Y_0^{\rm T}, \ldots, Y_k^{\rm T})$ obeys an autonomous dynamics up to the extinction time of $Y_0^{\rm T}$; for $k=0$ this is evident from \eqref{bestdown} and \eqref{bestup}. For later reference we note here that $(Y_0^{\rm T}, Y_1^{\rm T})$ has, asymptotically as $N\to \infty$, the center of attraction
\begin{equation}\label{centattY}
(\mathfrak a, \mathfrak b):= \left(N (1-\rho), N\sqrt{1-\rho}\right)
\end{equation}
provided $Nm\to \infty$ and $\rho \to 1$. To see this, note that the dynamics of $(Y_0^{\rm T}, Y_1^{\rm T})$ is autonomous up to the first hitting of $\{0\}\times \{0,\ldots, N\}$, and that the states of $(Y_0^{\rm T}, Y_1^{\rm T})$ for which the upward jump rates are asymptotically equal to the downward jump rates have the asymptotic $(Np_0, Np_1)$, with $(p_0,p_1)$ given by \eqref{systeq} and \eqref{systeqrhs}. In addition to $p_0 = 1-\rho$, this leads to the equation $$p_1(1-p_1 -2(1-\rho)) = \rho(p_1-(1-\rho)),$$
\\ with the solution $p_1 = \sqrt{1-\rho}\left(\sqrt{\rho+\frac 14(1-\rho)}-\frac 12 \sqrt{1-\rho}\right) \sim \sqrt{1-\rho}$ as $\rho \uparrow 1$.

In contrast to the tournament ratchet, the rates \eqref{bestupclass1} depend not only on the size of the best class but also on the profile $\left(X_k\right(t))_{k\ge 0}$ (via its first moment $M(t)$). There are various ways to predict $M(t)$ on the basis of~$Y^{\rm C}_0(t)$, and thereby to replace \eqref{bestupclass1} by a rate which is autonomous. One of them will be described in the remainder of this section, a second one will be addressed in Remark \ref{RPPA}.  As observed already by John Haigh \cite{H78}, such a strategy requires a regime in which ``genetic drift'', i.e. the fluctuations due to neutral reproduction, needs a time to take  $Y_0^{\rm C}$ to extinction which is large compared to the time  which the noiseless classical ratchet needs to ``relax'' towards its (new) equilibrium. 
The dynamics of the latter is
\begin{equation}\label{detdyn}
 \dif x_k(t) = \left(\mathfrak s\sum_{\ell}x_\ell(\ell-k) + m(x_{k-1}(t)-x_k(t))\right)\dif t, \qquad k= 0,1, \ldots 
 \end{equation}
(with  $x_{-1} \equiv 0$). As already indicated  after \eqref{mutseleqclass}, the unique vector of probability weights on $\N_0$ which has a non-vanishing weight at $0$ and is a stationary point of \eqref{detdyn} is given by the {\it Poisson profile}
\begin{equation}\label{PoPr}
 \pi_k = e^{-\theta}\frac {\theta^k}{k!}, \quad k\ge 0. 
 \end{equation}
 For the initial profile 
\begin{equation*}
x(0) := \tfrac{1}{1-\pi_0} \big(\pi_1, \pi_2, \ldots \big),
\end{equation*} 
the {\it relaxation time} $\tau$ which it takes for $x_0(t)$ to come down from $\frac 1{1-\pi_0}\pi_1$   to $\frac e{e-1} \pi_0$ turns out to be
\begin{equation*}
\tau = \frac{\log \theta}{\mathfrak s},
\end{equation*} 
(see \cite[Remark 4.3]{EPW09}\footnote{In order to ease the look-up we use here and below the numbering of the arxiv version of \cite{EPW09}, which otherwise is identical in content with the version published in the LMS Lecture Note Series.}).
The time to extinction of a neutral Moran$(N)$-process starting in $N\pi_0 = Ne^{-\theta}$ is of the order $Ne^{-\theta}$. Haigh's requirement can thus  be formulated as
\begin{equation*}
Ne^{-\theta} \gg \frac {\log \theta}{\mathfrak s}, 
\end{equation*}
which in the $(\beta,\delta)$-scaling
\eqref{bdclassi} just means that $\beta + \delta < 1$.
\subsection{The Poisson profile approximation for the classical ratchet}\label{secPPA}~ \\ 
Here the idea is to think of the
 profile $\left(X_k\right)_{k\ge 1}$ as (nearly) proportional to the  Poisson profile~\eqref{PoPr}, 
 and as the mass $\pi_0-X_0$ being distributed proportionally upon this profile. This leads to the so-called {\it Poisson profile approximation} of $\left(X_k\right)_{k\ge 1}$ based on $X_0$, given by 
 \begin{equation*}
 \Pi(X_0) := \left( X_0, \frac{1-X_0} {1-\pi_0}\big(\pi_1, \pi_2, \ldots\big) \right).
 \end{equation*}
 (cf  \cite[(2.5)]{EPW09}).
 The first moment of $ \Pi(X_0) $ is
  \begin{equation*}
 M(X_0) := \left(1- X_0 \right) \frac{\theta} {1-\pi_0},
  \end{equation*}
  in accordance with \cite[(5.3a)]{EPW09}.
  Plugging this into \eqref{bestupclass1} in place of $M$  leads to the following {\it Poisson profile approximation} of the upward jump rates \eqref{bestupclass1}:
\begin{equation}\label{bestupclass2}
n\left(\frac 12\left(1-\frac nN\right)+\frac{m}{1-e^{-\theta}}\left(1-   \frac n{N}\right)\right).
\end{equation} 
We denote the birth-and death-process on $\N_0$ with downward jump rates  \eqref{bestdown} and upward jump rates \eqref{bestupclass2} by $Y_{\rm PPA}$;  this process  can be seen as an approximation of $Y_0^{\rm C}$. 
\begin{rem}\label{parallels}
    A~crucial observation is that the upward jump rates
 \eqref{bestupclass2} and \eqref{bestup} are equal if and only if $\rho = 1-e^{-\theta}$. In other words, under the ``dictionary''  \eqref{rhodelta},
the jump rates \eqref{bestdown} and \eqref{bestup}
of the size of the best class of the $(m,s)$-tournament ratchet are  equal to the jump rates \eqref{bestdown} and \eqref{bestupclass2} of the Poisson profile approximation for the size of the best class of the classical $(m, \mathfrak s)$-ratchet. 
\end{rem}
\begin{rem}\label{RPPA}
Not least to provide a systematic framework for previous approaches (\cite{Stephan1993TheAO, Gordo2000TheDO}) to the approximation of the size of the ratchet's best class, \cite{EPW09} embedded the Poisson profile approximation (PPA) into a one-parameter family RPPA($A$), $A\ge 0$, the so-called {\it relaxed Poisson profile approximations}.
Roughly, the idea was to take some {\it delay} into account for the prediction of $M$ based on $X_0$. For $A=1$, this results (see \cite[(5.3b)]{EPW09}) in 
\begin{equation} \label{MRPPA}
M(X_0):= \theta+\frac 1{e-1}\left(1-\frac{X_0}{\pi_0}\right),
\end{equation}
which then is plugged into the upward jump rate \eqref{bestupclass1} in place of $M$. In Figure \ref{fig:6R} we compare the quality of the PPA and RPPA(1) approximations for the rate of the classical ratchet in the light of simulations of our Moran model.
\end{rem}

 \subsection{On the expected time to extinction of the best class} 
 In this subsection we focus on the birth-and-death process $Y:=Y^{\rm T}_0$ with jump rates \eqref{bestdown} and~\eqref{bestup}. As observed in Remark \ref{parallels}, this process has the same dynamics as the process $Y_{\rm PPA}$ defined in Section~\ref{secPPA}, provided the mutation rates are equal and the selection coefficients are translated through the ``dictionary'' ~\eqref{rhodelta}.
\begin{rem} \label{remrttandrtc} Before turning to a rigorous analysis, let us give a heuristics for the long-term behaviour of $Y$, which also points towards (RTT) as well as part of (RTC). The rates  \eqref{bestdown} and \eqref{bestup} display 3 parts: the fluctuation terms $\pm \frac n2\left(1-\frac nN\right)$, the net linear birth rate $n \frac m\rho(1-\rho)$ and the quadratic death rate $\frac m\rho \frac {n^2}N$. The center of attraction of $Y$ (which we encountered already in \eqref{centattY}) is that (asymptotic) value of $n$ for which the net linear birth rate equals the quadratic death rate and thus equals
$$\mathfrak a = N(1-\rho).$$
As long as $Y$ is below $\mathfrak a /2$, it is stochastically bounded from below by a binary Galton-Watson process $Y^{\ell}$ with supercriticality $m (1-\rho)/2$, and stochastically bounded from above by a binary Galton-Watson process $Y^{u}$ with supercriticality $m (1-\rho)$. By Haldane's formula (which in this case coincides with the formula for the escape probability of a simple random walk with constant drift), the survival probability of the offspring of one individual in $Y^{\ell}$ (resp $Y^u$) is $\sim N^{-\beta-\delta}$ (resp. $\sim  2N^{-\beta-\delta}$). Hence the probability that $Y$ when starting in $\mathfrak a/4$ hits $0$ before reaching $\mathfrak a/2$, is asymptotically between $\left(1-2N^{-\beta-\delta}\right)^{N^{1-\delta}/4}$ and $\left(1-N^{-\beta-\delta}\right)^{N^{1-\delta}/4}$, which converge to $0$ if and only if $1-\beta -2\delta > 0$, i.e. $\gamma > \frac 12$. In this case the number of excursions which $Y$ makes from $\mathfrak a/4$ up to $\mathfrak a/2$ before going extinct is geometric with expectation asymptotically between $\exp\left(\frac 14 N^{1-\beta-2\delta}\right)$ and $\exp\left(\frac 12N^{1-\beta-2\delta}\right)$. 
 This gives an intuitive explanation why $\gamma = \frac 12$ is the bound between the exponential and the polynomial regime, and also sheds light on the result of Theorem \ref{mainth} 
 
In the case $\gamma > \frac 12$, the center of attraction  plays a negligible role. What becomes relevant then is that threshold for $n$ above which the quadratic death rate $\frac m\rho \frac {n^2}N$ becomes large. Obviously, the order of magnitude of this threshold is $\sqrt{\frac Nm} =N^{\frac {1+\beta}2}$. Above this threshold,  $Y$ is strongly pushed downwards, making the time spent above the threshold negligible. Below the threshold, $Y$ behaves similar to a (driftless linear) birth-and-death process with upward and downward jump rates \eqref{bestdown}.  This gives a qualitative explanation of the orders of magnitude of the expected times to extinction that are obtained in Theorem~\ref{mainth} also for the polynomial regime.
\end{rem}
 \begin{figure}[t]
  \begin{subfigure}{0.45\textwidth}
  \centering
     {
%
%
\definecolor{mycolor1}{rgb}{0.00000,0.44700,0.74100}%
\definecolor{mycolor2}{rgb}{0.85000,0.32500,0.09800}%
\definecolor{mycolor3}{rgb}{0.92900,0.69400,0.12500}%
\begin{tikzpicture}[scale=0.4]

\begin{axis}[%
width=6.028in,
height=4.754in,
at={(1.011in,0.642in)},
scale only axis,
xmode=log,
xmin=500,
xmax=10000,
xminorticks=true,
xlabel style={font=\color{white!15!black}},
xlabel={\scalebox{2}{$N$}},
ymode=log,
ymin=87.2553635649065,
ymax=10000,
yminorticks=true,
axis background/.style={fill=white},
title style={below right,at={(0,0.75)},font=\bfseries},
align =center,
title={\scalebox{2.5}{$\gamma =0.7$,}\\\\\scalebox{2.5}{ $\beta =0.6$, 
$\delta= 0.28$}},
legend style={at={(0.03,0.97)}, anchor=north west, legend cell align=left, align=left, draw=white!15!black}
]
\addplot [color=mycolor1, line width=2.0pt, mark=x, mark size=8pt,mark options={solid, mycolor1}]
  table[row sep=crcr]{%
500	447.46198032435\\
1000	783.948231694432\\
2000	1305.34108460813\\
3000	1802.41693941717\\
4000	2258.39524318728\\
5000	2605.63089462677\\
10000	4563.81707897655\\
};
\addlegendentry{\scalebox{1.8}{avg. observed interclick times}}
\addplot [color=mycolor2, dotted, line width=2.0pt]
  table[row sep=crcr]{%
500	87.2553635649065\\
1000	159.755907309135\\
2000	289.552113250593\\
3000	408.397251358191\\
4000	520.438967634323\\
5000	627.588365753196\\
10000	1117.88122633073\\
};
\addlegendentry{\scalebox{1.5}{$ 2N^{1-\delta} (\log \sqrt{N/m}- \log N^{1-\delta}) $}}
\addplot [color=mycolor3, dashed, line width=2.0pt]
  table[row sep=crcr]{%
500	401.67131385448\\
1000	699.350377071824\\
2000	1217.63972940247\\
3000	1684.19284172473\\
4000	2120.03390464612\\
5000	2534.37465543493\\
10000	4412.6025677846\\
};
\addlegendentry{\scalebox{1.5}{$ \sqrt{N/m} \pi^{ \frac32}/2  $}}
\end{axis}
\end{tikzpicture}%
}
    \caption{
    }\label{subfig2b}
  \end{subfigure}
   \begin{subfigure}{0.45\textwidth}
  \centering
    {
%
%
\definecolor{mycolor1}{rgb}{0.00000,0.44700,0.74100}%
\definecolor{mycolor2}{rgb}{0.85000,0.32500,0.09800}%
\definecolor{mycolor3}{rgb}{0.92900,0.69400,0.12500}%
\begin{tikzpicture}[scale = 0.4] 
\begin{axis}[%
width=6.028in,
height=4.754in,
at={(1.011in,0.642in)},
scale only axis,
xmode=log,
xmin=500,
xmax=10000,
xminorticks=true,
xlabel style={font=\color{white!15!black}},
xlabel={\scalebox{2}{$N$}},
ymode=log,
ymin=10000,
ymax=138276.242996473,
yminorticks=true,
axis background/.style={fill=white},
title style={below right,at={(0,0.75)},font=\bfseries},
align =center,
title={\scalebox{2.5}{$\gamma =0.4$,}\\\\\scalebox{2.5}{ $\beta =0.8$, 
$\delta= 0.08$}},
legend style={at={(0.03,0.97)}, anchor=north west, legend cell align=left, align=left, draw=white!15!black}
]
\addplot [color=mycolor1, line width=2.0pt, mark=x, mark options={solid, mycolor1}]
  table[row sep=crcr]{%
500	10688.190382791\\
1000	16906.3160836274\\
2000	26698.9133341433\\
3000	35615.6033918222\\
4000	45079.0551541157\\
5000	52785.7531334099\\
10000	90031.0639383355\\
};
\addlegendentry{\scalebox{1.8}{avg. observed interclick times}}

\addplot [color=mycolor2, dashed, line width=2.0pt]
  table[row sep=crcr]{%
500	9247.84426817634\\
1000	14292.667668884\\
2000	22683.4742837356\\
3000	30086.1306945744\\
4000	36950.3243536778\\
5000	43455.4709300165\\
10000	72928.6162154962\\
};
\addlegendentry{\scalebox{1.5}{(3.13) with $j_N = N^{1-\delta}$}}
\addplot [color=mycolor3, dashed, line width=2.0pt]
  table[row sep=crcr]{%
500	9259.04473392658\\
1000	14317.455701037\\
2000	22735.6033918652\\
3000	30165.3683937354\\
4000	37056.3789303356\\
5000	43588.0655220037\\
10000	73190.5221218325\\
};
\addlegendentry{\scalebox{1.5}{(3.13)  with $j_N = N^{ 1- \frac{\delta}{2} }$}}
\end{axis}
\end{tikzpicture}%
}
    \caption{
    }\label{subfig2a}
  \end{subfigure}
   %
  \caption{This is an illustration of the {\it Rule of Thumb for the tournament ratchet} (RTT) in the light of Theorem ~\ref{mainth}. 
Each data point was obtained  by pooling the interclick times no.~50 to 150 from 100~simulations of the tournament ratchet for the corresponding value of $N$. Here, in panel~(\subref{subfig2b}) $(\beta,\delta) = (0.6,0.28)$, which belongs to the polynomial regime~$\mathcal P$, and in panel~(\subref{subfig2a})  $(\beta,\delta) = (0.8,0.08)$, which belongs to the exponential regime~$\mathcal E$.  Each panel shows two predictions based on the asymptotics of Theorem~\ref{mainth}, using the initial values $\mathfrak a = N^{1-\delta}$ and $\mathfrak b = N^{1-\delta/2}$, respectively. In the exponential regime the predictions using $\mathfrak a$ and $\mathfrak b$, respectively, are virtually indistinguishable, while in the polynomial regime the prediction using~$\mathfrak b$ is by far better than the one using~$\mathfrak a$.}\label{fig:2}
\end{figure}
 The proof of the following theorem is the content of
 Section~\ref{secGreenproof}. This proof relies on an asymptotic analysis of the Green function represented by formula \eqref{lem:green}. The fit of a numerical calculation of the Green function based on this formula with the empirical occupation times of the size of the best class of the tournament ratchet is displayed in Figure \ref{fig:green}. A heuristic explanation of the orders obtained in Theorem \ref{mainth} has been given in Remark \ref{remrttandrtc}.

  In the following we use the notation
  \begin{equation}
  \label{ll}
      f(N) \ll g(N) \Longleftrightarrow \lim\limits_{N\rightarrow \infty} \frac{f(N)}{g(N)}=0.
  \end{equation}
  Also, we will usually suppress the $N$-dependence in the notation, as for example in $Y$, $m$ and $\rho$  in the following theorem. Note that this theorem comprises a larger regime than the one  described by the $(\beta,\delta)$-scaling for $(\beta,\delta)\in \Delta$, see \eqref{defDelta}.
 \begin{thm} \label{mainth} Let $T_0$ be the extinction time of the birth-and-death process $Y$ with jump rates \eqref{bestdown} and~\eqref{bestup}, let $1\gg m \gg \frac 1N$, and let $\rho$ be a sequence in $[\rho_0,1)$ for some fixed $\rho_0 \in (0,1)$. 
 \begin{enumerate}[a)] 
     \item\label{th:mainth:a} {\rm [Polynomial regime]} Assume $ Nm(1-\rho)^2  \to 0$ as $N \to \infty$. Let $(j_N)$ be a sequence of natural numbers in $[N]$. If $j_N \ll \sqrt{\frac{N/m}{
     {\log(N/m)}}}$, then
     \begin{equation}\label{polysmall}
\E_{j_N}[T_0]\sim 2j_N\left(\log \sqrt{\frac Nm}-\log j_N\right),
     \end{equation}
     whereas if 
     {$j_N \gg \sqrt{\frac Nm}$}, 
     then
     \begin{equation}\label{polylarge}
         \E_{j_N}[T_0] \sim \frac{\pi^{3/2}}{2}\sqrt{\frac{N}{m}}.  
     \end{equation}
     The expected number of returns of the process $Y$ to $\lceil \mathfrak a \rceil $, when starting above $\mathfrak a=(1-\rho)N$, is asymptotically equivalent to $\frac{1}{m (1-\rho)}$ as $N\rightarrow\infty$.
\item \label{th:mainth:b} {\rm [Exponential regime]} Assume  $ Nm(1-\rho)^2  \to \infty$ and $1 \ll j_N\le N$ as $N \to \infty$. 
Then\footnote{This corrects a typo from the previous version v2 (published as~\cite{IGSW}) where the factor $\frac 1{1-\rho}$ was missing in formula~\eqref{expclr}.}
\begin{equation}\label{expclr}
   \E_{j_N}[T_0] \sim  \left(1-\exp\left(-2m\left(\frac 1\rho-1\right)j_N\right)\right) \frac 1{1-\rho}\sqrt{\frac{\pi}{m N}}\, v_N, 
\end{equation}
with
\begin{equation}\label{defvN}
v_N:= \frac{1}{m\left(\frac 1\rho-1\right)}\exp   \left( 2Nm(1-\rho)^2\eta(m,\rho) \right),
\end{equation}
\begin{equation*}
\eta(m,\rho):=  -\frac{1}{2m} \left[\frac{1}{1-\rho}\log \left( \frac{1+2m}{1+2m/\rho}  \right)+ \sum_{\ell=1}^{\infty} \left(1- \frac{1}{(1+2m)^\ell}\right) \frac{(1-\rho)^{\ell-1}}{\ell(\ell+1)}\right]. \end{equation*}
 In particular, with 
\begin{equation}\label{defeN}
e_N:= \frac 1{1-\rho}\sqrt{\frac{\pi}{m N}}\, v_N
\end{equation}
one has
\begin{equation}
\label{threecases}\E_{j_N}[T_0] \sim\begin{cases}e_N \phantom{AAAAAAAAAAAAi}\mbox{ if } j_N \gg \frac 1{m(1-\rho)}
\\
e_N(1-\exp(-2C/\rho)) 
\phantom{AA}\mbox{ if } j_N \sim \frac C{m(1-\rho)}
\\
e_N\, 2 j_N\,m(1/\rho-1) \phantom{AAAi}\mbox{ if } j_N \ll \frac 1{m(1-\rho)}\end{cases}.
\end{equation}
The expected number of returns of the process $Y$ to  $\lfloor \mathfrak a\rfloor$, when starting above   $\ \mathfrak a = (1-\rho)N$, is asymptotically equivalent to \eqref{defvN} as $N\to \infty$.
 \end{enumerate}
 \end{thm}
    \begin{rem} \label{missing}
 \begin{enumerate}[a)]
     \item \label{remark36a} Theorem \ref{mainth} constitutes an essential step on the way to a proof of the claim (RTT) formulated in \eqref{RT3} and \eqref{RT4}.
One way to complete this proof could lead via the analysis of 
the system $(Y_0^{N}, Y_1^{N})$ 
of the 
sizes of the best and the second-best class of the tournament ratchet; recall that this system is autonomous up to the time of extinction of its first component. Then, $Y_0^N(0)$ and $Y_1^N(0)$  stand for the (random) sizes of the {\it new} best and second best class at the time of a click. With $T_0^N$ denoting the extinction time of $Y_0^N$, we conjecture that both  in the polynomial and in the exponential regime $Y_1^N(T_0^N)$ will with high probability be $\gg \sqrt{\frac Nm}$, provided that both $Y_0^N(0)$ and $Y_1^N(0)$ are $\gg \sqrt{\frac Nm}$. 
 \item \label{remark36b} While the present work focuses on a forward-in-time approach, an alternative route for proving (RTT) is provided by the backward-in-time approach that was developed in \cite{GSW23} in terms of a hierarchical duality for the tournament ratchet. This requires the extension of the backward-in-time analysis  from the subcritical to the near-critical regime, and will be a subject of future research.
 \end{enumerate}
 \end{rem}
 \begin{rem} \label{remcentatt}
 \begin{enumerate}[a)]
\item Theorem \ref{mainth} \ref{th:mainth:b} suggests the conjecture that not only in the exponential regime of the near-critical case $\rho \uparrow 1$, but also in the entire subcritical case $\rho < 1$ the rate of the tournament ratchet is asymptotically equivalent to \eqref{defeN}. 
This would improve the logarithmic equivalence \eqref{subcras}  obtained in \cite[Theorem~2.2]{GSW23} to an asymptotic equivalence. Here it is worth noticing that (as we will show at the end of Section~\ref{rR})  the
   exponents in ~\eqref{subcras} and  \eqref{defvN} obey for all $\rho < 1$
     \begin{equation}
     \label{bridgeasympt}
         (1-\rho)^2\eta(m,\rho)\sim  \frac 1\rho-1+\log\rho \quad \mbox{ as } m \to 0.
     \end{equation}
    
     \item \label{rem:remcentatt_b} In the light of Remark \ref{parallels}, Theorem \ref{mainth} is relevant not only for the tournament ratchet, but also for the Poisson profile approximation of the classical ratchet.  Prominent starting values for $Y$ are 

      -- with regard to the classical ratchet:  $n_0^C := N \pi_1 = N \theta e^{-\theta}$, which in the $(\beta, \delta)$-scaling equals $N^{1-\delta}\delta \log N$,

     -- with regard to the tournament ratchet:  $n_0^T := N\sqrt{1-\rho}$, which according to~\eqref{centattY}   is the asymptotic center of attraction of the size of its second best class, and in the  $(\beta, \delta)$-scaling equals $N^{1-\delta/2}$. 
     
     Figures \ref{fig:4R} and \ref{fig:4F}
     illustrate that these asymptotics of the starting values can indeed be seen in simulations of the classical and the tournament ratchet. The starting values~$n_0^C$ and $n_0^T$  are used in  Figures~\ref{fig:6R} and   \ref{fig:6F}.

     For $(\beta, \delta) \in \mathcal P$ we have 
     $$ 1-\delta < \frac{1+\beta}2 < 1-\tfrac{\delta} 2.$$
     Hence Theorem \ref{mainth}.\ref{th:mainth:a} gives, in accordance with \eqref{RT3} and \eqref{RT1},
     \begin{equation*}
      \E_{n_0^T}[T_0] \asymp N^{\frac{1+\beta}2} \qquad \mbox {and} \qquad  \E_{n_0^C}[T_0] \asymp N^{1-\delta}.   
     \end{equation*} 
     \item Recalling that $m =  \rho s $ with $\rho < 1$ (and all these parameters depending on $N$),      the difference of the upward and downward jump rates \eqref{bestup} and \eqref{bestdown} is 
$$\lambda_n-\mu_n = n\left(s-m-s\frac n{N}\right)$$ and their sum is $\lambda_n+\mu_n\sim~n$ as long as $n\ll N$. Hence the dynamics of $Y^N$ (although its state space is $\{0,1,\ldots, N\}$ rather than $\N_0$) bears similarities to that of a logistic branching process. Indeed, we conjecture that a logistic branching process $\widehat Y^N$ with upward and downward  jump rates $\widehat\lambda_n$ and $\widehat \mu_n$ given by
\begin{equation}\label{logbran}\widehat\lambda_n = n\left(\frac 12 + s\right), \qquad \widehat\mu_n = n\left(\frac 12 + m+s\frac n{N}\right)
\end{equation}
will exhibit very similar asymptotics of the expected times to extinction as those obtained for the process $Y^N$ in Theorem~\ref{mainth}. This would complement results of  \cite{sagitov2013extinction} and \cite{chazottes2016sharp}, both of which do not cover the parameter regime given by  \eqref{logbran}. The paper \cite{sagitov2013extinction} considers jump rates of the form $\widehat\lambda_n = ns$, $\widehat\mu_n = nm + n(n-1)\theta $
with constant $s,m$ and small $\theta$; this corresponds to \eqref{logbran} but without the  fluctuation terms $\tfrac 12$ which are of dominant order in  \eqref{logbran}. (For conceptual clarification we point out that  \cite{sagitov2013extinction} addresses the case of a constant ratio $s/m > 1$ as supercritical, while in our context this corresponds to a subcritical mutation-selection ratio.)
      The paper \cite{chazottes2016sharp} considers quasi-equilibria and  extinction times of a class of birth-and-death processes that is more general than logistic branching processes, but imposes a scaling condition of the dynamics which is not fulfilled by  \eqref{bestdown} and~\eqref{bestup}. Still, both papers point to interesting  routes which may offer alternatives to our way of proving Theorem \ref{mainth}.  
     \end{enumerate}\end{rem}
 \begin{figure}[ht]
  \centering
  \begin{subfigure}{0.45\textwidth}
    \centering
\input{\currfiledir tikz/0_REVBrealandfancy_y0_beta_0.5_delta_0.1_N_100.tex}
  \caption{}\label{fig7f:a}
\end{subfigure}
\begin{subfigure}{0.45\textwidth}
  \centering
 \input{\currfiledir tikz/0_REVBrealandfancy_y0_beta_0.7_delta_0.2_N_500.tex}
  \caption{}\label{fig7f:b}
  \end{subfigure}
  \caption{The empirical occupation times of the size of the best class in a simulation of the tournament ratchet are 
  compared to the  Green functions $G(\mathfrak a, \cdot)$, $G(\mathfrak b, \cdot)$, which are computed numerically using formula \eqref{generalgreen1}. Panels (\subref{fig7f:a}) and (\subref{fig7f:b}) feature the exponential and the polynomial  regime, respectively, with $\gamma = 0.2$ in panel (\subref{fig7f:a}) and $\gamma = \tfrac23$   in panel (\subref{fig7f:b}). In panel~(\subref{fig7f:b}) the population size is $N=500$ and simulations were run up to the first $10^4 +1$ clicks, where the first click was ignored. In (\subref{fig7f:a}),  101 clicks were observed and the first one ignored. Here the population size was $N=100$. See \cite[Figure~5]{EPW09} for similar plots concerning the classical ratchet.}\label{fig:green}
\end{figure}
 \FloatBarrier
 \begin{figure}[ht]
   \centering
%
%
\definecolor{mycolor1}{rgb}{0.00000,0.44700,0.74100}%
\definecolor{mycolor2}{rgb}{0.85000,0.32500,0.09800}%
\definecolor{mycolor3}{rgb}{0.92900,0.69400,0.12500}%
\begin{tikzpicture}[scale=0.4]

\begin{axis}[%
width=6.028in,
height=4.754in,
at={(1.011in,0.642in)},
scale only axis,
xmin=0,
xmax=0.9,
xlabel style={font=\color{white!15!black}},
xlabel={\scalebox{1.5}{$\beta$}},
ymin=5,
ymax=11,
axis background/.style={fill=white},
title style={font=\bfseries},
title={\scalebox{2.5}{$\gamma=0.55$}},
axis x line*=bottom,
axis y line*=left,
legend style={at={(0.97,0.03)}, anchor=south east, legend cell align=left, align=left, draw=white!15!black}
]
\addplot [color=mycolor1, line width=2.0pt, mark=x, mark options={solid, mycolor1}]
  table[row sep=crcr]{%
0.05	6.09187968142371\\
0.15	6.64801506867117\\
0.25	7.2647624754112\\
0.35	7.9006793629806\\
0.45	8.53718132186763\\
0.55	9.13379757413336\\
0.65	9.67163454919853\\
0.75	10.2317321870145\\
0.85	10.722964809478\\
};
\addlegendentry{\scalebox{1.5}{$\log$ size of new best class}}
\addplot [color=mycolor2, dotted, line width=2.0pt]
  table[row sep=crcr]{%
0.05	7.29176197206217\\
0.06	7.34450095278897\\
0.07	7.39712675372956\\
0.08	7.44963692768268\\
0.09	7.50202894720782\\
0.1	7.55430020107857\\
0.11	7.60644799053779\\
0.12	7.65846952534119\\
0.13	7.7103619195749\\
0.14	7.76212218723116\\
0.15	7.81374723752531\\
0.16	7.86523386993564\\
0.17	7.91657876894626\\
0.18	7.96777849847125\\
0.19	8.01882949593677\\
0.2	8.06972806599555\\
0.21	8.12047037384603\\
0.22	8.17105243812594\\
0.23	8.22147012334736\\
0.24	8.27171913183735\\
0.25	8.32179499514466\\
0.26	8.37169306486986\\
0.27	8.42140850287142\\
0.28	8.47093627079642\\
0.29	8.52027111887901\\
0.3	8.56940757394439\\
0.31	8.61833992654963\\
0.32	8.66706221718581\\
0.33	8.71556822145801\\
0.34	8.7638514341508\\
0.35	8.81190505207735\\
0.36	8.85972195559872\\
0.37	8.90729468868792\\
0.38	8.95461543739882\\
0.39	9.00167600658437\\
0.4	9.0484677946905\\
0.41	9.09498176643145\\
0.42	9.14120842312949\\
0.43	9.18713777047495\\
0.44	9.23275928343289\\
0.45	9.27806186798755\\
0.46	9.32303381937669\\
0.47	9.36766277642187\\
0.48	9.41193567150851\\
0.49	9.45583867570875\\
0.5	9.49935713846991\\
0.51	9.54247552120972\\
0.52	9.58517732406432\\
0.53	9.62744500492383\\
0.54	9.6692598897602\\
0.55	9.71060207309876\\
0.56	9.75145030730404\\
0.57	9.79178187913668\\
0.58	9.83157247178382\\
0.59	9.87079601026209\\
0.6	9.90942448772906\\
0.61	9.94742776980211\\
0.62	9.98477337345618\\
0.63	10.0214262164314\\
0.64	10.0573483323006\\
0.65	10.0924985453912\\
0.66	10.1268320985753\\
0.67	10.160300225483\\
0.68	10.1928496568735\\
0.69	10.2244220486163\\
0.7	10.2549533158506\\
0.71	10.2843728542323\\
0.72	10.3126026244784\\
0.73	10.3395560703648\\
0.74	10.3651368324393\\
0.75	10.3892372093434\\
0.76	10.4117363048804\\
0.77	10.432497780519\\
0.78	10.4513671080055\\
0.79	10.4681681824279\\
0.8	10.4826991083158\\
0.81	10.4947269039856\\
0.82	10.5039807727727\\
0.83	10.5101434489901\\
0.84	10.512839917231\\
0.85	10.5116224861507\\
};
\addlegendentry{\scalebox{1.5}{$\log( N^{1-\delta}\delta \log N)$}}
\addplot [color=mycolor3, dashed, line width=2.0pt]
  table[row sep=crcr]{%
0.05	5.49742190952328\\
0.06	5.56074299958062\\
0.07	5.62406408963796\\
0.08	5.68738517969529\\
0.09	5.75070626975263\\
0.1	5.81402735980996\\
0.11	5.8773484498673\\
0.12	5.94066953992464\\
0.13	6.00399062998197\\
0.14	6.06731172003931\\
0.15	6.13063281009665\\
0.16	6.19395390015398\\
0.17	6.25727499021132\\
0.18	6.32059608026865\\
0.19	6.38391717032599\\
0.2	6.44723826038333\\
0.21	6.51055935044066\\
0.22	6.573880440498\\
0.23	6.63720153055534\\
0.24	6.70052262061267\\
0.25	6.76384371067001\\
0.26	6.82716480072735\\
0.27	6.89048589078468\\
0.28	6.95380698084202\\
0.29	7.01712807089935\\
0.3	7.08044916095669\\
0.31	7.14377025101403\\
0.32	7.20709134107136\\
0.33	7.2704124311287\\
0.34	7.33373352118604\\
0.35	7.39705461124337\\
0.36	7.46037570130071\\
0.37	7.52369679135804\\
0.38	7.58701788141538\\
0.39	7.65033897147272\\
0.4	7.71366006153005\\
0.41	7.77698115158739\\
0.42	7.84030224164473\\
0.43	7.90362333170206\\
0.44	7.9669444217594\\
0.45	8.03026551181673\\
0.46	8.09358660187407\\
0.47	8.15690769193141\\
0.48	8.22022878198874\\
0.49	8.28354987204608\\
0.5	8.34687096210341\\
0.51	8.41019205216075\\
0.52	8.47351314221809\\
0.53	8.53683423227542\\
0.54	8.60015532233276\\
0.55	8.6634764123901\\
0.56	8.72679750244743\\
0.57	8.79011859250477\\
0.58	8.85343968256211\\
0.59	8.91676077261944\\
0.6	8.98008186267678\\
0.61	9.04340295273412\\
0.62	9.10672404279145\\
0.63	9.17004513284879\\
0.64	9.23336622290612\\
0.65	9.29668731296346\\
0.66	9.3600084030208\\
0.67	9.42332949307813\\
0.68	9.48665058313547\\
0.69	9.5499716731928\\
0.7	9.61329276325014\\
0.71	9.67661385330748\\
0.72	9.73993494336481\\
0.73	9.80325603342215\\
0.74	9.86657712347949\\
0.75	9.92989821353682\\
0.76	9.99321930359416\\
0.77	10.0565403936515\\
0.78	10.1198614837088\\
0.79	10.1831825737662\\
0.8	10.2465036638235\\
0.81	10.3098247538808\\
0.82	10.3731458439382\\
0.83	10.4364669339955\\
0.84	10.4997880240528\\
0.85	10.5631091141102\\
};
\addlegendentry{\scalebox{1.5}{$\log( N^{1-\delta})$}}
\end{axis}
\end{tikzpicture}%
%
%
\definecolor{mycolor1}{rgb}{0.00000,0.44700,0.74100}%
\definecolor{mycolor2}{rgb}{0.85000,0.32500,0.09800}%
\definecolor{mycolor3}{rgb}{0.92900,0.69400,0.12500}%
\begin{tikzpicture}[scale=0.4]

\begin{axis}[%
width=6.028in,
height=4.754in,
at={(1.011in,0.642in)},
scale only axis,
xmin=0,
xmax=0.9,
xlabel style={font=\color{white!15!black}},
xlabel={\scalebox{1.5}{$\beta$}},
ymin=4,
ymax=11,
axis background/.style={fill=white},
title style={font=\bfseries},
title={\scalebox{2.5}{$\gamma=0.6$}},
axis x line*=bottom,
axis y line*=left,
legend style={at={(0.97,0.03)}, anchor=south east, legend cell align=left, align=left, draw=white!15!black}
]
\addplot [color=mycolor1, line width=2.0pt, mark=x, mark options={solid, mycolor1}]
  table[row sep=crcr]{%
0.05	5.77393621054081\\
0.15	6.33555194541294\\
0.25	7.01929493360898\\
0.35	7.66901661173496\\
0.45	8.36305398588212\\
0.55	8.99565160304676\\
0.65	9.61269860746795\\
0.75	10.1519215864613\\
0.85	10.6979874962892\\
};
\addlegendentry{\scalebox{1.5}{$\log$ size of new best class}}

\addplot [color=mycolor2, dashed, line width=2.0pt]
  table[row sep=crcr]{%
0.05	6.83190938946571\\
0.06	6.890404832925\\
0.07	6.94878709659807\\
0.08	7.00705373328368\\
0.09	7.06520221554131\\
0.1	7.12322993214454\\
0.11	7.18113418433624\\
0.12	7.23891218187213\\
0.13	7.29656103883833\\
0.14	7.35407776922707\\
0.15	7.4114592822537\\
0.16	7.46870237739652\\
0.17	7.52580373913963\\
0.18	7.5827599313971\\
0.19	7.63956739159511\\
0.2	7.69622242438637\\
0.21	7.75272119496934\\
0.22	7.80905972198173\\
0.23	7.86523386993564\\
0.24	7.92123934115811\\
0.25	7.97707166719791\\
0.26	8.03272619965559\\
0.27	8.08819810038963\\
0.28	8.14348233104712\\
0.29	8.1985736418622\\
0.3	8.25346655966007\\
0.31	8.30815537499779\\
0.32	8.36263412836646\\
0.33	8.41689659537114\\
0.34	8.47093627079642\\
0.35	8.52474635145545\\
0.36	8.57831971770931\\
0.37	8.63164891353099\\
0.38	8.68472612497437\\
0.39	8.73754315689241\\
0.4	8.79009140773102\\
0.41	8.84236184220446\\
0.42	8.89434496163498\\
0.43	8.94603077171293\\
0.44	8.99740874740335\\
0.45	9.0484677946905\\
0.46	9.09919620881212\\
0.47	9.14958162858979\\
0.48	9.19961098640892\\
0.49	9.24927045334164\\
0.5	9.29854537883528\\
0.51	9.34742022430758\\
0.52	9.39587848989467\\
0.53	9.44390263348666\\
0.54	9.49147398105551\\
0.55	9.53857262712656\\
0.56	9.58517732406432\\
0.57	9.63126535862945\\
0.58	9.67681241400907\\
0.59	9.72179241521983\\
0.6	9.76617735541928\\
0.61	9.80993710022481\\
0.62	9.85303916661138\\
0.63	9.89544847231904\\
0.64	9.93712705092074\\
0.65	9.97803372674387\\
0.66	10.0181237426604\\
0.67	10.0573483323006\\
0.68	10.0956542264236\\
0.69	10.1329830808989\\
0.7	10.1692708108657\\
0.71	10.2044468119799\\
0.72	10.2384330449584\\
0.73	10.2711429535774\\
0.74	10.3024801783843\\
0.75	10.3323370180209\\
0.76	10.3605925762904\\
0.77	10.3871105146615\\
0.78	10.4117363048804\\
0.79	10.4342938420354\\
0.8	10.4545812306558\\
0.81	10.472365489058\\
0.82	10.4873758205776\\
0.83	10.4992949595275\\
0.84	10.5077478905008\\
0.85	10.5122869221531\\
};
\addlegendentry{\scalebox{1.5}{$\log( N^{1-\delta}\delta \log N)$}}
\addplot [color=mycolor3, dashed, line width=2.0pt]
  table[row sep=crcr]{%
0.05	4.9505579499372\\
0.06	5.01963550272702\\
0.07	5.08871305551684\\
0.08	5.15779060830666\\
0.09	5.22686816109648\\
0.1	5.2959457138863\\
0.11	5.36502326667613\\
0.12	5.43410081946595\\
0.13	5.50317837225577\\
0.14	5.57225592504559\\
0.15	5.64133347783541\\
0.16	5.71041103062523\\
0.17	5.77948858341505\\
0.18	5.84856613620488\\
0.19	5.9176436889947\\
0.2	5.98672124178452\\
0.21	6.05579879457434\\
0.22	6.12487634736416\\
0.23	6.19395390015398\\
0.24	6.2630314529438\\
0.25	6.33210900573363\\
0.26	6.40118655852345\\
0.27	6.47026411131327\\
0.28	6.53934166410309\\
0.29	6.60841921689291\\
0.3	6.67749676968273\\
0.31	6.74657432247255\\
0.32	6.81565187526238\\
0.33	6.8847294280522\\
0.34	6.95380698084202\\
0.35	7.02288453363184\\
0.36	7.09196208642166\\
0.37	7.16103963921148\\
0.38	7.2301171920013\\
0.39	7.29919474479112\\
0.4	7.36827229758095\\
0.41	7.43734985037077\\
0.42	7.50642740316059\\
0.43	7.57550495595041\\
0.44	7.64458250874023\\
0.45	7.71366006153005\\
0.46	7.78273761431987\\
0.47	7.8518151671097\\
0.48	7.92089271989952\\
0.49	7.98997027268934\\
0.5	8.05904782547916\\
0.51	8.12812537826898\\
0.52	8.1972029310588\\
0.53	8.26628048384862\\
0.54	8.33535803663844\\
0.55	8.40443558942827\\
0.56	8.47351314221809\\
0.57	8.54259069500791\\
0.58	8.61166824779773\\
0.59	8.68074580058755\\
0.6	8.74982335337737\\
0.61	8.8189009061672\\
0.62	8.88797845895702\\
0.63	8.95705601174684\\
0.64	9.02613356453666\\
0.65	9.09521111732648\\
0.66	9.1642886701163\\
0.67	9.23336622290612\\
0.68	9.30244377569595\\
0.69	9.37152132848577\\
0.7	9.44059888127559\\
0.71	9.50967643406541\\
0.72	9.57875398685523\\
0.73	9.64783153964505\\
0.74	9.71690909243487\\
0.75	9.78598664522469\\
0.76	9.85506419801452\\
0.77	9.92414175080434\\
0.78	9.99321930359416\\
0.79	10.062296856384\\
0.8	10.1313744091738\\
0.81	10.2004519619636\\
0.82	10.2695295147534\\
0.83	10.3386070675433\\
0.84	10.4076846203331\\
0.85	10.4767621731229\\
};
\addlegendentry{\scalebox{1.5}{$\log( N^{1-\delta})$}}
\end{axis}
\end{tikzpicture}%
%
%
\definecolor{mycolor1}{rgb}{0.00000,0.44700,0.74100}%
\definecolor{mycolor2}{rgb}{0.85000,0.32500,0.09800}%
\definecolor{mycolor3}{rgb}{0.92900,0.69400,0.12500}%
\begin{tikzpicture}[scale=0.4]

\begin{axis}[%
width=6.028in,
height=4.754in,
at={(1.011in,0.642in)},
scale only axis,
xmin=0,
xmax=0.8,
xlabel style={font=\color{white!15!black}},
xlabel={\scalebox{1.5}{$\beta$}},
ymin=3,
ymax=11,
axis background/.style={fill=white},
title style={font=\bfseries},
title={\scalebox{2.5}{$\gamma=0.7$}},
axis x line*=bottom,
axis y line*=left,
legend style={at={(0.97,0.03)}, anchor=south east, legend cell align=left, align=left, draw=white!15!black}
]
\addplot [color=mycolor1, line width=2.0pt, mark=x, mark options={solid, mycolor1}]
  table[row sep=crcr]{%
0.05	5.2517497307317\\
0.15	5.88372406482289\\
0.25	6.56567658628485\\
0.35	7.2832844368143\\
0.45	8.03642412998171\\
0.55	8.74562182894885\\
0.65	9.43436491971444\\
0.75	10.039093210653\\
0.85	10.6074985383219\\
};
\addlegendentry{\scalebox{1.5}{$\log$ size of new best class}}
\addplot [color=mycolor2, dotted, line width=2.0pt]
  table[row sep=crcr]{%
0.05	5.8923321501208\\
0.06	5.96234051904506\\
0.07	6.0322357081831\\
0.08	6.10201527033367\\
0.09	6.17167667805628\\
0.1	6.24121732012448\\
0.11	6.31063449778115\\
0.12	6.37992542078201\\
0.13	6.44908720321318\\
0.14	6.51811685906689\\
0.15	6.58701129755849\\
0.16	6.65576731816628\\
0.17	6.72438160537436\\
0.18	6.7928507230968\\
0.19	6.86117110875978\\
0.2	6.92933906701601\\
0.21	6.99735076306395\\
0.22	7.06520221554131\\
0.23	7.13288928896019\\
0.24	7.20040768564763\\
0.25	7.2677529371524\\
0.26	7.33492039507505\\
0.27	7.40190522127407\\
0.28	7.46870237739652\\
0.29	7.53530661367657\\
0.3	7.60171245693941\\
0.31	7.6679141977421\\
0.32	7.73390587657574\\
0.33	7.79968126904539\\
0.34	7.86523386993564\\
0.35	7.93055687605964\\
0.36	7.99564316777847\\
0.37	8.06048528906512\\
0.38	8.12507542597347\\
0.39	8.18940538335648\\
0.4	8.25346655966007\\
0.41	8.31724991959848\\
0.42	8.38074596449397\\
0.43	8.44394470003689\\
0.44	8.50683560119228\\
0.45	8.56940757394439\\
0.46	8.63164891353099\\
0.47	8.69354725877363\\
0.48	8.75508954205773\\
0.49	8.81626193445542\\
0.5	8.87704978541403\\
0.51	8.9374375563513\\
0.52	8.99740874740336\\
0.53	9.05694581646032\\
0.54	9.11603008949414\\
0.55	9.17464166103016\\
0.56	9.23275928343289\\
0.57	9.29036024346298\\
0.58	9.34742022430758\\
0.59	9.40391315098331\\
0.6	9.45981101664773\\
0.61	9.51508368691823\\
0.62	9.56969867876977\\
0.63	9.6236209099424\\
0.64	9.67681241400907\\
0.65	9.72923201529717\\
0.66	9.78083495667871\\
0.67	9.83157247178382\\
0.68	9.88139129137186\\
0.69	9.93023307131207\\
0.7	9.97803372674387\\
0.71	10.024722653323\\
0.72	10.0702218117665\\
0.73	10.1144446458504\\
0.74	10.1572947961224\\
0.75	10.1986645612239\\
0.76	10.2384330449584\\
0.77	10.2764639087944\\
0.78	10.3126026244784\\
0.79	10.3466730870983\\
0.8	10.3784734011836\\
0.81	10.4077705850509\\
0.82	10.4342938420354\\
0.83	10.4577259064502\\
0.84	10.4776917628886\\
0.85	10.4937437200058\\
};
\addlegendentry{\scalebox{1.5}{$\log( N^{1-\delta}\delta \log N)$}}
\addplot [color=mycolor3, dashed, line width=2.0pt]
  table[row sep=crcr]{%
0.05	3.85683003076503\\
0.06	3.93742050901982\\
0.07	4.01801098727461\\
0.08	4.0986014655294\\
0.09	4.17919194378419\\
0.1	4.25978242203898\\
0.11	4.34037290029378\\
0.12	4.42096337854857\\
0.13	4.50155385680336\\
0.14	4.58214433505815\\
0.15	4.66273481331294\\
0.16	4.74332529156773\\
0.17	4.82391576982253\\
0.18	4.90450624807732\\
0.19	4.98509672633211\\
0.2	5.0656872045869\\
0.21	5.14627768284169\\
0.22	5.22686816109648\\
0.23	5.30745863935128\\
0.24	5.38804911760607\\
0.25	5.46863959586086\\
0.26	5.54923007411565\\
0.27	5.62982055237044\\
0.28	5.71041103062523\\
0.29	5.79100150888003\\
0.3	5.87159198713482\\
0.31	5.95218246538961\\
0.32	6.0327729436444\\
0.33	6.11336342189919\\
0.34	6.19395390015398\\
0.35	6.27454437840878\\
0.36	6.35513485666357\\
0.37	6.43572533491836\\
0.38	6.51631581317315\\
0.39	6.59690629142794\\
0.4	6.67749676968273\\
0.41	6.75808724793752\\
0.42	6.83867772619232\\
0.43	6.91926820444711\\
0.44	6.9998586827019\\
0.45	7.08044916095669\\
0.46	7.16103963921148\\
0.47	7.24163011746627\\
0.48	7.32222059572107\\
0.49	7.40281107397586\\
0.5	7.48340155223065\\
0.51	7.56399203048544\\
0.52	7.64458250874023\\
0.53	7.72517298699502\\
0.54	7.80576346524982\\
0.55	7.88635394350461\\
0.56	7.9669444217594\\
0.57	8.04753490001419\\
0.58	8.12812537826898\\
0.59	8.20871585652377\\
0.6	8.28930633477857\\
0.61	8.36989681303336\\
0.62	8.45048729128815\\
0.63	8.53107776954294\\
0.64	8.61166824779773\\
0.65	8.69225872605252\\
0.66	8.77284920430731\\
0.67	8.85343968256211\\
0.68	8.9340301608169\\
0.69	9.01462063907169\\
0.7	9.09521111732648\\
0.71	9.17580159558127\\
0.72	9.25639207383606\\
0.73	9.33698255209086\\
0.74	9.41757303034565\\
0.75	9.49816350860044\\
0.76	9.57875398685523\\
0.77	9.65934446511002\\
0.78	9.73993494336481\\
0.79	9.8205254216196\\
0.8	9.9011158998744\\
0.81	9.98170637812919\\
0.82	10.062296856384\\
0.83	10.1428873346388\\
0.84	10.2234778128936\\
0.85	10.3040682911484\\
};
\addlegendentry{\scalebox{1.5}{$\log( N^{1-\delta})$}}
\end{axis}
\end{tikzpicture}%
%
%
\definecolor{mycolor1}{rgb}{0.00000,0.44700,0.74100}%
\definecolor{mycolor2}{rgb}{0.85000,0.32500,0.09800}%
\definecolor{mycolor3}{rgb}{0.92900,0.69400,0.12500}%
\begin{tikzpicture}[scale=0.4]

\begin{axis}[%
width=6.028in,
height=4.754in,
at={(1.011in,0.642in)},
scale only axis,
xmin=0,
xmax=0.9,
xlabel style={font=\color{white!15!black}},
xlabel={\scalebox{1.5}{$\beta$}},
ymin=1,
ymax=11,
axis background/.style={fill=white},
title style={font=\bfseries},
title={\scalebox{2.5}{$\gamma=0.95$}},
axis x line*=bottom,
axis y line*=left,
legend style={at={(0.97,0.03)}, anchor=south east, legend cell align=left, align=left, draw=white!15!black}
]
\addplot [color=mycolor1, line width=2.0pt, mark=x, mark options={solid, mycolor1}]
  table[row sep=crcr]{%
0.05	4.49172149348051\\
0.15	5.08239527377659\\
0.25	5.84125640576507\\
0.35	6.68789239807139\\
0.45	7.46549898599674\\
0.55	8.33242633963162\\
0.65	9.03885032660786\\
0.75	9.75587395295472\\
0.85	10.4141346988375\\
};
\addlegendentry{\scalebox{1.5}{$\log$ size of new best class}}
\addplot [color=mycolor2, dashed, line width=2.0pt]
  table[row sep=crcr]{%
0.05	3.46339400174155\\
0.06	3.56218468432823\\
0.07	3.6608621871287\\
0.08	3.7594240629417\\
0.09	3.85786778432673\\
0.1	3.95619074005736\\
0.11	4.05439023137646\\
0.12	4.15246346803974\\
0.13	4.25040756413333\\
0.14	4.34821953364947\\
0.15	4.4458962858035\\
0.16	4.54343462007371\\
0.17	4.64083122094422\\
0.18	4.73808265232909\\
0.19	4.83518535165449\\
0.2	4.93213562357315\\
0.21	5.02892963328351\\
0.22	5.12556339942329\\
0.23	5.22203278650461\\
0.24	5.31833349685447\\
0.25	5.41446106202167\\
0.26	5.51041083360674\\
0.27	5.60617797346818\\
0.28	5.70175744325306\\
0.29	5.79714399319554\\
0.3	5.8923321501208\\
0.31	5.98731620458592\\
0.32	6.08209019708198\\
0.33	6.17664790321406\\
0.34	6.27098281776674\\
0.35	6.36508813755316\\
0.36	6.45895674293442\\
0.37	6.55258117788349\\
0.38	6.64595362845427\\
0.39	6.73906589949971\\
0.4	6.83190938946571\\
0.41	6.92447506306655\\
0.42	7.01675342162447\\
0.43	7.10873447082981\\
0.44	7.20040768564763\\
0.45	7.29176197206217\\
0.46	7.38278562531119\\
0.47	7.47346628421626\\
0.48	7.56379088116278\\
0.49	7.65374558722289\\
0.5	7.74331575184393\\
0.51	7.83248583644363\\
0.52	7.92123934115811\\
0.53	8.0095587238775\\
0.54	8.09742531057375\\
0.55	8.18481919577219\\
0.56	8.27171913183735\\
0.57	8.35810240552987\\
0.58	8.44394470003689\\
0.59	8.52921994037505\\
0.6	8.61390011970189\\
0.61	8.69795510363482\\
0.62	8.78135240914878\\
0.63	8.86405695398383\\
0.64	8.94603077171294\\
0.65	9.02723268666346\\
0.66	9.10761794170742\\
0.67	9.18713777047496\\
0.68	9.26573890372542\\
0.69	9.34336299732806\\
0.7	9.41994596642228\\
0.71	9.49541720666382\\
0.72	9.56969867876977\\
0.73	9.64270382651611\\
0.74	9.71433629045048\\
0.75	9.78448836921442\\
0.76	9.85303916661138\\
0.77	9.9198523441098\\
0.78	9.98477337345618\\
0.79	10.0476261497385\\
0.8	10.1082087774863\\
0.81	10.166288275016\\
0.82	10.2215938456629\\
0.83	10.2738082237402\\
0.84	10.3225563938409\\
0.85	10.3673906646206\\
};
\addlegendentry{\scalebox{1.5}{$\log( N^{1-\delta}\delta \log N)$}}
\addplot [color=mycolor3, dashed, line width=2.0pt]
  table[row sep=crcr]{%
0.05	1.1225102328346\\
0.06	1.23188302475182\\
0.07	1.34125581666903\\
0.08	1.45062860858625\\
0.09	1.56000140050347\\
0.1	1.66937419242068\\
0.11	1.7787469843379\\
0.12	1.88811977625512\\
0.13	1.99749256817233\\
0.14	2.10686536008955\\
0.15	2.21623815200677\\
0.16	2.32561094392399\\
0.17	2.4349837358412\\
0.18	2.54435652775842\\
0.19	2.65372931967564\\
0.2	2.76310211159285\\
0.21	2.87247490351007\\
0.22	2.98184769542729\\
0.23	3.09122048734451\\
0.24	3.20059327926172\\
0.25	3.30996607117894\\
0.26	3.41933886309616\\
0.27	3.52871165501338\\
0.28	3.63808444693059\\
0.29	3.74745723884781\\
0.3	3.85683003076503\\
0.31	3.96620282268224\\
0.32	4.07557561459946\\
0.33	4.18494840651668\\
0.34	4.2943211984339\\
0.35	4.40369399035111\\
0.36	4.51306678226833\\
0.37	4.62243957418555\\
0.38	4.73181236610276\\
0.39	4.84118515801998\\
0.4	4.9505579499372\\
0.41	5.05993074185442\\
0.42	5.16930353377163\\
0.43	5.27867632568885\\
0.44	5.38804911760607\\
0.45	5.49742190952328\\
0.46	5.6067947014405\\
0.47	5.71616749335772\\
0.48	5.82554028527494\\
0.49	5.93491307719215\\
0.5	6.04428586910937\\
0.51	6.15365866102659\\
0.52	6.2630314529438\\
0.53	6.37240424486102\\
0.54	6.48177703677824\\
0.55	6.59114982869546\\
0.56	6.70052262061267\\
0.57	6.80989541252989\\
0.58	6.91926820444711\\
0.59	7.02864099636432\\
0.6	7.13801378828154\\
0.61	7.24738658019876\\
0.62	7.35675937211598\\
0.63	7.46613216403319\\
0.64	7.57550495595041\\
0.65	7.68487774786763\\
0.66	7.79425053978485\\
0.67	7.90362333170206\\
0.68	8.01299612361928\\
0.69	8.1223689155365\\
0.7	8.23174170745371\\
0.71	8.34111449937093\\
0.72	8.45048729128815\\
0.73	8.55986008320536\\
0.74	8.66923287512258\\
0.75	8.7786056670398\\
0.76	8.88797845895702\\
0.77	8.99735125087424\\
0.78	9.10672404279145\\
0.79	9.21609683470867\\
0.8	9.32546962662589\\
0.81	9.4348424185431\\
0.82	9.54421521046032\\
0.83	9.65358800237754\\
0.84	9.76296079429475\\
0.85	9.87233358621197\\
};
\addlegendentry{\scalebox{1.5}{$\log( N^{1-\delta})$}}
\end{axis}
\end{tikzpicture}%
  \caption{For $N=10^5$ we compare the size of the ``new best class'' of the classical ratchet immediately after a click (observed in simulations) with the two theoretical predictions $N\pi_0= N^{1-\delta}$ and $N\pi_1= N^{1-\delta}\delta \log N$, cf. Remark \ref{remcentatt}.~\ref{rem:remcentatt_b}. For various values of $\gamma= \delta/(1-\beta)$, we consider (the logarithms of) these observed and predicted quantities as functions of $\beta$.
  Each data point was obtained  by pooling the interclick times no.~5 to 30 from 20 simulations of the classical ratchet for the corresponding parameter configuration.
Roughly, the average of the observed logarithmic sizes of the new best class seems to wander away from $N\pi_0$  towards $N\pi_1$ (and beyond) as $\gamma$ increases.} \label{fig:4R}
\end{figure}
 \begin{figure}[ht]
  \centering
%
%
\definecolor{mycolor1}{rgb}{0.00000,0.44700,0.74100}%
\definecolor{mycolor2}{rgb}{0.85000,0.32500,0.09800}%
\definecolor{mycolor3}{rgb}{0.92900,0.69400,0.12500}%
\begin{tikzpicture}[scale=0.4]

\begin{axis}[%
width=6.028in,
height=4.754in,
at={(1.011in,0.642in)},
scale only axis,
xmin=0,
xmax=0.9,
xlabel style={font=\color{white!15!black}},
xlabel={\scalebox{1.5}{$\beta$}},
ymin=5,
ymax=12,
axis background/.style={fill=white},
title style={font=\bfseries},
title={\scalebox{2.5}{$\gamma=0.55$}},
legend style={at={(0.97,0.03)}, anchor=south east, legend cell align=left, align=left, draw=white!15!black}
]
\addplot [color=mycolor1, line width=2.0pt, mark=x, mark options={solid, mycolor1}]
  table[row sep=crcr]{%
0.05	7.80738334103762\\
0.15	8.04878152509406\\
0.25	8.39041346043636\\
0.35	8.72627245654024\\
0.45	9.094052420048\\
0.55	9.49891853320508\\
0.65	9.85992472603823\\
0.75	10.2310280554294\\
0.85	10.6014337203476\\
};
\addlegendentry{\scalebox{1.5}{$\log$ size of new best class}}
\addplot [color=mycolor2, dotted, line width=2.0pt]
  table[row sep=crcr]{%
0.05	8.50517368724676\\
0.06	8.53683423227542\\
0.07	8.56849477730409\\
0.08	8.60015532233276\\
0.09	8.63181586736143\\
0.1	8.6634764123901\\
0.11	8.69513695741876\\
0.12	8.72679750244743\\
0.13	8.7584580474761\\
0.14	8.79011859250477\\
0.15	8.82177913753344\\
0.16	8.85343968256211\\
0.17	8.88510022759077\\
0.18	8.91676077261944\\
0.19	8.94842131764811\\
0.2	8.98008186267678\\
0.21	9.01174240770545\\
0.22	9.04340295273411\\
0.23	9.07506349776278\\
0.24	9.10672404279145\\
0.25	9.13838458782012\\
0.26	9.17004513284879\\
0.27	9.20170567787746\\
0.28	9.23336622290612\\
0.29	9.26502676793479\\
0.3	9.29668731296346\\
0.31	9.32834785799213\\
0.32	9.36000840302079\\
0.33	9.39166894804946\\
0.34	9.42332949307813\\
0.35	9.4549900381068\\
0.36	9.48665058313547\\
0.37	9.51831112816414\\
0.38	9.5499716731928\\
0.39	9.58163221822147\\
0.4	9.61329276325014\\
0.41	9.64495330827881\\
0.42	9.67661385330748\\
0.43	9.70827439833614\\
0.44	9.73993494336481\\
0.45	9.77159548839348\\
0.46	9.80325603342215\\
0.47	9.83491657845082\\
0.48	9.86657712347949\\
0.49	9.89823766850815\\
0.5	9.92989821353682\\
0.51	9.96155875856549\\
0.52	9.99321930359416\\
0.53	10.0248798486228\\
0.54	10.0565403936515\\
0.55	10.0882009386802\\
0.56	10.1198614837088\\
0.57	10.1515220287375\\
0.58	10.1831825737662\\
0.59	10.2148431187948\\
0.6	10.2465036638235\\
0.61	10.2781642088522\\
0.62	10.3098247538808\\
0.63	10.3414852989095\\
0.64	10.3731458439382\\
0.65	10.4048063889668\\
0.66	10.4364669339955\\
0.67	10.4681274790242\\
0.68	10.4997880240528\\
0.69	10.5314485690815\\
0.7	10.5631091141102\\
0.71	10.5947696591389\\
0.72	10.6264302041675\\
0.73	10.6580907491962\\
0.74	10.6897512942249\\
0.75	10.7214118392535\\
0.76	10.7530723842822\\
0.77	10.7847329293109\\
0.78	10.8163934743395\\
0.79	10.8480540193682\\
0.8	10.8797145643969\\
0.81	10.9113751094255\\
0.82	10.9430356544542\\
0.83	10.9746961994829\\
0.84	11.0063567445115\\
0.85	11.0380172895402\\
};
\addlegendentry{\scalebox{1.5}{$\log( N^{1-\delta/2})$}}
\addplot [color=mycolor3, dashed, line width=2.0pt]
  table[row sep=crcr]{%
0.05	5.49742190952328\\
0.06	5.56074299958062\\
0.07	5.62406408963796\\
0.08	5.68738517969529\\
0.09	5.75070626975263\\
0.1	5.81402735980996\\
0.11	5.8773484498673\\
0.12	5.94066953992464\\
0.13	6.00399062998197\\
0.14	6.06731172003931\\
0.15	6.13063281009665\\
0.16	6.19395390015398\\
0.17	6.25727499021132\\
0.18	6.32059608026865\\
0.19	6.38391717032599\\
0.2	6.44723826038333\\
0.21	6.51055935044066\\
0.22	6.573880440498\\
0.23	6.63720153055534\\
0.24	6.70052262061267\\
0.25	6.76384371067001\\
0.26	6.82716480072735\\
0.27	6.89048589078468\\
0.28	6.95380698084202\\
0.29	7.01712807089935\\
0.3	7.08044916095669\\
0.31	7.14377025101403\\
0.32	7.20709134107136\\
0.33	7.2704124311287\\
0.34	7.33373352118604\\
0.35	7.39705461124337\\
0.36	7.46037570130071\\
0.37	7.52369679135804\\
0.38	7.58701788141538\\
0.39	7.65033897147272\\
0.4	7.71366006153005\\
0.41	7.77698115158739\\
0.42	7.84030224164473\\
0.43	7.90362333170206\\
0.44	7.9669444217594\\
0.45	8.03026551181673\\
0.46	8.09358660187407\\
0.47	8.15690769193141\\
0.48	8.22022878198874\\
0.49	8.28354987204608\\
0.5	8.34687096210341\\
0.51	8.41019205216075\\
0.52	8.47351314221809\\
0.53	8.53683423227542\\
0.54	8.60015532233276\\
0.55	8.6634764123901\\
0.56	8.72679750244743\\
0.57	8.79011859250477\\
0.58	8.85343968256211\\
0.59	8.91676077261944\\
0.6	8.98008186267678\\
0.61	9.04340295273412\\
0.62	9.10672404279145\\
0.63	9.17004513284879\\
0.64	9.23336622290612\\
0.65	9.29668731296346\\
0.66	9.3600084030208\\
0.67	9.42332949307813\\
0.68	9.48665058313547\\
0.69	9.5499716731928\\
0.7	9.61329276325014\\
0.71	9.67661385330748\\
0.72	9.73993494336481\\
0.73	9.80325603342215\\
0.74	9.86657712347949\\
0.75	9.92989821353682\\
0.76	9.99321930359416\\
0.77	10.0565403936515\\
0.78	10.1198614837088\\
0.79	10.1831825737662\\
0.8	10.2465036638235\\
0.81	10.3098247538808\\
0.82	10.3731458439382\\
0.83	10.4364669339955\\
0.84	10.4997880240528\\
0.85	10.5631091141102\\
};
5\addlegendentry{$\text{log(N}^{\text{1-}\delta}\text{)}$}
\addlegendentry{\scalebox{1.5}{$\log( N^{1-\delta})$}}
\end{axis}
\end{tikzpicture}%
%
%
\definecolor{mycolor1}{rgb}{0.00000,0.44700,0.74100}%
\definecolor{mycolor2}{rgb}{0.85000,0.32500,0.09800}%
\definecolor{mycolor3}{rgb}{0.92900,0.69400,0.12500}%
\begin{tikzpicture}[scale=0.4]

\begin{axis}[%
width=6.028in,
height=4.754in,
at={(1.011in,0.642in)},
scale only axis,
xmin=0,
xmax=0.9,
xlabel style={font=\color{white!15!black}},
xlabel={\scalebox{1.5}{$\beta$}},
ymin=4,
ymax=11,
axis background/.style={fill=white},
title style={font=\bfseries},
title={\scalebox{2.5}{$\gamma=0.6$}},
legend style={at={(0.97,0.03)}, anchor=south east, legend cell align=left, align=left, draw=white!15!black}
]
\addplot [color=mycolor1, line width=2.0pt, mark=x, mark options={solid, mycolor1}]
  table[row sep=crcr]{%
0.05	7.77519426786785\\
0.15	8.01527140036628\\
0.25	8.33033891581756\\
0.35	8.70039673094392\\
0.45	9.06792590515108\\
0.55	9.45902135715515\\
0.65	9.82387943791335\\
0.75	10.1828370000868\\
0.85	10.556683571412\\
};
\addlegendentry{\scalebox{1.5}{$\log$ size of new best class}}
\addplot [color=mycolor2, dashed, line width=2.0pt]
  table[row sep=crcr]{%
0.05	8.23174170745371\\
0.06	8.26628048384862\\
0.07	8.30081926024354\\
0.08	8.33535803663844\\
0.09	8.36989681303336\\
0.1	8.40443558942827\\
0.11	8.43897436582318\\
0.12	8.47351314221809\\
0.13	8.508051918613\\
0.14	8.54259069500791\\
0.15	8.57712947140282\\
0.16	8.61166824779773\\
0.17	8.64620702419264\\
0.18	8.68074580058755\\
0.19	8.71528457698246\\
0.2	8.74982335337737\\
0.21	8.78436212977229\\
0.22	8.8189009061672\\
0.23	8.85343968256211\\
0.24	8.88797845895702\\
0.25	8.92251723535193\\
0.26	8.95705601174684\\
0.27	8.99159478814175\\
0.28	9.02613356453666\\
0.29	9.06067234093157\\
0.3	9.09521111732648\\
0.31	9.12974989372139\\
0.32	9.1642886701163\\
0.33	9.19882744651121\\
0.34	9.23336622290612\\
0.35	9.26790499930103\\
0.36	9.30244377569595\\
0.37	9.33698255209085\\
0.38	9.37152132848577\\
0.39	9.40606010488068\\
0.4	9.44059888127559\\
0.41	9.4751376576705\\
0.42	9.50967643406541\\
0.43	9.54421521046032\\
0.44	9.57875398685523\\
0.45	9.61329276325014\\
0.46	9.64783153964505\\
0.47	9.68237031603996\\
0.48	9.71690909243487\\
0.49	9.75144786882978\\
0.5	9.78598664522469\\
0.51	9.8205254216196\\
0.52	9.85506419801452\\
0.53	9.88960297440943\\
0.54	9.92414175080434\\
0.55	9.95868052719925\\
0.56	9.99321930359416\\
0.57	10.0277580799891\\
0.58	10.062296856384\\
0.59	10.0968356327789\\
0.6	10.1313744091738\\
0.61	10.1659131855687\\
0.62	10.2004519619636\\
0.63	10.2349907383585\\
0.64	10.2695295147534\\
0.65	10.3040682911484\\
0.66	10.3386070675433\\
0.67	10.3731458439382\\
0.68	10.4076846203331\\
0.69	10.442223396728\\
0.7	10.4767621731229\\
0.71	10.5113009495178\\
0.72	10.5458397259127\\
0.73	10.5803785023076\\
0.74	10.6149172787026\\
0.75	10.6494560550975\\
0.76	10.6839948314924\\
0.77	10.7185336078873\\
0.78	10.7530723842822\\
0.79	10.7876111606771\\
0.8	10.822149937072\\
0.81	10.8566887134669\\
0.82	10.8912274898618\\
0.83	10.9257662662567\\
0.84	10.9603050426517\\
0.85	10.9948438190466\\
};
\addlegendentry{\scalebox{1.5}{$\log( N^{1-\delta/2})$}}
\addplot [color=mycolor3, dashed, line width=2.0pt]
  table[row sep=crcr]{%
0.05	4.9505579499372\\
0.06	5.01963550272702\\
0.07	5.08871305551684\\
0.08	5.15779060830666\\
0.09	5.22686816109648\\
0.1	5.2959457138863\\
0.11	5.36502326667613\\
0.12	5.43410081946595\\
0.13	5.50317837225577\\
0.14	5.57225592504559\\
0.15	5.64133347783541\\
0.16	5.71041103062523\\
0.17	5.77948858341505\\
0.18	5.84856613620488\\
0.19	5.9176436889947\\
0.2	5.98672124178452\\
0.21	6.05579879457434\\
0.22	6.12487634736416\\
0.23	6.19395390015398\\
0.24	6.2630314529438\\
0.25	6.33210900573363\\
0.26	6.40118655852345\\
0.27	6.47026411131327\\
0.28	6.53934166410309\\
0.29	6.60841921689291\\
0.3	6.67749676968273\\
0.31	6.74657432247255\\
0.32	6.81565187526238\\
0.33	6.8847294280522\\
0.34	6.95380698084202\\
0.35	7.02288453363184\\
0.36	7.09196208642166\\
0.37	7.16103963921148\\
0.38	7.2301171920013\\
0.39	7.29919474479112\\
0.4	7.36827229758095\\
0.41	7.43734985037077\\
0.42	7.50642740316059\\
0.43	7.57550495595041\\
0.44	7.64458250874023\\
0.45	7.71366006153005\\
0.46	7.78273761431987\\
0.47	7.8518151671097\\
0.48	7.92089271989952\\
0.49	7.98997027268934\\
0.5	8.05904782547916\\
0.51	8.12812537826898\\
0.52	8.1972029310588\\
0.53	8.26628048384862\\
0.54	8.33535803663844\\
0.55	8.40443558942827\\
0.56	8.47351314221809\\
0.57	8.54259069500791\\
0.58	8.61166824779773\\
0.59	8.68074580058755\\
0.6	8.74982335337737\\
0.61	8.8189009061672\\
0.62	8.88797845895702\\
0.63	8.95705601174684\\
0.64	9.02613356453666\\
0.65	9.09521111732648\\
0.66	9.1642886701163\\
0.67	9.23336622290612\\
0.68	9.30244377569595\\
0.69	9.37152132848577\\
0.7	9.44059888127559\\
0.71	9.50967643406541\\
0.72	9.57875398685523\\
0.73	9.64783153964505\\
0.74	9.71690909243487\\
0.75	9.78598664522469\\
0.76	9.85506419801452\\
0.77	9.92414175080434\\
0.78	9.99321930359416\\
0.79	10.062296856384\\
0.8	10.1313744091738\\
0.81	10.2004519619636\\
0.82	10.2695295147534\\
0.83	10.3386070675433\\
0.84	10.4076846203331\\
0.85	10.4767621731229\\
};
\addlegendentry{\scalebox{1.5}{$\log( N^{1-\delta})$}}
\end{axis}
\end{tikzpicture}%
%
%
\definecolor{mycolor1}{rgb}{0.00000,0.44700,0.74100}%
\definecolor{mycolor2}{rgb}{0.85000,0.32500,0.09800}%
\definecolor{mycolor3}{rgb}{0.92900,0.69400,0.12500}%
\begin{tikzpicture}[scale=0.4]

\begin{axis}[%
width=6.028in,
height=4.754in,
at={(1.011in,0.642in)},
scale only axis,
xmin=0,
xmax=0.9,
xlabel style={font=\color{white!15!black}},
xlabel={\scalebox{1.5}{$\beta$}},
ymin=3,
ymax=11,
axis background/.style={fill=white},
title style={font=\bfseries},
title={\scalebox{2.5}{$\gamma=0.7$}},
legend style={at={(0.97,0.03)}, anchor=south east, legend cell align=left, align=left, draw=white!15!black}
]
\addplot [color=mycolor1, line width=2.0pt, mark=x, mark options={solid, mycolor1}]
  table[row sep=crcr]{%
0.05	7.73982207401453\\
0.15	7.96808241672212\\
0.25	8.29217095339785\\
0.35	8.64212191775654\\
0.45	9.0253143922943\\
0.55	9.38674735673751\\
0.65	9.74631460191264\\
0.75	10.0954227657184\\
0.85	10.4812846917085\\
};
\addlegendentry{\scalebox{1.5}{$\log$ size of new best class}}

\addplot [color=mycolor2, dotted, line width=2.0pt]
  table[row sep=crcr]{%
0.05	7.68487774786763\\
0.06	7.72517298699502\\
0.07	7.76546822612242\\
0.08	7.80576346524981\\
0.09	7.84605870437721\\
0.1	7.88635394350461\\
0.11	7.926649182632\\
0.12	7.9669444217594\\
0.13	8.00723966088679\\
0.14	8.04753490001419\\
0.15	8.08783013914159\\
0.16	8.12812537826898\\
0.17	8.16842061739638\\
0.18	8.20871585652377\\
0.19	8.24901109565117\\
0.2	8.28930633477856\\
0.21	8.32960157390596\\
0.22	8.36989681303336\\
0.23	8.41019205216075\\
0.24	8.45048729128815\\
0.25	8.49078253041554\\
0.26	8.53107776954294\\
0.27	8.57137300867034\\
0.28	8.61166824779773\\
0.29	8.65196348692513\\
0.3	8.69225872605252\\
0.31	8.73255396517992\\
0.32	8.77284920430731\\
0.33	8.81314444343471\\
0.34	8.85343968256211\\
0.35	8.8937349216895\\
0.36	8.9340301608169\\
0.37	8.97432539994429\\
0.38	9.01462063907169\\
0.39	9.05491587819908\\
0.4	9.09521111732648\\
0.41	9.13550635645388\\
0.42	9.17580159558127\\
0.43	9.21609683470867\\
0.44	9.25639207383606\\
0.45	9.29668731296346\\
0.46	9.33698255209086\\
0.47	9.37727779121825\\
0.48	9.41757303034565\\
0.49	9.45786826947304\\
0.5	9.49816350860044\\
0.51	9.53845874772783\\
0.52	9.57875398685523\\
0.53	9.61904922598263\\
0.54	9.65934446511002\\
0.55	9.69963970423742\\
0.56	9.73993494336481\\
0.57	9.78023018249221\\
0.58	9.8205254216196\\
0.59	9.860820660747\\
0.6	9.9011158998744\\
0.61	9.94141113900179\\
0.62	9.98170637812919\\
0.63	10.0220016172566\\
0.64	10.062296856384\\
0.65	10.1025920955114\\
0.66	10.1428873346388\\
0.67	10.1831825737662\\
0.68	10.2234778128936\\
0.69	10.263773052021\\
0.7	10.3040682911484\\
0.71	10.3443635302758\\
0.72	10.3846587694031\\
0.73	10.4249540085305\\
0.74	10.4652492476579\\
0.75	10.5055444867853\\
0.76	10.5458397259127\\
0.77	10.5861349650401\\
0.78	10.6264302041675\\
0.79	10.6667254432949\\
0.8	10.7070206824223\\
0.81	10.7473159215497\\
0.82	10.7876111606771\\
0.83	10.8279063998045\\
0.84	10.8682016389319\\
0.85	10.9084968780593\\
};
\addlegendentry{\scalebox{1.5}{$\log( N^{1-\delta/2})$}}
\addplot [color=mycolor3, dashed, line width=2.0pt]
  table[row sep=crcr]{%
0.05	3.85683003076503\\
0.06	3.93742050901982\\
0.07	4.01801098727461\\
0.08	4.0986014655294\\
0.09	4.17919194378419\\
0.1	4.25978242203898\\
0.11	4.34037290029378\\
0.12	4.42096337854857\\
0.13	4.50155385680336\\
0.14	4.58214433505815\\
0.15	4.66273481331294\\
0.16	4.74332529156773\\
0.17	4.82391576982253\\
0.18	4.90450624807732\\
0.19	4.98509672633211\\
0.2	5.0656872045869\\
0.21	5.14627768284169\\
0.22	5.22686816109648\\
0.23	5.30745863935128\\
0.24	5.38804911760607\\
0.25	5.46863959586086\\
0.26	5.54923007411565\\
0.27	5.62982055237044\\
0.28	5.71041103062523\\
0.29	5.79100150888003\\
0.3	5.87159198713482\\
0.31	5.95218246538961\\
0.32	6.0327729436444\\
0.33	6.11336342189919\\
0.34	6.19395390015398\\
0.35	6.27454437840878\\
0.36	6.35513485666357\\
0.37	6.43572533491836\\
0.38	6.51631581317315\\
0.39	6.59690629142794\\
0.4	6.67749676968273\\
0.41	6.75808724793753\\
0.42	6.83867772619232\\
0.43	6.91926820444711\\
0.44	6.9998586827019\\
0.45	7.08044916095669\\
0.46	7.16103963921148\\
0.47	7.24163011746627\\
0.48	7.32222059572107\\
0.49	7.40281107397586\\
0.5	7.48340155223065\\
0.51	7.56399203048544\\
0.52	7.64458250874023\\
0.53	7.72517298699502\\
0.54	7.80576346524982\\
0.55	7.88635394350461\\
0.56	7.9669444217594\\
0.57	8.04753490001419\\
0.58	8.12812537826898\\
0.59	8.20871585652377\\
0.6	8.28930633477857\\
0.61	8.36989681303336\\
0.62	8.45048729128815\\
0.63	8.53107776954294\\
0.64	8.61166824779773\\
0.65	8.69225872605252\\
0.66	8.77284920430731\\
0.67	8.85343968256211\\
0.68	8.9340301608169\\
0.69	9.01462063907169\\
0.7	9.09521111732648\\
0.71	9.17580159558127\\
0.72	9.25639207383606\\
0.73	9.33698255209086\\
0.74	9.41757303034565\\
0.75	9.49816350860044\\
0.76	9.57875398685523\\
0.77	9.65934446511002\\
0.78	9.73993494336481\\
0.79	9.8205254216196\\
0.8	9.9011158998744\\
0.81	9.98170637812919\\
0.82	10.062296856384\\
0.83	10.1428873346388\\
0.84	10.2234778128936\\
0.85	10.3040682911484\\
};
\addlegendentry{\scalebox{1.5}{$\log( N^{1-\delta})$}}
\end{axis}
\end{tikzpicture}%
%
%
\definecolor{mycolor1}{rgb}{0.00000,0.44700,0.74100}%
\definecolor{mycolor2}{rgb}{0.85000,0.32500,0.09800}%
\definecolor{mycolor3}{rgb}{0.92900,0.69400,0.12500}%
\begin{tikzpicture}[scale=0.4]

\begin{axis}[%
width=6.028in,
height=4.754in,
at={(1.011in,0.642in)},
scale only axis,
xmin=0,
xmax=0.9,
xlabel style={font=\color{white!15!black}},
xlabel={\scalebox{1.5}{$\beta$}},
ymin=1,
ymax=11,
axis background/.style={fill=white},
title style={font=\bfseries},
title={\scalebox{2.5}{$\gamma=0.95$}},
legend style={at={(0.97,0.03)}, anchor=south east, legend cell align=left, align=left, draw=white!15!black}
]
\addplot [color=mycolor1, line width=2.0pt, mark=x, mark options={solid, mycolor1}]
  table[row sep=crcr]{%
0.05	7.73908623418847\\
0.15	7.96970890297937\\
0.25	8.27886376227547\\
0.35	8.60737872415714\\
0.45	8.98883946117365\\
0.55	9.31139938017087\\
0.65	9.63078313749563\\
0.75	9.96697088975906\\
0.85	10.356906700023\\
};
\addlegendentry{\scalebox{1.5}{$\log$ size of new best class}}
\addplot [color=mycolor2, dashed, line width=2.0pt]
  table[row sep=crcr]{%
0.05	6.31771784890241\\
0.06	6.37240424486102\\
0.07	6.42709064081963\\
0.08	6.48177703677824\\
0.09	6.53646343273685\\
0.1	6.59114982869546\\
0.11	6.64583622465407\\
0.12	6.70052262061267\\
0.13	6.75520901657128\\
0.14	6.80989541252989\\
0.15	6.8645818084885\\
0.16	6.91926820444711\\
0.17	6.97395460040572\\
0.18	7.02864099636432\\
0.19	7.08332739232293\\
0.2	7.13801378828154\\
0.21	7.19270018424015\\
0.22	7.24738658019876\\
0.23	7.30207297615737\\
0.24	7.35675937211598\\
0.25	7.41144576807458\\
0.26	7.46613216403319\\
0.27	7.5208185599918\\
0.28	7.57550495595041\\
0.29	7.63019135190902\\
0.3	7.68487774786763\\
0.31	7.73956414382624\\
0.32	7.79425053978485\\
0.33	7.84893693574345\\
0.34	7.90362333170206\\
0.35	7.95830972766067\\
0.36	8.01299612361928\\
0.37	8.06768251957789\\
0.38	8.1223689155365\\
0.39	8.1770553114951\\
0.4	8.23174170745371\\
0.41	8.28642810341232\\
0.42	8.34111449937093\\
0.43	8.39580089532954\\
0.44	8.45048729128815\\
0.45	8.50517368724676\\
0.46	8.55986008320536\\
0.47	8.61454647916397\\
0.48	8.66923287512258\\
0.49	8.72391927108119\\
0.5	8.7786056670398\\
0.51	8.83329206299841\\
0.52	8.88797845895702\\
0.53	8.94266485491563\\
0.54	8.99735125087424\\
0.55	9.05203764683284\\
0.56	9.10672404279145\\
0.57	9.16141043875006\\
0.58	9.21609683470867\\
0.59	9.27078323066728\\
0.6	9.32546962662589\\
0.61	9.38015602258449\\
0.62	9.4348424185431\\
0.63	9.48952881450171\\
0.64	9.54421521046032\\
0.65	9.59890160641893\\
0.66	9.65358800237754\\
0.67	9.70827439833615\\
0.68	9.76296079429475\\
0.69	9.81764719025336\\
0.7	9.87233358621197\\
0.71	9.92701998217058\\
0.72	9.98170637812919\\
0.73	10.0363927740878\\
0.74	10.0910791700464\\
0.75	10.145765566005\\
0.76	10.2004519619636\\
0.77	10.2551383579222\\
0.78	10.3098247538808\\
0.79	10.3645111498394\\
0.8	10.4191975457981\\
0.81	10.4738839417567\\
0.82	10.5285703377153\\
0.83	10.5832567336739\\
0.84	10.6379431296325\\
0.85	10.6926295255911\\
};
\addlegendentry{\scalebox{1.5}{$\log( N^{1-\delta/2})$}}
\addplot [color=mycolor3, dashed, line width=2.0pt]
  table[row sep=crcr]{%
0.05	1.1225102328346\\
0.06	1.23188302475182\\
0.07	1.34125581666903\\
0.08	1.45062860858625\\
0.09	1.56000140050347\\
0.1	1.66937419242068\\
0.11	1.7787469843379\\
0.12	1.88811977625512\\
0.13	1.99749256817233\\
0.14	2.10686536008955\\
0.15	2.21623815200677\\
0.16	2.32561094392399\\
0.17	2.4349837358412\\
0.18	2.54435652775842\\
0.19	2.65372931967564\\
0.2	2.76310211159285\\
0.21	2.87247490351007\\
0.22	2.98184769542729\\
0.23	3.09122048734451\\
0.24	3.20059327926172\\
0.25	3.30996607117894\\
0.26	3.41933886309616\\
0.27	3.52871165501338\\
0.28	3.63808444693059\\
0.29	3.74745723884781\\
0.3	3.85683003076503\\
0.31	3.96620282268224\\
0.32	4.07557561459946\\
0.33	4.18494840651668\\
0.34	4.2943211984339\\
0.35	4.40369399035111\\
0.36	4.51306678226833\\
0.37	4.62243957418555\\
0.38	4.73181236610276\\
0.39	4.84118515801998\\
0.4	4.9505579499372\\
0.41	5.05993074185442\\
0.42	5.16930353377163\\
0.43	5.27867632568885\\
0.44	5.38804911760607\\
0.45	5.49742190952328\\
0.46	5.6067947014405\\
0.47	5.71616749335772\\
0.48	5.82554028527494\\
0.49	5.93491307719215\\
0.5	6.04428586910937\\
0.51	6.15365866102659\\
0.52	6.2630314529438\\
0.53	6.37240424486102\\
0.54	6.48177703677824\\
0.55	6.59114982869546\\
0.56	6.70052262061267\\
0.57	6.80989541252989\\
0.58	6.91926820444711\\
0.59	7.02864099636432\\
0.6	7.13801378828154\\
0.61	7.24738658019876\\
0.62	7.35675937211598\\
0.63	7.46613216403319\\
0.64	7.57550495595041\\
0.65	7.68487774786763\\
0.66	7.79425053978485\\
0.67	7.90362333170206\\
0.68	8.01299612361928\\
0.69	8.1223689155365\\
0.7	8.23174170745371\\
0.71	8.34111449937093\\
0.72	8.45048729128815\\
0.73	8.55986008320536\\
0.74	8.66923287512258\\
0.75	8.7786056670398\\
0.76	8.88797845895702\\
0.77	8.99735125087424\\
0.78	9.10672404279145\\
0.79	9.21609683470867\\
0.8	9.32546962662589\\
0.81	9.4348424185431\\
0.82	9.54421521046032\\
0.83	9.65358800237754\\
0.84	9.76296079429475\\
0.85	9.87233358621197\\
};
\addlegendentry{\scalebox{1.5}{$\log( N^{1-\delta})$}}

\end{axis}
\end{tikzpicture}%
  \caption{For $N=10^5$ we compare the size of the ``new best class'' of the tournament ratchet immediately after a click (as observed by simulations) with $\mathfrak a = N^{1-\delta}$ and $\mathfrak b = N^{1-\delta/2}$, which are the centers of attraction of the best and the second best class of the tournament ratchet(cf.~Remark~\ref{remcentatt}.~\ref{rem:remcentatt_b}). For various values of $\gamma= \delta/(1-\beta)$, we consider (the logarithms of) these quantities as functions of $\beta$.
  Each data point was obtained  by pooling the interclick times no.~5 to 30 from 20 simulations of the tournament  ratchet for the corresponding parameter configuration. For a wide range of parameters with $\gamma$ between $1/2$ and $1$, $\mathfrak b$ is a better fit for the size of the new best class than $\mathfrak a$.} \label{fig:4F}
\end{figure}
\begin{figure}[ht]
  \centering
%
%
\definecolor{mycolor1}{rgb}{0.00000,0.44700,0.74100}%
\definecolor{mycolor2}{rgb}{0.85000,0.32500,0.09800}%
\definecolor{mycolor3}{rgb}{0.92900,0.69400,0.12500}%
\begin{tikzpicture}[scale=0.4]

\begin{axis}[%
width=6.028in,
height=4.754in,
at={(1.011in,0.642in)},
scale only axis,
xmin=0,
xmax=0.9,
xlabel style={font=\color{white!15!black}},
xlabel={\scalebox{1.5}{$\beta$}},
ymin=6,
ymax=13,
axis background/.style={fill=white},
title style={font=\bfseries},
title={\scalebox{2.5}{$\gamma=0.55$}},
legend style={at={(0.03,0.97)}, anchor=north west, legend cell align=left, align=left, draw=white!15!black}
]
\addplot [color=mycolor1, line width=2.0pt, mark=x, mark options={solid, mycolor1}]
  table[row sep=crcr]{%
0.05	8.21182443394317\\
0.15	9.54311389960177\\
0.25	9.85886863286466\\
0.35	10.0035905260744\\
0.45	10.2241678242936\\
0.55	10.5720426015684\\
0.65	11.1084775282745\\
0.75	11.7746089263095\\
0.85	12.721266946156\\
};
\addlegendentry{\scalebox{1.5}{$\log$(avg. obs. interclick times)}}

\addplot [color=mycolor2, dashed, line width=2.0pt]
  table[row sep=crcr]{%
0.05	6.7852217435251\\
0.06	6.88599062093243\\
0.07	6.98481389176185\\
0.08	7.08157437541199\\
0.09	7.17626343615926\\
0.1	7.26895010014874\\
0.11	7.35969982351498\\
0.12	7.44841509550641\\
0.13	7.53524635308631\\
0.14	7.62018620122767\\
0.15	7.70338439692744\\
0.16	7.78485329981093\\
0.17	7.86466422256076\\
0.18	7.94293138665046\\
0.19	8.01973076797991\\
0.2	8.09517469217054\\
0.21	8.16930944393642\\
0.22	8.24227635976266\\
0.23	8.31409461955921\\
0.24	8.38492438983914\\
0.25	8.45480921904503\\
0.26	8.52381970769632\\
0.27	8.59206844892041\\
0.28	8.6595719578527\\
0.29	8.72645317493092\\
0.3	8.79274351360453\\
0.31	8.85849381255556\\
0.32	8.92376954428756\\
0.33	8.9886302057717\\
0.34	9.05309776083602\\
0.35	9.11724071082293\\
0.36	9.181090583513\\
0.37	9.24467706067733\\
0.38	9.30804161848545\\
0.39	9.37120894962761\\
0.4	9.43421492446396\\
0.41	9.49708028145106\\
0.42	9.55983641000163\\
0.43	9.62248992745941\\
0.44	9.6850807211233\\
0.45	9.74762437358351\\
0.46	9.81013629481716\\
0.47	9.87264146623991\\
0.48	9.93515430876799\\
0.49	9.9976895195563\\
0.5	10.0602708968329\\
0.51	10.1229047904352\\
0.52	10.1856156799824\\
0.53	10.2484199337463\\
0.54	10.3113349629263\\
0.55	10.3743793943226\\
0.56	10.4375732637997\\
0.57	10.5009311679989\\
0.58	10.564477198451\\
0.59	10.6282308293741\\
0.6	10.6922277027635\\
0.61	10.7564800095755\\
0.62	10.8210236509373\\
0.63	10.8858919904332\\
0.64	10.9511170755396\\
0.65	11.0167426053399\\
0.66	11.0828067217135\\
0.67	11.1493609546505\\
0.68	11.2164595260652\\
0.69	11.2841720874951\\
0.7	11.3525630103471\\
0.71	11.4217267180992\\
0.72	11.4917515829025\\
0.73	11.5627557776504\\
0.74	11.6348639569866\\
0.75	11.7082387349275\\
0.76	11.7830652360159\\
0.77	11.8595641085837\\
0.78	11.9379960334463\\
0.79	12.0186960247487\\
0.8	12.1020641110101\\
0.81	12.1885992697389\\
0.82	12.2789314696703\\
0.83	12.3738606872677\\
0.84	12.4744317041167\\
0.85	12.5820149681962\\
};
\addlegendentry{\scalebox{1.5}{$\log \E_{N\pi_1}[T^{\rm PPA}_0]$}}
\addplot [color=mycolor3, dashed, line width=2.0pt]
  table[row sep=crcr]{%
0.05	8.95220787331932\\
0.06	9.15904698738116\\
0.07	9.35017898165389\\
0.08	9.52470573253438\\
0.09	9.68202805921535\\
0.1	9.82185719193587\\
0.11	9.94421318472556\\
0.12	10.0494108323005\\
0.13	10.1380365527475\\
0.14	10.2109141817304\\
0.15	10.2690668779706\\
0.16	10.3136734973669\\
0.17	10.3460255757997\\
0.18	10.3674848555486\\
0.19	10.3794434361808\\
0.2	10.3832889243002\\
0.21	10.3803737611642\\
0.22	10.3719917054944\\
0.23	10.359357307013\\
0.24	10.3435943184673\\
0.25	10.3257246402725\\
0.26	10.3066648433609\\
0.27	10.2872255575327\\
0.28	10.2681112225394\\
0.29	10.2499273697241\\
0.3	10.2331830260021\\
0.31	10.2182994194807\\
0.32	10.2056177820756\\
0.33	10.1954069416657\\
0.34	10.1878700629785\\
0.35	10.1831555271022\\
0.36	10.1813614338662\\
0.37	10.1825437605967\\
0.38	10.1867237338713\\
0.39	10.1938921854301\\
0.4	10.2040165550384\\
0.41	10.2170438171487\\
0.42	10.2329065384836\\
0.43	10.2515232742321\\
0.44	10.2728072015726\\
0.45	10.2966632158607\\
0.46	10.3229922411601\\
0.47	10.3516945097889\\
0.48	10.3826684387942\\
0.49	10.4158131197878\\
0.5	10.4510309610575\\
0.51	10.4882237723509\\
0.52	10.5272997764033\\
0.53	10.5681693243373\\
0.54	10.6107468578959\\
0.55	10.6549512282971\\
0.56	10.700705949117\\
0.57	10.7479373074966\\
0.58	10.796578517628\\
0.59	10.8465657541591\\
0.6	10.8978448709832\\
0.61	10.9503580026775\\
0.62	11.004059465188\\
0.63	11.0589069552921\\
0.64	11.1148614340056\\
0.65	11.1718924371765\\
0.66	11.2299705365412\\
0.67	11.2890755790415\\
0.68	11.3491917881308\\
0.69	11.4103141549618\\
0.7	11.4724371865028\\
0.71	11.5355741017791\\
0.72	11.5997361290296\\
0.73	11.6649529784346\\
0.74	11.7312579588905\\
0.75	11.7987075582517\\
0.76	11.8673704614404\\
0.77	11.9373344669095\\
0.78	12.008706862525\\
0.79	12.081637591767\\
0.8	12.1563072046072\\
0.81	12.2329462138995\\
0.82	12.3118509879279\\
0.83	12.3934005631172\\
0.84	12.478098625313\\
0.85	12.5666090411147\\
};
\addlegendentry{\scalebox{1.5}{$\log \E_{N\pi_1}[T^{\rm RPPA(1)}_0]$}}
\end{axis}
\end{tikzpicture}%
%
%
\definecolor{mycolor1}{rgb}{0.00000,0.44700,0.74100}%
\definecolor{mycolor2}{rgb}{0.85000,0.32500,0.09800}%
\definecolor{mycolor3}{rgb}{0.92900,0.69400,0.12500}%
\begin{tikzpicture}[scale=0.4]

\begin{axis}[%
width=6.028in,
height=4.754in,
at={(1.011in,0.642in)},
scale only axis,
xmin=0,
xmax=0.9,
xlabel style={font=\color{white!15!black}},
xlabel={\scalebox{1.5}{$\beta$}},
ymin=6,
ymax=13,
axis background/.style={fill=white},
title style={font=\bfseries},
title={\scalebox{2.5}{$\gamma=0.6$}},
legend style={at={(0.03,0.97)}, anchor=north west, legend cell align=left, align=left, draw=white!15!black}
]
\addplot [color=mycolor1, line width=2.0pt, mark=x, mark options={solid, mycolor1}]
  table[row sep=crcr]{%
0.05	6.1794926447466\\
0.15	7.52372453178012\\
0.25	8.33181477874763\\
0.35	8.84333464034647\\
0.45	9.44552703098637\\
0.55	10.1626907771674\\
0.65	10.7481337196684\\
0.75	11.4849007879469\\
0.85	12.4946819596088\\
};
\addlegendentry{\scalebox{1.5}{$\log$(avg. obs. interclick times)}}
\addplot [color=mycolor2, dashed, line width=2.0pt]
  table[row sep=crcr]{%
0.05	6.59100696151915\\
0.06	6.69117768402409\\
0.07	6.78912699455862\\
0.08	6.88503317193443\\
0.09	6.9786995436211\\
0.1	7.07031531375598\\
0.11	7.15989351269974\\
0.12	7.2474659434268\\
0.13	7.33320280053324\\
0.14	7.41702490284453\\
0.15	7.49911183708394\\
0.16	7.57962181530047\\
0.17	7.65851301730488\\
0.18	7.73594940799269\\
0.19	7.81191792743069\\
0.2	7.88672937847555\\
0.21	7.96022063182263\\
0.22	8.03267905216073\\
0.23	8.10415800335348\\
0.24	8.17465633221448\\
0.25	8.24429608050181\\
0.26	8.31318518989609\\
0.27	8.38132361639206\\
0.28	8.4488574378352\\
0.29	8.51582548514985\\
0.3	8.58226655914921\\
0.31	8.64825675114687\\
0.32	8.71382717147263\\
0.33	8.7790423919037\\
0.34	8.84392710854786\\
0.35	8.9085362019748\\
0.36	8.97291787205397\\
0.37	9.03706007575634\\
0.38	9.10100851222943\\
0.39	9.16482798864273\\
0.4	9.2285038682536\\
0.41	9.29209356099199\\
0.42	9.3555831107818\\
0.43	9.41902388196901\\
0.44	9.4824220743325\\
0.45	9.54578514259731\\
0.46	9.60915634933931\\
0.47	9.67252401769646\\
0.48	9.73592838995991\\
0.49	9.79935998516783\\
0.5	9.86287132850943\\
0.51	9.92643976048348\\
0.52	9.99008895140716\\
0.53	10.0538418677904\\
0.54	10.1177211146492\\
0.55	10.1817259397649\\
0.56	10.2458707699094\\
0.57	10.3101828349951\\
0.58	10.3746795204389\\
0.59	10.439380658064\\
0.6	10.5042894566182\\
0.61	10.5694518574771\\
0.62	10.6348872756117\\
0.63	10.7006024889185\\
0.64	10.7666535332768\\
0.65	10.8330495927895\\
0.66	10.8998483147944\\
0.67	10.9670734493466\\
0.68	11.0347811192379\\
0.69	11.1030210625759\\
0.7	11.1718540044199\\
0.71	11.2413460860459\\
0.72	11.3115782695339\\
0.73	11.3826418784601\\
0.74	11.4546553516042\\
0.75	11.527734888493\\
0.76	11.6020397608276\\
0.77	11.677752168472\\
0.78	11.7550924474356\\
0.79	11.8343301267242\\
0.8	11.9157859428396\\
0.81	11.9998895838278\\
0.82	12.0871497885602\\
0.83	12.1782364853168\\
0.84	12.2740090325675\\
0.85	12.3756069757186\\
};
\addlegendentry{\scalebox{1.5}{$\log \E_{N\pi_1}[T^{\rm PPA}_0]$}}
\addplot [color=mycolor3, dashed, line width=2.0pt]
  table[row sep=crcr]{%
0.05	6.62988656521202\\
0.06	6.79848477653022\\
0.07	6.96001291876\\
0.08	7.11394311991923\\
0.09	7.25986060006529\\
0.1	7.39749172925497\\
0.11	7.52669865786435\\
0.12	7.64748313728199\\
0.13	7.75998293218763\\
0.14	7.8644523479012\\
0.15	7.96125766176353\\
0.16	8.05085329567773\\
0.17	8.13375772738113\\
0.18	8.21054310710276\\
0.19	8.28180672149642\\
0.2	8.34816698678694\\
0.21	8.4102256007922\\
0.22	8.4685814162466\\
0.23	8.52380004842811\\
0.24	8.5764105057986\\
0.25	8.62690649799616\\
0.26	8.67573874308622\\
0.27	8.72330880912606\\
0.28	8.76998072032613\\
0.29	8.8160705202936\\
0.3	8.86185286500499\\
0.31	8.9075661650358\\
0.32	8.95341069453767\\
0.33	8.99955680376912\\
0.34	9.04614311327117\\
0.35	9.09328490865514\\
0.36	9.14107471599782\\
0.37	9.18957963129533\\
0.38	9.23885489597162\\
0.39	9.28894392181919\\
0.4	9.33986903772335\\
0.41	9.39165059691014\\
0.42	9.44429134420611\\
0.43	9.49779559944307\\
0.44	9.55215580113644\\
0.45	9.60735952578976\\
0.46	9.66339682235564\\
0.47	9.72024549975102\\
0.48	9.77789070651183\\
0.49	9.83630654586289\\
0.5	9.89547919221806\\
0.51	9.95537809382737\\
0.52	10.0159824613204\\
0.53	10.0772719575651\\
0.54	10.1392270298491\\
0.55	10.2018222871449\\
0.56	10.2650369337106\\
0.57	10.3288551965747\\
0.58	10.3932596246975\\
0.59	10.4582347725586\\
0.6	10.5237601923504\\
0.61	10.5898323734202\\
0.62	10.6564398136788\\
0.63	10.7235665250515\\
0.64	10.7912185426877\\
0.65	10.8593831635533\\
0.66	10.928070859108\\
0.67	10.9972776087971\\
0.68	11.0670162528526\\
0.69	11.137297779112\\
0.7	11.2081398990002\\
0.71	11.2795639323147\\
0.72	11.351600031357\\
0.73	11.4242844070313\\
0.74	11.4976697675423\\
0.75	11.5718054678418\\
0.76	11.6467664921557\\
0.77	11.7226391873232\\
0.78	11.7995298837976\\
0.79	11.8775705151188\\
0.8	11.9569155898427\\
0.81	12.0377832580702\\
0.82	12.1204221001154\\
0.83	12.2051667161157\\
0.84	12.2924428207036\\
0.85	12.3828175811634\\
};
\addlegendentry{\scalebox{1.5}{$\log \E_{N\pi_1}[T^{\rm RPPA(1)}_0]$}}
\end{axis}
\end{tikzpicture}%
%
%
\definecolor{mycolor1}{rgb}{0.00000,0.44700,0.74100}%
\definecolor{mycolor2}{rgb}{0.85000,0.32500,0.09800}%
\definecolor{mycolor3}{rgb}{0.92900,0.69400,0.12500}%
\begin{tikzpicture}[scale=0.4]

\begin{axis}[%
width=6.028in,
height=4.754in,
at={(1.011in,0.642in)},
scale only axis,
xmin=0,
xmax=0.9,
xlabel style={font=\color{white!15!black}},
xlabel={\scalebox{1.5}{$\beta$}},
ymin=4,
ymax=13,
axis background/.style={fill=white},
title style={font=\bfseries},
title={\scalebox{2.5}{$\gamma=0.7$}},
legend style={at={(0.03,0.97)}, anchor=north west, legend cell align=left, align=left, draw=white!15!black}
]
\addplot [color=mycolor1, line width=2.0pt, mark=x, mark options={solid, mycolor1}]
  table[row sep=crcr]{%
0.05	4.67846077621733\\
0.15	5.87324563597626\\
0.25	6.98172193358042\\
0.35	7.85068877948564\\
0.45	8.6030145013879\\
0.55	9.44427943668403\\
0.65	10.2476167128869\\
0.75	11.1114110701174\\
0.85	12.1169634008261\\
};
\addlegendentry{\scalebox{1.5}{$\log$(avg. obs. interclick times)}}
\addplot [color=mycolor2, dotted, line width=2.0pt]
  table[row sep=crcr]{%
0.05	6.2019612046576\\
0.06	6.30840815377719\\
0.07	6.41240394694229\\
0.08	6.51383944187054\\
0.09	6.61265681184343\\
0.1	6.70956385350776\\
0.11	6.80373914653787\\
0.12	6.89525140229724\\
0.13	6.9853015932235\\
0.14	7.07270405051051\\
0.15	7.15807125715106\\
0.16	7.24187002243084\\
0.17	7.32366128786421\\
0.18	7.40389521720965\\
0.19	7.48258337466835\\
0.2	7.55975982406389\\
0.21	7.63576056839305\\
0.22	7.71058326154399\\
0.23	7.78424547591182\\
0.24	7.85677992121038\\
0.25	7.92864528315497\\
0.26	7.99941682822116\\
0.27	8.0696923548559\\
0.28	8.1390929034624\\
0.29	8.20796530431519\\
0.3	8.27614857341331\\
0.31	8.34392902557347\\
0.32	8.41128042692847\\
0.33	8.4781896410469\\
0.34	8.54486387305234\\
0.35	8.61106968381846\\
0.36	8.67709501722066\\
0.37	8.74290393611404\\
0.38	8.8085512870654\\
0.39	8.87393429729877\\
0.4	8.93918530600913\\
0.41	9.00434295344722\\
0.42	9.06943834438252\\
0.43	9.13438603798285\\
0.44	9.19933173980359\\
0.45	9.26424249368614\\
0.46	9.32909543406832\\
0.47	9.3939605478703\\
0.48	9.45881445659779\\
0.49	9.52375307029392\\
0.5	9.58867836108616\\
0.51	9.65368103650794\\
0.52	9.71877276740099\\
0.53	9.78390822102359\\
0.54	9.84919082943493\\
0.55	9.9145523657464\\
0.56	9.98003941342611\\
0.57	10.0456930131439\\
0.58	10.111486610597\\
0.59	10.1774448402182\\
0.6	10.2436121486461\\
0.61	10.3099767879006\\
0.62	10.3765873745752\\
0.63	10.4434268172113\\
0.64	10.5105510789157\\
0.65	10.5779719700334\\
0.66	10.6457237371991\\
0.67	10.7138334682485\\
0.68	10.7823245863921\\
0.69	10.8512673205311\\
0.7	10.9206689211315\\
0.71	10.9906092155923\\
0.72	11.0611354532499\\
0.73	11.1323226498083\\
0.74	11.204234377174\\
0.75	11.2769769288024\\
0.76	11.350653613685\\
0.77	11.425399584477\\
0.78	11.5013614541731\\
0.79	11.5787496506212\\
0.8	11.6577810415563\\
0.81	11.7387620906533\\
0.82	11.8220669145675\\
0.83	11.9081507184879\\
0.84	11.9976373835682\\
0.85	12.0913373120441\\
};
\addlegendentry{\scalebox{1.5}{$\log \E_{N\pi_1}[T^{\rm PPA}_0]$}}
\addplot [color=mycolor3, dashed, line width=2.0pt]
  table[row sep=crcr]{%
0.05	4.8151614382661\\
0.06	4.95635053197337\\
0.07	5.09468340932912\\
0.08	5.23002685741849\\
0.09	5.36227830265487\\
0.1	5.49141695552652\\
0.11	5.61734980311245\\
0.12	5.74007512119846\\
0.13	5.85970181765767\\
0.14	5.97620135399343\\
0.15	6.08969718062595\\
0.16	6.20032768648144\\
0.17	6.30818351364947\\
0.18	6.41343669429034\\
0.19	6.51623761280755\\
0.2	6.61674498481421\\
0.21	6.71514651722972\\
0.22	6.81160276091242\\
0.23	6.90627276396352\\
0.24	6.99931152665022\\
0.25	7.09090919624235\\
0.26	7.18116274108906\\
0.27	7.27026240136769\\
0.28	7.35828913605596\\
0.29	7.44539099576809\\
0.3	7.53165058759487\\
0.31	7.61719327112749\\
0.32	7.70209940629357\\
0.33	7.78644126182628\\
0.34	7.87031317480921\\
0.35	7.95374096631204\\
0.36	8.03681740750766\\
0.37	8.11958347238139\\
0.38	8.20208803066401\\
0.39	8.28434819685364\\
0.4	8.36641550654843\\
0.41	8.44832284658544\\
0.42	8.53009854720576\\
0.43	8.6117442003259\\
0.44	8.69330451890254\\
0.45	8.77478634219482\\
0.46	8.85619497335345\\
0.47	8.93755481191041\\
0.48	9.01886726972697\\
0.49	9.10016209915112\\
0.5	9.1814185006825\\
0.51	9.26266341711549\\
0.52	9.3439030993368\\
0.53	9.42512566995362\\
0.54	9.50636336387184\\
0.55	9.58759562882167\\
0.56	9.66883705984143\\
0.57	9.75010179534137\\
0.58	9.83138030327041\\
0.59	9.91268052927953\\
0.6	9.99401894305709\\
0.61	10.0753905307798\\
0.62	10.1568147432524\\
0.63	10.2382838480097\\
0.64	10.3198218513412\\
0.65	10.4014341517056\\
0.66	10.4831368352841\\
0.67	10.5649431848286\\
0.68	10.6468646707545\\
0.69	10.7289381184168\\
0.7	10.811167021191\\
0.71	10.8935950239977\\
0.72	10.9762480926467\\
0.73	11.0591684689306\\
0.74	11.1423907267441\\
0.75	11.2259755003959\\
0.76	11.3099801713927\\
0.77	11.3944808324241\\
0.78	11.4795570132126\\
0.79	11.5653280634174\\
0.8	11.6519084004651\\
0.81	11.7394664792878\\
0.82	11.8282038233567\\
0.83	11.9183557984678\\
0.84	12.010252040909\\
0.85	12.1043130393963\\
};
\addlegendentry{\scalebox{1.5}{$\log \E_{N\pi_1}[T^{\rm RPPA(1)}_0]$}}
\end{axis}
\end{tikzpicture}%
%
%
\definecolor{mycolor1}{rgb}{0.00000,0.44700,0.74100}%
\definecolor{mycolor2}{rgb}{0.85000,0.32500,0.09800}%
\definecolor{mycolor3}{rgb}{0.92900,0.69400,0.12500}%
\begin{tikzpicture}[scale=0.4]

\begin{axis}[%
width=6.028in,
height=4.754in,
at={(1.011in,0.642in)},
scale only axis,
unbounded coords=jump,
xmin=0,
xmax=0.9,
xlabel style={font=\color{white!15!black}},
xlabel={\scalebox{1.5}{$\beta$}},
ymin=3,
ymax=12,
axis background/.style={fill=white},
title style={font=\bfseries},
title={\scalebox{2.5}{$\gamma=0.95$}},
legend style={at={(0.03,0.97)}, anchor=north west, legend cell align=left, align=left, draw=white!15!black}
]
\addplot [color=mycolor1, line width=2.0pt, mark=x, mark options={solid, mycolor1}]
  table[row sep=crcr]{%
0.05	3.43096287604115\\
0.15	4.56323097420136\\
0.25	5.67849907365384\\
0.35	6.69118879940998\\
0.45	7.64398379854549\\
0.55	8.675992043498\\
0.65	9.60050885554145\\
0.75	10.5389401279975\\
0.85	11.6409971906482\\
};
\addlegendentry{\scalebox{1.5}{$\log$(avg. obs. interclick times)}}
\addplot [color=mycolor2, dotted, line width=2.0pt]
  table[row sep=crcr]{%
0.05	4.72000333476808\\
0.06	4.87157419432275\\
0.07	4.99303374011725\\
0.08	5.12398199208242\\
0.09	5.2605515589529\\
0.1	5.38639932409574\\
0.11	5.50311731103396\\
0.12	5.62276582804744\\
0.13	5.7433250716721\\
0.14	5.85457699305781\\
0.15	5.96578579941661\\
0.16	6.07587686421291\\
0.17	6.17771954506507\\
0.18	6.28411460998061\\
0.19	6.38248705793289\\
0.2	6.48350226618633\\
0.21	6.58134715370665\\
0.22	6.6798330750434\\
0.23	6.7745217270654\\
0.24	6.86873815048199\\
0.25	6.95908553425047\\
0.26	7.05092283782108\\
0.27	7.14089345999236\\
0.28	7.22883214135937\\
0.29	7.31649006442186\\
0.3	7.40332841673059\\
0.31	7.48897018886876\\
0.32	7.57316718913559\\
0.33	7.65818525962145\\
0.34	7.74107475933053\\
0.35	7.8248426165295\\
0.36	7.90696123829885\\
0.37	7.98898005583099\\
0.38	8.06972700353494\\
0.39	8.1503892935169\\
0.4	8.23054566193711\\
0.41	8.31042839156973\\
0.42	8.3901647590621\\
0.43	8.4689491649587\\
0.44	8.54742067178695\\
0.45	8.62598476859655\\
0.46	8.70392517542793\\
0.47	8.78133555095957\\
0.48	8.85879005539057\\
0.49	8.93571008198397\\
0.5	9.01255132065509\\
0.51	9.08902692460275\\
0.52	9.1652843542725\\
0.53	9.24125664191228\\
0.54	9.31703476311669\\
0.55	9.39279984742681\\
0.56	9.46819767714079\\
0.57	9.54351738767568\\
0.58	9.61866254587222\\
0.59	9.69373981922103\\
0.6	9.76866227401515\\
0.61	9.84351604640137\\
0.62	9.91829708712491\\
0.63	9.9930658189055\\
0.64	10.0677613105829\\
0.65	10.1424489032375\\
0.66	10.217183694821\\
0.67	10.2919742423591\\
0.68	10.3668404403038\\
0.69	10.4418128766016\\
0.7	10.5169323842344\\
0.71	10.5922498340241\\
0.72	10.6677999590485\\
0.73	10.7435879868849\\
0.74	10.8197409920851\\
0.75	10.8962639434868\\
0.76	10.9732424272578\\
0.77	11.0507358798939\\
0.78	11.1288793091129\\
0.79	11.2077478267751\\
0.8	11.2874914193434\\
0.81	11.3682636163448\\
0.82	11.4502566817865\\
0.83	11.5337212076971\\
0.84	11.6189906197734\\
0.85	11.7064813287884\\
};
\addlegendentry{\scalebox{1.5}{$\log \E_{N\pi_1}[T^{\rm PPA}_0]$}}
\addplot [color=mycolor3, dashed, line width=2.0pt]
  table[row sep=crcr]{%
0.05	inf\\
0.06	inf\\
0.07	inf\\
0.08	inf\\
0.09	inf\\
0.1	inf\\
0.11	inf\\
0.12	inf\\
0.13	3.99376418724219\\
0.14	4.12148240443039\\
0.15	4.24792947949318\\
0.16	4.37295983803448\\
0.17	4.49527679074632\\
0.18	4.61737149174577\\
0.19	4.73689945019554\\
0.2	4.8559748216888\\
0.21	4.9735349388693\\
0.22	5.09039032012185\\
0.23	5.20571995790834\\
0.24	5.32024089725504\\
0.25	5.43331303297375\\
0.26	5.54608462473939\\
0.27	5.65791402450766\\
0.28	5.76880293151339\\
0.29	5.87916384305939\\
0.3	5.98892104851383\\
0.31	6.0980243029075\\
0.32	6.20644337642874\\
0.33	6.31472598174945\\
0.34	6.42221640421421\\
0.35	6.5296445525133\\
0.36	6.63642888712014\\
0.37	6.74296083062495\\
0.38	6.84896328383031\\
0.39	6.95474770708281\\
0.4	7.06022016496023\\
0.41	7.16545306603815\\
0.42	7.27049298016537\\
0.43	7.37512434949752\\
0.44	7.47952761966301\\
0.45	7.583828205642\\
0.46	7.68782262432834\\
0.47	7.79153693392379\\
0.48	7.8951496000025\\
0.49	7.99848260537866\\
0.5	8.1016812499627\\
0.51	8.2046528295607\\
0.52	8.30744492788209\\
0.53	8.41003300809194\\
0.54	8.5124466900941\\
0.55	8.61475143679632\\
0.56	8.71681460157606\\
0.57	8.81874087785748\\
0.58	8.92049081274124\\
0.59	9.02210386233559\\
0.6	9.12354156910648\\
0.61	9.22483589019098\\
0.62	9.32598178274967\\
0.63	9.42700265391963\\
0.64	9.52786723302633\\
0.65	9.62860108874397\\
0.66	9.72922709674286\\
0.67	9.82974649141766\\
0.68	9.93016586351924\\
0.69	10.0304974696797\\
0.7	10.1307596242612\\
0.71	10.2309772055491\\
0.72	10.3311670587199\\
0.73	10.4313284771824\\
0.74	10.5315332223469\\
0.75	10.6317813057465\\
0.76	10.7321213466661\\
0.77	10.8325854084195\\
0.78	10.9332550533302\\
0.79	11.0341710274706\\
0.8	11.1354226455541\\
0.81	11.2370982355784\\
0.82	11.3393071893385\\
0.83	11.4421917284608\\
0.84	11.5459417122824\\
0.85	11.6507868426019\\
};
\addlegendentry{\scalebox{1.5}{$\log \E_{N\pi_1}[T^{\rm RPPA(1)}_0]$}}
\end{axis}
\end{tikzpicture}%
    \caption{ 
    For fixed population size $N=10^5$ the predictions for the expected interclick time of the classical ratchet based on a numerical calculation of the Green function (i) of the PPA and (ii) of the RPPA(1) approximation  
    are compared with simulations. Here, 
formula \eqref{eq:generalgreen} is used (i) for the jump rates \eqref{bestdown} and~\eqref{bestup}, and (ii) for the downward jump rate \eqref{bestdown} and the upward jump rate resulting from \eqref{bestupclass1} and \eqref{MRPPA}.
    Each data point was obtained  by pooling the interclick times no.~5 to 30 from 20 simulations of the tournament ratchet for the corresponding parameter configuration. Each plot shows this for one fixed value of $\gamma$ with varying $\beta$.}
\label{fig:6R}  
\end{figure}
\begin{figure}[ht]
  \centering
%
%
\definecolor{mycolor1}{rgb}{0.00000,0.44700,0.74100}%
\definecolor{mycolor2}{rgb}{0.85000,0.32500,0.09800}%
\definecolor{mycolor3}{rgb}{0.92900,0.69400,0.12500}%
\begin{tikzpicture}[scale=0.4]

\begin{axis}[%
width=6.028in,
height=4.754in,
at={(1.011in,0.642in)},
scale only axis,
xmin=0,
xmax=0.9,
xlabel style={font=\color{white!15!black}},
xlabel={\scalebox{1.5}{$\beta$}},
ymin=6,
ymax=13,
axis background/.style={fill=white},
title style={font=\bfseries},
title={\scalebox{2.5}{$\gamma=0.55$}},
legend style={at={(0.03,0.97)}, anchor=north west, legend cell align=left, align=left, draw=white!15!black}
]
\addplot [color=mycolor1, line width=2.0pt, mark=x, mark options={solid, mycolor1}]
  table[row sep=crcr]{%
0.05	6.8390816792528\\
0.15	7.70966853154643\\
0.25	8.49629970734424\\
0.35	9.0767082800799\\
0.45	9.72529279344019\\
0.55	10.3436233816793\\
0.65	11.0238568051869\\
0.75	11.6260538835497\\
0.85	12.5854254478517\\
};
\addlegendentry{\scalebox{1.5}{$\log$(avg. obs. interclick times)}}

\addplot [color=mycolor2, dashed, line width=2.0pt]
  table[row sep=crcr]{%
0.05	6.87942989495697\\
0.06	6.97647292359747\\
0.07	7.07177202641445\\
0.08	7.16529610783316\\
0.09	7.25703331302059\\
0.1	7.34697115350473\\
0.11	7.43512326354195\\
0.12	7.52150907276427\\
0.13	7.60616563154898\\
0.14	7.68912906440134\\
0.15	7.77045126553292\\
0.16	7.85019768173933\\
0.17	7.92843491239588\\
0.18	8.00523494280464\\
0.19	8.08067859248592\\
0.2	8.15484397965423\\
0.21	8.22780578106554\\
0.22	8.29965306998364\\
0.23	8.37045969923319\\
0.24	8.44030712628415\\
0.25	8.5092697350347\\
0.26	8.57742338376481\\
0.27	8.64483182639712\\
0.28	8.71155975907719\\
0.29	8.77768073579623\\
0.3	8.84323911909602\\
0.31	8.90830066896228\\
0.32	8.97291073714727\\
0.33	9.03711932712051\\
0.34	9.10096496949331\\
0.35	9.16449493770001\\
0.36	9.22774543748317\\
0.37	9.29074991413716\\
0.38	9.35354309254661\\
0.39	9.41615346518479\\
0.4	9.4786074413899\\
0.41	9.54093328815615\\
0.42	9.60315750295274\\
0.43	9.66530137322117\\
0.44	9.72738496615593\\
0.45	9.78943092671622\\
0.46	9.85146095715172\\
0.47	9.91349247104537\\
0.48	9.97554594938707\\
0.49	10.0376380185459\\
0.5	10.0997886864517\\
0.51	10.1620145380655\\
0.52	10.2243392546142\\
0.53	10.2867800576357\\
0.54	10.3493548773055\\
0.55	10.4120891809919\\
0.56	10.4750028022667\\
0.57	10.538120225579\\
0.58	10.6014642786001\\
0.59	10.6650661356844\\
0.6	10.7289527364963\\
0.61	10.7931599527474\\
0.62	10.8577203542163\\
0.63	10.9226731828066\\
0.64	10.9880587254288\\
0.65	11.0539310625729\\
0.66	11.1203435970303\\
0.67	11.1873503355937\\
0.68	11.2550216765855\\
0.69	11.3234366650455\\
0.7	11.3926817794599\\
0.71	11.462853055662\\
0.72	11.5340667140027\\
0.73	11.6064563210336\\
0.74	11.6801712149952\\
0.75	11.7553916679115\\
0.76	11.8323291596583\\
0.77	11.9112313178644\\
0.78	11.992394112871\\
0.79	12.0761743997883\\
0.8	12.1630132066514\\
0.81	12.2534421132087\\
0.82	12.3481256824671\\
0.83	12.4479076891957\\
0.84	12.5538552393169\\
0.85	12.6673679212883\\
};
\addlegendentry{\scalebox{1.5}{$\log \mathbb E_{N^{1-\delta/2}}[\hat T_0]$}}
\addplot [color=mycolor3, dashed, line width=2.0pt]
  table[row sep=crcr]{%
0.05	7.06823351732353\\
0.06	7.12579814464838\\
0.07	7.18336277197323\\
0.08	7.24092739929808\\
0.09	7.29849202662293\\
0.1	7.35605665394778\\
0.11	7.41362128127263\\
0.12	7.47118590859748\\
0.13	7.52875053592233\\
0.14	7.58631516324719\\
0.15	7.64387979057204\\
0.16	7.70144441789689\\
0.17	7.75900904522174\\
0.18	7.81657367254659\\
0.19	7.87413829987144\\
0.2	7.93170292719629\\
0.21	7.98926755452114\\
0.22	8.04683218184599\\
0.23	8.10439680917085\\
0.24	8.1619614364957\\
0.25	8.21952606382055\\
0.26	8.2770906911454\\
0.27	8.33465531847025\\
0.28	8.3922199457951\\
0.29	8.44978457311995\\
0.3	8.5073492004448\\
0.31	8.56491382776965\\
0.32	8.62247845509451\\
0.33	8.68004308241936\\
0.34	8.73760770974421\\
0.35	8.79517233706906\\
0.36	8.85273696439391\\
0.37	8.91030159171876\\
0.38	8.96786621904361\\
0.39	9.02543084636846\\
0.4	9.08299547369331\\
0.41	9.14056010101817\\
0.42	9.19812472834302\\
0.43	9.25568935566787\\
0.44	9.31325398299272\\
0.45	9.37081861031757\\
0.46	9.42838323764242\\
0.47	9.48594786496727\\
0.48	9.54351249229212\\
0.49	9.60107711961698\\
0.5	9.65864174694183\\
0.51	9.71620637426668\\
0.52	9.77377100159153\\
0.53	9.83133562891638\\
0.54	9.88890025624123\\
0.55	9.94646488356608\\
0.56	10.0040295108909\\
0.57	10.0615941382158\\
0.58	10.1191587655406\\
0.59	10.1767233928655\\
0.6	10.2342880201903\\
0.61	10.2918526475152\\
0.62	10.34941727484\\
0.63	10.4069819021649\\
0.64	10.4645465294897\\
0.65	10.5221111568146\\
0.66	10.5796757841394\\
0.67	10.6372404114643\\
0.68	10.6948050387891\\
0.69	10.752369666114\\
0.7	10.8099342934389\\
0.71	10.8674989207637\\
0.72	10.9250635480886\\
0.73	10.9826281754134\\
0.74	11.0401928027383\\
0.75	11.0977574300631\\
0.76	11.155322057388\\
0.77	11.2128866847128\\
0.78	11.2704513120377\\
0.79	11.3280159393625\\
0.8	11.3855805666874\\
0.81	11.4431451940122\\
0.82	11.5007098213371\\
0.83	11.5582744486619\\
0.84	11.6158390759868\\
0.85	11.6734037033116\\
};
\addlegendentry{\scalebox{1.5}{$ \log( \pi^{3/2 }/2 N^{\tfrac{1+\beta}{2}})  $}}
\end{axis}
\end{tikzpicture}%
%
%
\definecolor{mycolor1}{rgb}{0.00000,0.44700,0.74100}%
\definecolor{mycolor2}{rgb}{0.85000,0.32500,0.09800}%
\definecolor{mycolor3}{rgb}{0.92900,0.69400,0.12500}%
\begin{tikzpicture}[scale=0.4]

\begin{axis}[%
width=6.028in,
height=4.754in,
at={(1.011in,0.642in)},
scale only axis,
xmin=0,
xmax=0.9,
xlabel style={font=\color{white!15!black}},
xlabel={\scalebox{1.5}{$\beta$}},
ymin=6,
ymax=13,
axis background/.style={fill=white},
title style={font=\bfseries},
title={\scalebox{2.5}{$\gamma=0.6$}},
legend style={at={(0.03,0.97)}, anchor=north west, legend cell align=left, align=left, draw=white!15!black}
]
\addplot [color=mycolor1, line width=2.0pt, mark=x, mark options={solid, mycolor1}]
  table[row sep=crcr]{%
0.05	6.7350205576417\\
0.15	7.58097842775967\\
0.25	8.26045817640827\\
0.35	8.89448470845425\\
0.45	9.5351560520885\\
0.55	10.1356898321932\\
0.65	10.7374704419284\\
0.75	11.4228784163809\\
0.85	12.2459745777304\\
};
\addlegendentry{\scalebox{1.5}{$\log$(avg. obs. interclick times)}}

\addplot [color=mycolor2, dashed, line width=2.0pt]
  table[row sep=crcr]{%
0.05	6.76082592990213\\
0.06	6.85430522631215\\
0.07	6.94607683511279\\
0.08	7.03611449913198\\
0.09	7.12444314202491\\
0.1	7.21104792090272\\
0.11	7.29597504434501\\
0.12	7.379255937027\\
0.13	7.4609213705618\\
0.14	7.54103307100022\\
0.15	7.61964891563876\\
0.16	7.69684277207869\\
0.17	7.7726733671808\\
0.18	7.84722249287451\\
0.19	7.92056425281414\\
0.2	7.99277389170992\\
0.21	8.06393526903174\\
0.22	8.1341139106799\\
0.23	8.20339901024824\\
0.24	8.27185211825922\\
0.25	8.33954061216433\\
0.26	8.40654415024891\\
0.27	8.47291526151617\\
0.28	8.53871101519579\\
0.29	8.60399221892285\\
0.3	8.66881564755666\\
0.31	8.73322010741178\\
0.32	8.79725538263154\\
0.33	8.86096761838625\\
0.34	8.92438615341898\\
0.35	8.98755770112848\\
0.36	9.05050610259552\\
0.37	9.11327236267421\\
0.38	9.17586938719596\\
0.39	9.23833983398959\\
0.4	9.30070518117646\\
0.41	9.3629793540207\\
0.42	9.42519303631743\\
0.43	9.48736337191629\\
0.44	9.54951227179193\\
0.45	9.61165483998557\\
0.46	9.67381113086996\\
0.47	9.73599490710743\\
0.48	9.79822505189036\\
0.49	9.86052533027064\\
0.5	9.92289786537199\\
0.51	9.98537150994812\\
0.52	10.0479592285458\\
0.53	10.1106795467945\\
0.54	10.1735464460834\\
0.55	10.2365796824099\\
0.56	10.2998046075424\\
0.57	10.3632378456884\\
0.58	10.4269022284997\\
0.59	10.4908221095949\\
0.6	10.5550191625973\\
0.61	10.6195307699458\\
0.62	10.6843831965377\\
0.63	10.7496102268974\\
0.64	10.8152490875065\\
0.65	10.8813410198827\\
0.66	10.9479359935566\\
0.67	11.0150851814369\\
0.68	11.0828500550248\\
0.69	11.1512878754953\\
0.7	11.220484065714\\
0.71	11.2905190107841\\
0.72	11.361489499716\\
0.73	11.4335177658135\\
0.74	11.5067260772507\\
0.75	11.5812692177397\\
0.76	11.6573316273427\\
0.77	11.7351237667699\\
0.78	11.8149002196931\\
0.79	11.8969668600696\\
0.8	11.9816920385781\\
0.81	12.0695369730562\\
0.82	12.1610615724385\\
0.83	12.2569802131373\\
0.84	12.3582014131172\\
0.85	12.4659089538433\\
};
\addlegendentry{\scalebox{1.5}{$\log \mathbb E_{N^{1-\delta/2}}[\hat T_0]$}}
\addplot [color=mycolor3, dashed, line width=2.0pt]
  table[row sep=crcr]{%
0.05	7.06823351732353\\
0.06	7.12579814464838\\
0.07	7.18336277197323\\
0.08	7.24092739929808\\
0.09	7.29849202662293\\
0.1	7.35605665394778\\
0.11	7.41362128127263\\
0.12	7.47118590859748\\
0.13	7.52875053592233\\
0.14	7.58631516324719\\
0.15	7.64387979057204\\
0.16	7.70144441789689\\
0.17	7.75900904522174\\
0.18	7.81657367254659\\
0.19	7.87413829987144\\
0.2	7.93170292719629\\
0.21	7.98926755452114\\
0.22	8.04683218184599\\
0.23	8.10439680917085\\
0.24	8.1619614364957\\
0.25	8.21952606382055\\
0.26	8.2770906911454\\
0.27	8.33465531847025\\
0.28	8.3922199457951\\
0.29	8.44978457311995\\
0.3	8.5073492004448\\
0.31	8.56491382776965\\
0.32	8.62247845509451\\
0.33	8.68004308241936\\
0.34	8.73760770974421\\
0.35	8.79517233706906\\
0.36	8.85273696439391\\
0.37	8.91030159171876\\
0.38	8.96786621904361\\
0.39	9.02543084636846\\
0.4	9.08299547369331\\
0.41	9.14056010101817\\
0.42	9.19812472834302\\
0.43	9.25568935566787\\
0.44	9.31325398299272\\
0.45	9.37081861031757\\
0.46	9.42838323764242\\
0.47	9.48594786496727\\
0.48	9.54351249229212\\
0.49	9.60107711961698\\
0.5	9.65864174694183\\
0.51	9.71620637426668\\
0.52	9.77377100159153\\
0.53	9.83133562891638\\
0.54	9.88890025624123\\
0.55	9.94646488356608\\
0.56	10.0040295108909\\
0.57	10.0615941382158\\
0.58	10.1191587655406\\
0.59	10.1767233928655\\
0.6	10.2342880201903\\
0.61	10.2918526475152\\
0.62	10.34941727484\\
0.63	10.4069819021649\\
0.64	10.4645465294897\\
0.65	10.5221111568146\\
0.66	10.5796757841394\\
0.67	10.6372404114643\\
0.68	10.6948050387891\\
0.69	10.752369666114\\
0.7	10.8099342934389\\
0.71	10.8674989207637\\
0.72	10.9250635480886\\
0.73	10.9826281754134\\
0.74	11.0401928027383\\
0.75	11.0977574300631\\
0.76	11.155322057388\\
0.77	11.2128866847128\\
0.78	11.2704513120377\\
0.79	11.3280159393625\\
0.8	11.3855805666874\\
0.81	11.4431451940122\\
0.82	11.5007098213371\\
0.83	11.5582744486619\\
0.84	11.6158390759868\\
0.85	11.6734037033116\\
};
\addlegendentry{\scalebox{1.5}{$ \log( \pi^{3/2 }/2 N^{\tfrac{1+\beta}{2}})  $}}
\end{axis}
\end{tikzpicture}%
%
%
\definecolor{mycolor1}{rgb}{0.00000,0.44700,0.74100}%
\definecolor{mycolor2}{rgb}{0.85000,0.32500,0.09800}%
\definecolor{mycolor3}{rgb}{0.92900,0.69400,0.12500}%
\begin{tikzpicture}[scale=0.4]

\begin{axis}[%
width=6.028in,
height=4.754in,
at={(1.011in,0.642in)},
scale only axis,
xmin=0,
xmax=0.9,
xlabel style={font=\color{white!15!black}},
xlabel={\scalebox{1.5}{$\beta$}},
ymin=6,
ymax=13,
axis background/.style={fill=white},
title style={font=\bfseries},
title={\scalebox{2.5}{$\gamma=0.7$}},
legend style={at={(0.03,0.97)}, anchor=north west, legend cell align=left, align=left, draw=white!15!black}
]
\addplot [color=mycolor1, line width=2.0pt, mark=x, mark options={solid, mycolor1}]
  table[row sep=crcr]{%
0.05	6.64945713499831\\
0.15	7.43473595362258\\
0.25	8.14996097833968\\
0.35	8.76922051027996\\
0.45	9.35366219112873\\
0.55	9.95849933883303\\
0.65	10.5182053018315\\
0.75	11.2620375649125\\
0.85	11.9516705773308\\
};
\addlegendentry{\scalebox{1.5}{$\log$(avg. obs. interclick times)}}

\addplot [color=mycolor2, dotted, line width=2.0pt]
  table[row sep=crcr]{%
0.05	6.63020699756499\\
0.06	6.72107028146999\\
0.07	6.81016838119338\\
0.08	6.89749983651816\\
0.09	6.98307548867924\\
0.1	7.06691793394421\\
0.11	7.14909558913194\\
0.12	7.22961410019257\\
0.13	7.30852614425985\\
0.14	7.38595313022724\\
0.15	7.46189525108318\\
0.16	7.53648064588653\\
0.17	7.60975069012026\\
0.18	7.681806135921\\
0.19	7.75271997548154\\
0.2	7.8225901846564\\
0.21	7.89144000790137\\
0.22	7.95941224842182\\
0.23	8.02652787095026\\
0.24	8.09289717800206\\
0.25	8.15855977997987\\
0.26	8.22359532290192\\
0.27	8.28807782418606\\
0.28	8.35203772240943\\
0.29	8.41554155661561\\
0.3	8.47863257368632\\
0.31	8.54136850439878\\
0.32	8.6037852383819\\
0.33	8.66593256300364\\
0.34	8.72782375886393\\
0.35	8.78951862035406\\
0.36	8.85102566559532\\
0.37	8.9123977916777\\
0.38	8.97365411745557\\
0.39	9.03481236911836\\
0.4	9.09590276566275\\
0.41	9.15693966540691\\
0.42	9.21796260930967\\
0.43	9.27896969882904\\
0.44	9.3399969472637\\
0.45	9.4010413537965\\
0.46	9.4621478827682\\
0.47	9.52330073639484\\
0.48	9.58454224384868\\
0.49	9.64587901333922\\
0.5	9.7073286561482\\
0.51	9.76889799647372\\
0.52	9.83061457102984\\
0.53	9.89248520525196\\
0.54	9.95453670820382\\
0.55	10.0167766177139\\
0.56	10.079231793627\\
0.57	10.1419112451809\\
0.58	10.2048427275855\\
0.59	10.2680458342135\\
0.6	10.3315414666544\\
0.61	10.3953599277583\\
0.62	10.4595173652802\\
0.63	10.5240478240698\\
0.64	10.5889871718832\\
0.65	10.6543666326772\\
0.66	10.7202211304533\\
0.67	10.7866030507792\\
0.68	10.853555531547\\
0.69	10.921127677835\\
0.7	10.9893875963249\\
0.71	11.0584043679258\\
0.72	11.1282505455415\\
0.73	11.1990158549866\\
0.74	11.2708139000188\\
0.75	11.3437527695603\\
0.76	11.4179766874358\\
0.77	11.4936585886757\\
0.78	11.5709824813306\\
0.79	11.6501943928642\\
0.8	11.731570507944\\
0.81	11.8154630237843\\
0.82	11.9022973469404\\
0.83	11.9926152097626\\
0.84	12.0870964819685\\
0.85	12.1866319576844\\
};
\addlegendentry{\scalebox{1.5}{$\log \mathbb E_{N^{1-\delta/2}}[\hat T_0]$}}
\addplot [color=mycolor3, dashed, line width=2.0pt]
  table[row sep=crcr]{%
0.05	7.06823351732353\\
0.06	7.12579814464838\\
0.07	7.18336277197323\\
0.08	7.24092739929808\\
0.09	7.29849202662293\\
0.1	7.35605665394778\\
0.11	7.41362128127263\\
0.12	7.47118590859748\\
0.13	7.52875053592233\\
0.14	7.58631516324719\\
0.15	7.64387979057204\\
0.16	7.70144441789689\\
0.17	7.75900904522174\\
0.18	7.81657367254659\\
0.19	7.87413829987144\\
0.2	7.93170292719629\\
0.21	7.98926755452114\\
0.22	8.04683218184599\\
0.23	8.10439680917085\\
0.24	8.1619614364957\\
0.25	8.21952606382055\\
0.26	8.2770906911454\\
0.27	8.33465531847025\\
0.28	8.3922199457951\\
0.29	8.44978457311995\\
0.3	8.5073492004448\\
0.31	8.56491382776965\\
0.32	8.62247845509451\\
0.33	8.68004308241936\\
0.34	8.73760770974421\\
0.35	8.79517233706906\\
0.36	8.85273696439391\\
0.37	8.91030159171876\\
0.38	8.96786621904361\\
0.39	9.02543084636846\\
0.4	9.08299547369331\\
0.41	9.14056010101817\\
0.42	9.19812472834302\\
0.43	9.25568935566787\\
0.44	9.31325398299272\\
0.45	9.37081861031757\\
0.46	9.42838323764242\\
0.47	9.48594786496727\\
0.48	9.54351249229212\\
0.49	9.60107711961698\\
0.5	9.65864174694183\\
0.51	9.71620637426668\\
0.52	9.77377100159153\\
0.53	9.83133562891638\\
0.54	9.88890025624123\\
0.55	9.94646488356608\\
0.56	10.0040295108909\\
0.57	10.0615941382158\\
0.58	10.1191587655406\\
0.59	10.1767233928655\\
0.6	10.2342880201903\\
0.61	10.2918526475152\\
0.62	10.34941727484\\
0.63	10.4069819021649\\
0.64	10.4645465294897\\
0.65	10.5221111568146\\
0.66	10.5796757841394\\
0.67	10.6372404114643\\
0.68	10.6948050387891\\
0.69	10.752369666114\\
0.7	10.8099342934389\\
0.71	10.8674989207637\\
0.72	10.9250635480886\\
0.73	10.9826281754134\\
0.74	11.0401928027383\\
0.75	11.0977574300631\\
0.76	11.155322057388\\
0.77	11.2128866847128\\
0.78	11.2704513120377\\
0.79	11.3280159393625\\
0.8	11.3855805666874\\
0.81	11.4431451940122\\
0.82	11.5007098213371\\
0.83	11.5582744486619\\
0.84	11.6158390759868\\
0.85	11.6734037033116\\
};
\addlegendentry{\scalebox{1.5}{$ \log( \pi^{3/2 }/2 N^{\tfrac{1+\beta}{2}})  $}}
\end{axis}
\end{tikzpicture}%
%
%
\definecolor{mycolor1}{rgb}{0.00000,0.44700,0.74100}%
\definecolor{mycolor2}{rgb}{0.85000,0.32500,0.09800}%
\definecolor{mycolor3}{rgb}{0.92900,0.69400,0.12500}%
\begin{tikzpicture}[scale=0.4]

\begin{axis}[%
width=6.028in,
height=4.754in,
at={(1.011in,0.642in)},
scale only axis,
xmin=0,
xmax=0.9,
xlabel style={font=\color{white!15!black}},
xlabel={\scalebox{1.5}{$\beta$}},
ymin=6,
ymax=12,
axis background/.style={fill=white},
title style={font=\bfseries},
title={\scalebox{2.5}{$\gamma=0.95$}},
legend style={at={(0.03,0.97)}, anchor=north west, legend cell align=left, align=left, draw=white!15!black}
]
\addplot [color=mycolor1, line width=2.0pt, mark=x, mark options={solid, mycolor1}]
  table[row sep=crcr]{%
0.05	6.60650294807161\\
0.15	7.43205368827541\\
0.25	8.01719451774266\\
0.35	8.63031104680734\\
0.45	9.19879736689773\\
0.55	9.81729238762713\\
0.65	10.2819284517045\\
0.75	10.9125518131438\\
0.85	11.6200637091203\\
};
\addlegendentry{\scalebox{1.5}{$\log$(avg. obs. interclick times)}}
\addplot [color=mycolor2, dashed, line width=2.0pt]
  table[row sep=crcr]{%
0.05	6.31530031247186\\
0.06	6.4112651762419\\
0.07	6.50507406001348\\
0.08	6.59668636732584\\
0.09	6.68564820202284\\
0.1	6.77287681434165\\
0.11	6.85793050781559\\
0.12	6.94085607697991\\
0.13	7.02205207444767\\
0.14	7.10121256335998\\
0.15	7.17872163930502\\
0.16	7.25461913648549\\
0.17	7.32895715937761\\
0.18	7.40179716600798\\
0.19	7.47320732515219\\
0.2	7.54347913657959\\
0.21	7.61265049844659\\
0.22	7.68076547707487\\
0.23	7.74787246577206\\
0.24	7.81402257194168\\
0.25	7.87943209623791\\
0.26	7.94412716925842\\
0.27	8.00828325225875\\
0.28	8.07163114072362\\
0.29	8.13461989227475\\
0.3	8.19711680465821\\
0.31	8.25913810523312\\
0.32	8.320813032612\\
0.33	8.38224886419554\\
0.34	8.44333467700367\\
0.35	8.5041737034347\\
0.36	8.56476214332186\\
0.37	8.62526913075468\\
0.38	8.68559051390864\\
0.39	8.74587278038816\\
0.4	8.80601783678261\\
0.41	8.86608513666089\\
0.42	8.92612388210383\\
0.43	8.98617451136797\\
0.44	9.0462116978633\\
0.45	9.10627119058508\\
0.46	9.16633010695409\\
0.47	9.22652040107372\\
0.48	9.28675970756781\\
0.49	9.34707193982172\\
0.5	9.40747747466646\\
0.51	9.46803410539853\\
0.52	9.52875097256412\\
0.53	9.5896000318899\\
0.54	9.6506288277003\\
0.55	9.71184455994776\\
0.56	9.77328557535247\\
0.57	9.83495515475286\\
0.58	9.89685797868923\\
0.59	9.95905307538544\\
0.6	10.0215155157674\\
0.61	10.084275889799\\
0.62	10.1473850975753\\
0.63	10.2108667536955\\
0.64	10.2747240888894\\
0.65	10.3390034057583\\
0.66	10.4037298232588\\
0.67	10.4689476931336\\
0.68	10.5346842112647\\
0.69	10.6009703856362\\
0.7	10.6678858366219\\
0.71	10.7354648461196\\
0.72	10.8037621677076\\
0.73	10.8728382728414\\
0.74	10.9427847099768\\
0.75	11.0136752171125\\
0.76	11.0856170405887\\
0.77	11.1587082559188\\
0.78	11.233085252244\\
0.79	11.3088955238458\\
0.8	11.3863404416442\\
0.81	11.4656289079112\\
0.82	11.5470334039548\\
0.83	11.6308798410558\\
0.84	11.7175912162391\\
0.85	11.8077080966206\\
};
\addlegendentry{\scalebox{1.5}{$\log \mathbb E_{N^{1-\delta/2}}[\hat T_0]$}}
\addplot [color=mycolor3, dashed, line width=2.0pt]
  table[row sep=crcr]{%
0.05	7.06823351732353\\
0.06	7.12579814464838\\
0.07	7.18336277197323\\
0.08	7.24092739929808\\
0.09	7.29849202662293\\
0.1	7.35605665394778\\
0.11	7.41362128127263\\
0.12	7.47118590859748\\
0.13	7.52875053592233\\
0.14	7.58631516324719\\
0.15	7.64387979057204\\
0.16	7.70144441789689\\
0.17	7.75900904522174\\
0.18	7.81657367254659\\
0.19	7.87413829987144\\
0.2	7.93170292719629\\
0.21	7.98926755452114\\
0.22	8.04683218184599\\
0.23	8.10439680917085\\
0.24	8.1619614364957\\
0.25	8.21952606382055\\
0.26	8.2770906911454\\
0.27	8.33465531847025\\
0.28	8.3922199457951\\
0.29	8.44978457311995\\
0.3	8.5073492004448\\
0.31	8.56491382776965\\
0.32	8.62247845509451\\
0.33	8.68004308241936\\
0.34	8.73760770974421\\
0.35	8.79517233706906\\
0.36	8.85273696439391\\
0.37	8.91030159171876\\
0.38	8.96786621904361\\
0.39	9.02543084636846\\
0.4	9.08299547369331\\
0.41	9.14056010101817\\
0.42	9.19812472834302\\
0.43	9.25568935566787\\
0.44	9.31325398299272\\
0.45	9.37081861031757\\
0.46	9.42838323764242\\
0.47	9.48594786496727\\
0.48	9.54351249229212\\
0.49	9.60107711961698\\
0.5	9.65864174694183\\
0.51	9.71620637426668\\
0.52	9.77377100159153\\
0.53	9.83133562891638\\
0.54	9.88890025624123\\
0.55	9.94646488356608\\
0.56	10.0040295108909\\
0.57	10.0615941382158\\
0.58	10.1191587655406\\
0.59	10.1767233928655\\
0.6	10.2342880201903\\
0.61	10.2918526475152\\
0.62	10.34941727484\\
0.63	10.4069819021649\\
0.64	10.4645465294897\\
0.65	10.5221111568146\\
0.66	10.5796757841394\\
0.67	10.6372404114643\\
0.68	10.6948050387891\\
0.69	10.752369666114\\
0.7	10.8099342934389\\
0.71	10.8674989207637\\
0.72	10.9250635480886\\
0.73	10.9826281754134\\
0.74	11.0401928027383\\
0.75	11.0977574300631\\
0.76	11.155322057388\\
0.77	11.2128866847128\\
0.78	11.2704513120377\\
0.79	11.3280159393625\\
0.8	11.3855805666874\\
0.81	11.4431451940122\\
0.82	11.5007098213371\\
0.83	11.5582744486619\\
0.84	11.6158390759868\\
0.85	11.6734037033116\\
};
\addlegendentry{\scalebox{1.5}{$ \log( \pi^{3/2 }/2 N^{\tfrac{1+\beta}{2}})  $}}
\end{axis}
\end{tikzpicture}%
    \caption{
    For fixed population size $N=10^5$ the predictions for the expected interclick time of the tournament ratchet based on i) a numerical calculation of the Green function (using formula \eqref{eq:generalgreen}) and ii) the asymptotics provided by Theorem~\ref{mainth}
    are compared with simulations.
    Each data point was obtained  by pooling the interclick times no.~5 to 30 from 20 simulations of the tournament ratchet for the corresponding parameter configuration. Each plot shows this for one fixed value of $\gamma$ and for varying~$\beta$.}
\label{fig:6F}  
\end{figure}
\FloatBarrier
\section{Proof of Theorem~\ref{mainth}}\label{secGreenproof}
\subsection{Green function}
The proof is based on an asymptotic analysis of the Green function of $Y=Y^N$,
$$G(j,n):= G^N(j,n) =  \E_j\left[\int_0^{T_0} I_{\{Y^N_t=n\}} \dif t\right], \quad 1\le j,n \le N$$
as $N\to \infty$.
By assumption the upward and downward jump rates of $Y$ from $n$ are given by 
\begin{equation}\label{bestupproof}
\lambda_n:=n\left(\frac 12\left(1-\frac nN\right) + \frac m\rho \left(1-\frac nN\right)\right),
\end{equation}
  \begin{equation*}
\mu_n: =n\left(\frac 12\left(1-\frac nN\right)+m\right).
\end{equation*}
Recall that all quantities, including $\lambda_n$ and $\mu_n$, depend on $N$, even if we suppress this in the notation for the sake of readability. 
 We express the Green function in terms of the {\it oddsratio products}
\begin{equation}\label{oddsrpr}
    r_0:=1, \quad r_k:= \prod_{l=1}^k \frac{\mu_l}{\lambda_l} , \qquad k\in \left\{1, \ldots , N-1\right\}.
\end{equation}
The following lemma is well known, see e.g.  \cite[(2.4)]{sagitov2013extinction} for a proof of \eqref{expr_text} via 
a decomposition with respect to excursions from $j$.
For convenience we include a derivation of \eqref{eq:generalgreen} in Section \ref{sec:greenproof}.  See also \cite[(15)]{doering} for a similar representation of $G(j, n)$. 
\begin{lem}\label{lem:green}
  For $1\le j, n \le N$, 
    \begin{equation}\displaystyle
 G(j, n)=\frac{1}{\mu_n} \sum_{l=0}^{j-1 \wedge n-1} \prod_{k=l+1}^{n-1} \frac{\lambda_k}{\mu_k}.\label{eq:generalgreen} 
\end{equation}
\end{lem}
In Figure~\ref{fig:green}, formula \eqref{eq:generalgreen} is compared to empirical occupation times from simulations of the process $Y$.\\
%
With
  \begin{equation*}
    R_k:= \sum_{i=0}^{k-1}r_i, \qquad k \in \left\{ 1,\ldots, N \right\},
  \end{equation*}
  we obtain 
  from \eqref{eq:generalgreen}:
  \begin{equation}\label{generalgreen1}
  G(j, n)=
 \begin{cases} \frac{R_{j\wedge n}}{\lambda_n r_n} \mbox{ if } n < N,
 \\ 
 \phantom . \vspace{-0.3cm}
 \\  \frac{R_{j}}{\mu_N r_{N-1}}  \mbox{ if } n = N.\end{cases}
  \end{equation}
Consequently,
\begin{equation} \label{expr_text} \E_j[T_0]= \sum_{n=1}^N G(j, n)= \sum_{n=1}^{N-1} \frac{R_{n\wedge j}}{\lambda_n r_n} +   \frac{R_{j}}{\mu_N r_{N-1}} .\end{equation}
 Note that
  \begin{equation}\label{potential}
    U(j):=\log r_{j} 
  \end{equation}
  (sometimes also referred to as {\it potential}, cf. \cite[(16)]{doering}) is an additive functional, and
 \eqref{eq:generalgreen} translates into
  \begin{equation*}
    G(j,n) =  \frac{1}{\mu_n }  \sum_{l=0}^{j-1 \wedge n-1}  e^{-(U(n-1)-U(\ell))}.
  \end{equation*}
\FloatBarrier
    \subsection{Asymptotics for the cumulated oddsratio products}
    \label{rR}
In view of \eqref{expr_text} we are going to find
  asymptotics for the terms $r_k$ and $R_k$ as $N\rightarrow\infty$. 
\\
  Our analysis, see Lemmas~~\ref{lemma_tech3} and \ref{lem_equiv_Pij}, shows that, as $j$ increases, $r_j$ is essentially constant on a large interval, before it starts to decrease  as $j$ approaches the center of attraction $N(1-\rho)$. The asymptotics of the cumulated oddsratio products $R_j$ and of the terms $G(j,n)$  will be analysed, depending on the order of magnitude of $j$, in Lemmas~\ref{lemma_tech3}, \ref{lemma_tech10} and~\ref{Rrhighj}  for the polynomial regime, and in  Lemmas~\ref{lem_equiv_Pij} and \ref{lem_tech2} for the exponential regime.\\

We recall the notation $f(N) \ll g(N)$ from \eqref{ll}. Also, we recall that we 
usually suppress the $N$-dependence in the notation, as for example in $m, \rho$ and $j$.

We can express $\log r_j$ as
\begin{equation}
    \log r_j = \sum_{k=1}^j \log \left(\frac{\mu_k}{\lambda_k}\right)= j \log\left( \frac{1+2m}{1+2m/\rho} \right)+ \sum_{k=1}^j \log \left(\frac{1-k/((1+2m)N)}{1-k/N}\right).\label{ln_pij}
\end{equation}
This expression allows us the following asymptotic description which is  key in what follows.
\begin{lem}\label{lemma:betterpi}
  Let $\xi = \xi_N$ be a sequence converging to $0$ so slowly  that $\xi \gg m$. Then for $N$ large enough and $j \leq (1-\xi)N$
  \begin{equation}\label{newsandwich}
    0\leq \log r_j - j \log\left( \frac{1+2m}{1+2m/\rho} \right)- \sum_{l=1}^{\infty}\left(1- \frac{1}{(1+2m)^l}\right)  \frac{1}{l(l+1)}  \frac{j^{l+1}}{N^l}  \leq \mathrm{const}\cdot \frac{m}{\xi}.
  \end{equation}
\end{lem}
\begin{lem} \label{lemma_tech3} Let $K:=K_N  > 0 $ either be constant or a diverging sequence. 
Then for all $k\leq K \sqrt{N/m}$
\begin{equation} \label{piPismallj1} e^{k^2 m/((1+2m)N)-4(1-\rho)K\sqrt{mN}} \leq r_k \leq e^{k^2 m/N +K^3/(\sqrt{mN}-K) }\end{equation}
and
\begin{equation} \label{piPismallj2}  e^{-4(1-\rho)K\sqrt{mN}} \int_0^{k-1} e^{x^2 m/((1+2m)N)}\dif x\leq R_k \leq  e^{K^3/(\sqrt{mN}-K) } \int_1^k e^{x^2 m/N}\dif x.\end{equation}
\end{lem}
Lemma~\ref{lemma:betterpi} and Lemma~\ref{lemma_tech3}
will be proved in 
Section~\ref{sec:prooflemmabetterpi}. 
We conclude this subsection by showing \eqref{bridgeasympt}. To this end, note the two asymptotic equivalences
     $$\log \left( \frac{1+2m}{1+2m/\rho}  \right)= \log\left(1+\frac{2m(\rho-1)}{\rho+2m}\right)\sim 2m\left(1-\frac 1\rho\right)$$
     and
     $$\sum_{\ell=1}^{\infty} \left(1- \frac{1}{(1+2m)^\ell}\right) \frac{(1-\rho)^{\ell-1}}{\ell(\ell+1)}\sim2m\sum_{\ell=1}^{\infty} \frac{(1-\rho)^{\ell-1}}{\ell+1},$$
     which combine to
     \begin{equation*}
(1-\rho)^2\eta(m,\rho)\sim  \frac 1\rho -1 -(1-\rho) - \sum_{\ell=2}^\infty \frac{(1-\rho)^{\ell}}{\ell} = \frac 1\rho-1+\log\rho.
\end{equation*}
\subsection{The polynomial regime: Proof of Theorem~\ref{mainth}.\ref{th:mainth:a}} Throughout this subsection we assume $ Nm(1-\rho)^2  \to 0$ as $N \to \infty$.\\

We start with the expected number of returns to $a := \lfloor N(1-\rho)\rfloor$ of the process $Y$ when starting in $a$.
\eqref{generalgreen1} together with Lemma~\ref{lemma_tech3} gives
\begin{equation*}
    G(a, a) = \frac{R_{ a} }{\lambda_{ a}r_{ a}} \sim \frac{ \int_1^{ a} e^{ x^2 \frac{m}{N}}\dif{x} }{\lambda_{ a}r_{a}}.
\end{equation*}
From \cite[(1), (9)]{wolframDwason} we get
\begin{equation*}
    \int_1^{a}  e^{ x^2 \frac{m}{N}}\dif{x} \sim \frac{ e^{ a^2 m/N}  }{2 \frac{m}{N} a } ,
\end{equation*}
and hence by using \eqref{piPismallj1} as well as $\lambda_a \sim \tfrac12 \rho (1-\rho)  N $ we get
\begin{equation*}
    G( a, a) \sim \frac{1}{m \rho (1-\rho)^2 N} \qquad \mbox{ as }N\rightarrow\infty,
\end{equation*}
which together with the asymptotics $\lambda_a+\mu_a \sim \rho(1-\rho)N$ gives 
\begin{equation}\label{expret}
    G(a, a) (\lambda_a+\mu_a)\sim  \frac{1}{m (1-\rho)} \qquad \mbox{ as }N\rightarrow\infty. 
\end{equation}
In order to prove the rest
the following two lemmas will be proved in Section~\ref{sec:proofstechlemmapolynomial}.
Here and below, we will omit the Gauss brackets in the summation bounds for better readability.
\begin{lem}\label{lemma_tech10}
Let $\zeta:=\zeta_N\to 0$ such that $\zeta \gg \left[ N (1-\rho)^2 m \right]^{1/4}$,  
and $K=K_N$ such that $K\to \infty$ and $$K\left((1-\rho)\sqrt{Nm}\vee  (Nm)^{-1/6} \right)\to 0.$$
Then
$$\sum_{k = \zeta \sqrt{N/m}}^{K\sqrt{N/m}} \frac{R_k}{\lambda_k r_k} =  \sqrt{\frac{N}{m}} \left(\frac{\pi^{3/2}}{2}+O({\zeta})+O\left(\frac 1K\right)\right) \mbox{ as }N \to \infty.$$
\end{lem}
For the sake of readability, let us introduce the function $f(k)$ via
\begin{equation}\label{def_K}
 \frac{m}{N}k^2f(k):= k \log\left( \frac{1+2m}{1+2m/\rho} \right)+ \sum_{l=1}^{\infty} \left(1- \frac{1}{(1+2m)^l}\right)  \frac{k^{l+1}}{l(l+1)N^l}.
\end{equation}
\begin{lem}\label{Rrhighj}\footnote{This corrects a mistake from the previous version v2 (published as \cite{IGSW}): There the estimates of $F4$ and $F5$, see Section~\ref{subsection:ratchetcorrectedmistakes}, and the statement of Lemma~\ref{Rrhighj} were not entirely correct. This is amended in the present version.}
Let $K= K_N$ and $\xi = \xi_N$ be sequences with $K_N\to \infty$ and  $1\gg \xi \gg m$. Then there exists a constant $C>0$ such that for all  $j$ with $K\sqrt{N/m} \le
 j\leq N(1-\xi)$ 
\begin{equation}\label{Rrest}
  \frac{R_j}{r_j} \le \mathrm{const} \frac 1m  \frac N{jf(j)}.
\end{equation}
\end{lem}
With these two lemmas we have the tools for proving Theorem~\ref{mainth}.\ref{th:mainth:a}, which concerns the polynomial regime. We will distinguish between the cases $j\ll \sqrt{\tfrac{N/m}{\log{(N/m)}}}$ and $j\gg \sqrt{{N/m}}$, since the potential $U$ (given by \eqref{potential}) turns out to be essentially flat far below $\sqrt{N/m}$.
\subsubsection{Proof of \eqref{polysmall}}\label{subsecsmallE4ref} Abbreviating $\gamma := \log(1/m)$ and recalling that we are  in the case 
\begin{equation}\label{condsmallj}
  j \ll \sqrt{\frac{N/m}{\log{(N/m)}}},  
\end{equation}
we decompose the mean extinction time from state $j$ given by  \eqref{expr_text} as follows 
\begin{eqnarray}
  \E_j[T_0]&=&\sum_{k=1}^{j-1} \frac{R_{k\wedge j}}{\lambda_k r_k}+\sum_{k=j}^{\gamma\sqrt{N/m}} \frac{R_{k\wedge j}}{\lambda_k r_k}+\sum_{k=\gamma\sqrt{N/m}+1}^{N} \frac{R_{k\wedge j}}{\lambda_k r_k}\nonumber\\
 &=&\sum_{k=1}^{j-1} \frac{R_{k}}{\lambda_k r_k}+R_{j}\sum_{k=j}^{\gamma\sqrt{N/m}} \frac{1}{\lambda_k r_k}+R_{j}\sum_{k=\gamma\sqrt{N/m}+1}^{N-1} \frac{1}{\lambda_k r_k}+ \frac{R_j}{\mu_N r_{N-1}}\nonumber\\
 &=:&E_1(j)+E_2(j)+E_3(j)+E_4(j).\label{eq:defe2}
\end{eqnarray}
The term $E_4(j)$ is bounded by  $\tfrac {R_{N-1}}{\mu_Nr_{N-1}}$. Since $ R_{N-1}/r_{N-1}  \le N$ and $\mu_N = m N$ we have $\tfrac {R_{N-1}}{\mu_Nr_{N-1}}  \le 1/m$,  which is $o\left(\sqrt{N/m}\right)$ because of the standing assumption that $Nm \to \infty$.\\
In view of the asymptotics 
\begin{equation} \label{eq:RKasymptstar}
    R_k \sim k \quad \mbox{ for } k \ll \sqrt{N/m}
\end{equation} and $\lambda_k \sim k/2$ for $k \ll N$, and because of the inequality $\lambda_k r_k = \mu_k r_{k-1}\geq mk r_{k-1}$ for any $k \leq N-1$, we have 
$$ E_1(j)+ E_3(j) \leq 4\sum_{k=1}^{j-1} \frac{k}{k}+\frac{2j}{m}\sum_{k=\gamma\sqrt{N/m}+1}^{N} \frac{1}{k r_{k-1}}. $$ 
Recall the definition of $f$ from \eqref{def_K}.
We see that $f(k)\geq 1/2$ when $k \gg \sqrt{N/m}$. 
Hence there exists a finite constant $C$ such that 
$$\sum_{k=\gamma\sqrt{N/m}+1}^{N} \frac{1}{k r_{k-1}} \leq\sum_{k=\gamma\sqrt{N/m}}^{N} \frac{e^{-\frac{mk^2}{2N}}}{k}\leq
\int_\gamma^\infty \frac{e^{-x^2/2}}{x}\dif{x} \leq \frac{e^{-\gamma^2/2}}{\gamma^2}. $$
In order to see the first inequality we argue as follows: 
From Lemma~\ref{lemma:betterpi}
we get that for any $k$, 
\begin{eqnarray*} \log r_k &\geq& k \log\left( \frac{1+2m}{1+2m/\rho} \right)+ \sum_{l=1}^{\infty}\left(1- \frac{1}{(1+2m)^l}\right)  \frac{1}{l(l+1)}  \frac{k^{l+1}}{N^l} \\
& \geq&  k \log\left( \frac{1+2m}{1+2m/\rho} \right)+ \left(1- \frac{1}{(1+2m)}\right)  \frac{1}{2}  \frac{k^{2}}{N}.
\end{eqnarray*}
From the observation that 
$$  \log\left( \frac{1+2m}{1+2m/\rho} \right)\sim - \frac{2m}{\rho}(1-\rho), \quad
 \left(1- \frac{1}{(1+2m)}\right)  \frac{1}{2}  \frac{k}{N} \sim m \frac{k}{N} $$
and
$$ \frac{k}{N} \gg (1-\rho) $$
for $k$ at least of order $\sqrt{N/m}$ we see that $r_{k-1}\ge e^{mk^2/(2N)}$. 
From this we get
 $$E_1(j)+ E_3(j) \leq 4j+2j e^\gamma \frac{e^{-\gamma^2/2}}{\gamma^2}, $$
which is of lower order than the r.h.s of \eqref{polysmall}.
We will now analyse $E_2(j)$, which turns out to be the dominant term. 
First, as $j\ll \sqrt{N/m}$, it may be simplified as follows, 
$$ E_2(j)\sim 2j\sum_{k=j}^{\log(1/m)\sqrt{N/m}} \frac{1}{k r_k}. $$
By sandwiching arguments using 
\eqref{newsandwich} and \eqref{def_K}  
we conclude that
\begin{equation*}
  \sum_{k=j}^{\gamma\sqrt{N/m}} \frac{1}{k r_k}  \sim \int_j^{\gamma\sqrt{N/m}} \frac{e^{-\frac{m}{N}x^2f(x)}}{x}\dif{x}.
\end{equation*}
Thanks to \eqref{condsmallj} there exists a sequence  $\xi = \xi_N\to 0$  such that
$\xi^2 \gg 1/\log\left( \sqrt{N/m}\right)$ and $j_N \le \xi_N\sqrt{N/m}.$ 
From \eqref{piPismallj1}, 
if $k \leq \xi \sqrt{N/m}$ and $N$ is large enough, then
$ |  (m/N)k^2f(k)|\leq 2\xi^2. $
Hence
$$e^{-2\xi^2} \left( \log \left( \xi \sqrt{N/m} \right) - \log j \right)\leq \int_j^{\xi\sqrt{N/m}} \frac{e^{-\frac{m}{N}x^2f(x)}}{x}\dif{x} 
\leq e^{2\xi^2} \left( \log \left( \xi \sqrt{N/m} \right) - \log j \right). $$
 Moreover, $f(k) \geq \xi^2/2$ for $k \geq \xi \sqrt{N/m}$. We deduce 
$$ \int_{\xi\sqrt{N/m}}^{\gamma\sqrt{N/m}} \frac{e^{-\frac{m}{N}x^2f(x)}}{x}\dif{x}\leq \int_{\xi\sqrt{N/m}}^{\gamma\sqrt{N/m}} \frac{e^{-\frac{m}{N}x^2\xi^2/2}}{x}\dif{x}\leq \int_{\xi}^{\infty} \frac{e^{-y^2\xi^2/2}}{y}\dif{y}.  $$
By substituting $t= \xi y$ in the integral, the right hand side can be written as
\begin{equation*}
    \int_{\xi^2}^\infty \frac{ e^{-t^2/2} }{t} \dif{t},
\end{equation*}
which is of order $\xi^{-2}$. Since this is of lower order than $\log\left( \sqrt{N/m}\right)$
 we deduce that 
 \begin{equation}\label{boundofE2} E_2(j)\sim 2j \left( \log \left( \sqrt{N/m} \right) - \log j \right).
 \end{equation}
This ends the proof of of \eqref{polysmall}. \hfill $\qed$
\subsubsection{Proof of \eqref{polylarge}}\label{subsection:ratchetcorrectedmistakes} We recall that this concerns the case $j=j_N \gg \sqrt{N/m}$. Let $K=K_N$ be a sequence which converges to $\infty$ so slowly that $K\sqrt{N/m} \le j$ and that $K$ satisfies the requirements of Lemma \ref{lemma_tech3}. Moreover, let $\xi= \xi_N$ be a sequence with $\xi\to 0$ and $\xi \gg m$.
In the {\bf first part of the proof} we impose the condition
\begin{equation}\label{jandxi} j_N \leq N(1-\xi_N). 
\end{equation}
Let $\zeta = \zeta_N$  be a  sequence converging to $0$. 
Using again \eqref{expr_text}, 
we decompose the mean extinction time from state $j$ as follows:
\begin{eqnarray*}
   &&\E_j[T_0]\\
   &=&\sum_{k=1}^{\zeta \sqrt{N/m}} \frac{R_{k}}{\lambda_k r_k}+\sum_{k=\zeta \sqrt{N/m}+1}^{K\sqrt{N/m}} \frac{R_{k}}{\lambda_k r_k}+\sum_{k=K\sqrt{N/m}+1}^{ N(1-\xi) } \frac{R_{k\wedge j}}{\lambda_k r_k} + \sum_{ k=N(1-\xi)+1}^{N-1} \frac{R_j}{\lambda_k r_k} + \frac{R_j}{\mu_Nr_{N-1}} \\
 &=:&F_1+F_2+F_3(j)+F_4(j)+F_5(j).
\end{eqnarray*} 
The asymptotic of the second sum,  $F_2$, has been derived in Lemma~\ref{lemma_tech10} and leads to the r.h.s. of \eqref{polylarge}. It thus suffices to show that $F_1+ F_3(j) + F_4(j)+F_5(j)=o\left(\sqrt{N/m}\right)$.
To bound $F_1$ we need the following lemma, see Section~\ref{sec:proofstechlemmapolynomial} for a proof. 
\begin{lem}\label{lemma:antonnov}
  Assume $1\ll Nm \ll (1-\rho)^{-2}$. Then
  \begin{equation*}\label{antonnov:Rrbound}
\lim\limits_{N\to \infty} \sup\limits_{1\le j \le N/2} \frac{R_j}{r_j \lambda_j} \le 4.
\end{equation*}
\end{lem}
For any sequence for  $\zeta=\zeta_N \to 0$ we have\begin{equation*}\label{antonnov:zetaest}
\zeta \sqrt{\frac Nm} \ll \frac N2.
\end{equation*} 
and hence by Lemma~\ref{lemma:antonnov} we have that
\begin{equation*}
F_1  =  O\left(\zeta \sqrt{\frac Nm}\right)  = o\left(\sqrt{\frac Nm}\right).
\end{equation*}
Like $E_4(j)$ the term $F_5(j)$ is bounded by  $1/m$,  which is $o\left(\sqrt{N/m}\right)$ because of the standing assumption that $Nm \to \infty$.\\ 
Let us now turn to the analysis of $F_3(j)$. 
For this we have the upper bound 
\begin{equation}\label{estF3new}
    \sum_{k=K\sqrt{N/m}+1}^{ N(1-\xi)} \frac{R_{k}}{\lambda_k r_k}.
\end{equation} 
By using 
\begin{equation*}
    \lambda_k \sim \frac{k}{2} \left(1- \frac{k}{N}\right) = \frac12 \frac{k}{N}\left(N-k\right)
\end{equation*}
and the bound \eqref{Rrest}, the term
 \eqref{estF3new} is asymptotically bounded from above by 
\begin{equation*}
    \mathrm{const} \sum_{k=K\sqrt{N/m}+1}^{N-1} \frac{N}{k (N-k)} \frac{N}{2mk} 
    = \frac{N^2}{m} \sum_{k=K\sqrt{N/m}+1}^{N-1} \frac{1}{k^2 (N-k)}.
\end{equation*}
We claim that this is $o\left(\sqrt{N/m}\right)$, which is equivalent to
\begin{equation}\label{F3estimatenew}
    \frac{N^2}{m} \sqrt{\frac{m}{N}} \sum_{k=K\sqrt{N/m}+1}^{N-1} \frac{1}{k^2 (N-k)}
\end{equation}
converging to zero. This term we approximate by an integral
\begin{eqnarray*}
   &&\frac{N^2}{m} \sqrt{\frac{m}{N}} \sum_{k=K\sqrt{N/m}+1}^{N-1} \frac{1}{k^2 (N-k)} \\
   &=&\frac{N^2}{m} \sqrt{\frac{m}{N}} \cdot \frac{1}{N^3} \cdot N\cdot \frac{1}{N}  \sum_{k=K\sqrt{N/m}+1}^{N-1}  \frac{1}{(k/N)^2 (1-k/N)} \\&\sim& 
 \frac{N^2}{m} \sqrt{\frac{m}{N}} \cdot \frac{1}{N^3}\cdot N \cdot \int\limits_{ K/\sqrt{mN} }^{1-\frac{1}N} \frac{1}{x^2(1-x)}\dif{x}\\
 &=& \frac{1}{\sqrt{Nm}}\int\limits_{ K/\sqrt{mN} }^{1-\frac{1}N} \frac{1}{x^2(1-x)}\dif{x}.
\end{eqnarray*}
The integral is of order 
$$  \left(K/\sqrt{mN}\right)^{-1} \vee \log N  =(\frac 1K \sqrt{mN} ) \vee \log N.  $$
Thus \eqref{F3estimatenew} is of order $\tfrac{( \sqrt{Nm}/K) \vee \log N}{\sqrt{Nm}}$, which converges to zero as $N\to \infty$.\\
We are left with the analysis of
\begin{equation*}
  F_4(j)= \sum_{k=N(1- \xi)+1 }^{N-1} \frac{R_{k\wedge j}}{\lambda_k r_k}.
\end{equation*}
This sum is bounded by 
\begin{equation*}
    R_{j} \sum_{k=N(1- \xi)+1}^{N-1} \frac{1}{\lambda_k r_k}.
\end{equation*}
Since we assumed $j\leq N- \xi N$ this is again bounded from above by
\begin{equation*}
    \frac{R_{N(1- \xi)}}{r_{N(1- \xi)}} \sum_{k=N(1- \xi)+1 }^{N-1} \frac{r_{N(1- \xi)}}{\lambda_k r_k}.
\end{equation*}
Recall from \eqref{ln_pij}  that for $j,\ell \in \N$,
$$  \log r_{j+l}-\log r_j = \ell \log\left( \frac{1+2m}{1+2m/\rho} \right)+ \sum_{k=j+1}^{j+\ell} \log \left(\frac{1-k/((1+2m)N)}{1-k/N}\right). $$
Noticing that the term 
$$ \log \left(\frac{1-k/((1+2m)N)}{1-k/N}\right) $$
is increasing with $k$, and performing a Taylor expansion, we obtain that for
\begin{equation*}
  k= N(1- \xi) +\ell
\end{equation*}
we have the inequality
\begin{equation}\label{eq:rismontonedefacto}
  \log r_k \geq \log r_{N(1-\xi) }.
\end{equation}
Together with Lemma~\ref{Rrhighj} 
we obtain
\begin{eqnarray*}
  && \frac{ R_{N(1-\xi)} }{r_{N(1-\xi )}} \sum_{k=  N (1-\xi )+1 }^{N-1} \frac{r_{N (1-\xi )}  }{\lambda_k r_k}\\
  &\leq& \frac{\mathrm{const}}{(1-\xi)mf(N(1-\xi))} \cdot \sum_{k=  N (1-\xi )+1 }^{N-1}  \frac{r_{N (1-\xi )}  }{\lambda_k r_k}.
\end{eqnarray*}
By \eqref{eq:rismontonedefacto} and since for $\ell=0,\ldots, \xi N$
\begin{eqnarray*}\lambda_{(1-\xi)N+\ell} &\geq&\frac12 \left((1-\xi)N+\ell\right)\left(1-(1-\xi)-\frac \ell N\right)
   \nonumber   \\ &=&\left((1-\xi)+\frac \ell N\right)(\xi N-\ell)  \sim  \xi N -\ell\nonumber
\end{eqnarray*}
we obtain
\begin{equation}\label{eq:sumwhichislogNxi}
  \sum_{k=  N (1-\xi )+1 }^{N-1}  \frac{r_{N (1-\xi )}  }{\lambda_k r_k}= O\left( \log (N\xi) \right).
\end{equation}
Now observe that for $N$ large enough
\begin{eqnarray*}
  f\left((1-\xi)N\right)&\geq&\frac{N}{m ((1-\xi)N)^2} \sum_{l=1}^{\infty} \left(1- \frac{1}{(1+2m)^l}\right)  \frac{N^{l+1} (1-\xi)^{l+1} }{l(l+1)N^l}
                               \nonumber\\
  &&+\frac{N}{m(1-\xi)N} \log\left( \frac{1+2m}{1+2m/\rho} \right)
                               \nonumber\\
      &=& \frac{1}{m (1-\xi)^2 } \sum_{l=1}^{\infty}  \frac{ (1+2m)^l-1 }{(1+2m)^l}   \frac{ (1-\xi)^{l+1} }{l(l+1)}+\frac{1}{m(1-\xi)} \log\left( \frac{1+2m}{1+2m/\rho} \right)\nonumber
      \\
      \\
      &\geq& \frac{1}{m (1-\xi)^2} \sum_{l=1}^\infty \frac{2m l}{(1+2m)^l}\frac{ (1-\xi)^{l+1} }{l(l+1)}-\mathrm{const}\nonumber\\
                        &=& \frac{2}{(1-\xi)} \sum_{l=1}^{\infty} \frac{(1-\xi)^{l}}{(1+2m)^l}\frac 1{l+1}-\mathrm{const}\nonumber\\
&=& \frac{2 (1+2m)}{(1-\xi)^2} \sum_{l=2}^\infty \frac{(1-\xi)^l}{(1+2m)^l } \frac{1}{l}-\mathrm{const}\nonumber.
\end{eqnarray*}
    By the identity
    \begin{equation*}
      \sum_{k=2}^\infty \frac{z^k}k = -z - \log(1-z), \qquad |z| < 1
    \end{equation*}
     and $\xi \gg m$ we get that there exists a constant $c>0$, such that for $N$ large enough
    \begin{equation*}
f\left((1-\xi)N\right) \ge  c|\log \xi|.
\end{equation*}
If  $m\ll N^{-\epsilon}$ for some $\epsilon$, then
together with \eqref{eq:sumwhichislogNxi} we  obtain 
\begin{equation}\label{eq:F4jisofrightorder}
  F_4(j)=
o \left( \sqrt{\frac{N}{m}}\right)
\end{equation}
for some sequence $\xi$ obeying $m\ll N^{-\epsilon}  \ll \xi$. If $m \gg  (\log N)^2/N$,
then by \eqref{eq:sumwhichislogNxi} we again obtain \eqref{eq:F4jisofrightorder}. If neither $m\ll N^{-\epsilon}$ for some $\epsilon > 0$ nor $m \gg  (\log N)^2/N$ holds, then both are true along two subsequences, which together cover the entire sequence $(m_N)_N$. 
This finishes the proof of \eqref{polylarge} in the case $j \le N(1-\xi).$

In the {\bf remaining part of the proof} we consider sequences which not necessarily satisfy the restriction \eqref{jandxi}.
In view of the first part it suffices to show that the expected time which $Y$ needs to come down from $N$ to  $N(1- \xi)$ is of lower order than $\sqrt{N/m}$. For this, we impose an additional condition on the sequence $\xi$, and will show the following claim:
Let $\xi = \xi_N$ be a sequence converging to $0$ and obeying $m  \ll \xi \ll m (N/m)^{1/4}$ as $N\to \infty$. Then  $\mathbb E_N[T_{N(1-\xi)}]=o(\sqrt{N/m})$.

To prove this claim, let  $\mathcal Y$ be the time-discrete  birth-and-death process corresponding to~$Y$.  By \eqref{bestdown} and \eqref{bestup} the probability of $\mathcal Y$ to go down in the next step when starting in $k$ is given by
\begin{equation*}
  \frac{ \frac12 \left(1-\frac{k}{N}\right)+m }{ \left(1-\frac{k}{N}\right) + m + \frac{m}{\rho} \left(1-\frac{k}{N}\right) } .
\end{equation*}
 which for $N(1-\xi) \le k \le N$ is bounded from below by
\begin{equation*}
q:=  \frac12 \frac{ 1+ \frac{2m}{\xi} }{1 + \frac{m}{\xi}+ \frac{m}{\rho} }.
\end{equation*}
Let us put $p:= 1-q$, and consider the $(p,q)$-random walk $W$ on $\mathbb Z$ as well as the random walk $\widehat W$ on $\mathbb Z \cap \{\ldots, N-2, N-1, N\}$ that is obtained by reflecting $W$ at $N$, i.e. by putting $\mathbb P_N(\widehat W_1=N) := p$,   $\mathbb P_N(\widehat W_1=N-1) := q$. A suitable coupling of $\mathcal Y$ and $\widehat W$ (both starting in $N$) shows that for $N(1-\xi) \le k \le N$ the expected number visits of $\mathcal Y$ to $k$ before $\mathcal Y$ reaches $N(1-\xi)$ is not larger than the expected number of visits of  $\widehat W$ to $k$  before $\widehat W$ reaches $N(1-\xi)$.
The expected number of visits of the transient random walk $W$ to its starting point is $\frac q{q-p}\sim \frac{\xi}m$, and the same is true for $\widehat W$.  The jump rates \eqref{bestdown} and \eqref{bestup} from  state $k\ge N(1-\xi)$ add up to
\begin{equation*}
  k \left[ \frac{N-k}{N} + m + \frac{N-k}{N} \frac{m}\rho \right]\ge  \frac12 Nm.
\end{equation*}
Altogether, we obtain the estimate
\begin{equation} \label{ext_from_N} \mathbb E_N[T_{N(1-\xi)}] \le \frac 2{Nm} \xi N \frac q{p-q}\sim \frac{2\xi^2}{m^2},\end{equation} 
whose r.h.s. is $o(\sqrt{N/m})$ due to our assumption on $\xi$.

This concludes the proof of Theorem~\ref{mainth}.\ref{th:mainth:a}. \hfill $\qed$
\subsection{The exponential regime: Proof of Theorem~\ref{mainth}.\ref{th:mainth:b}}
Throughout this section we assume $ Nm(1-\rho)^2  \to \infty$. In this  regime the process $Y$ should spend a long time around its center of attraction $(1-\rho)N$, which makes the following decomposition of \eqref{expr_text} natural:  for a small $\zeta>0$ write
\begin{eqnarray}
  \E_j[T_0] &= &\sum_{k=1}^{(1-\zeta)(1-\rho)N} \frac{R_{k\wedge j}}{\lambda_k r_k} +\sum_{k=(1-\zeta)(1-\rho)N+1}^{(1+\zeta)(1-\rho)N} \frac{R_{k\wedge j}}{\lambda_k r_k} +\sum_{k=(1+\zeta)(1-\rho)N+1}^{N-1} \frac{R_{k\wedge j}}{\lambda_k r_k}+ \frac{R_j}{\mu_N r_{N-1}} \nonumber \\
&=:& A(\zeta)+B(\zeta)+C(\zeta)+ \frac{R_j}{\mu_N r_{N-1}}.
  \label{eq:sumsplitup}
\end{eqnarray}
The assertion of Theorem~\ref{mainth}.\ref{th:mainth:b} will be derived at the end of this section from Proposition~\ref{lem_tech2}, whose proof, in turn, will rely on the following lemma.
The proof of this lemma as well as that of Proposition~\ref{lem_tech2} will be given in Section \ref{Prooftec2}.
\begin{lem}
     \label{lem_equiv_Pij}
Let $j =j_N$ be a sequence of natural numbers converging to $\infty$, and let $\xi < 1/2$. 
Then
\begin{itemize}
\item  If $j \leq   \xi(1-\rho)N $, then for sufficiently large $N$
\begin{equation*}
 \frac{1- e^{-(1+2\xi)2m(1-\rho)j/\rho} }{(1+2\xi)2m(1-\rho)/\rho}\leq R_j
 \leq 
\frac{1- e^{-(1-2\xi)2m(1-\rho)j/\rho} }{(1+2\xi)2m(1-\rho)/\rho}. \end{equation*}
\item If $1/(m (1-\rho)) \ll j \leq (2-\xi)(1-\rho)N \wedge N(1-\sqrt{m})$, then, under the assumption $\xi \geq 2 \log (mN(1-\rho)^2)/(mN(1-\rho)^2)$, 
$$ R_j \sim \frac{{\rho}}{2m(1-\rho)}  \quad \mbox{ as } N\to \infty. $$
\item If $j = C (1-\rho)N\le N(1-\sqrt m)$, 
with $\frac 1{1-\rho} \ge C> 2/\rho$ (implying $\rho > \tfrac23$), then  
$$R_j \sim \rho(1-C(1-\rho)) \frac{\exp \left( -2m(1-\rho)^2N H(C)\right)}{2m(C-1)(1-\rho)}  \quad \mbox{ as } N\to \infty, $$
where the function $H(.) = H((m,\rho), .)$ on $\R_+$ is defined by
 \begin{equation} \label{def_H} H(y):=  -\frac{y}{2m} \left[\frac{1}{1-\rho}\log \left( \frac{1+2m}{1+2m/\rho}  \right)+ \sum_{l=1}^{\infty} \left(1- \frac{1}{(1+2m)^l}\right) \frac{(1-\rho)^{l-1}y^{l}}{l(l+1)}\right]. \end{equation}
\end{itemize}
\end{lem}
This is the central ingredient for the proof of the following proposition, which, in turn, will be key for the proof of Theorem~\ref{mainth}.\ref{th:mainth:b}.
\begin{prop}\label{lem_tech2}
Let $A(\zeta)$, $B(\zeta)$ and $C(\zeta)$ be defined by~\eqref{eq:sumsplitup}. Then for $\zeta =\zeta_N$ converging to 0 so slowly that $ \zeta \sqrt{mN}(1-\rho) \to \infty $ , we have
\begin{equation} \label{simBeps}
B(\zeta) \sim \left( R_j \wedge \frac{\rho}{2m(1-\rho)} \right) \sqrt{\frac{\pi}{m N}}\frac{2}{1-\rho}\exp   \left( 2m(1-\rho)^2N H(1) \right) \quad \mbox{ as } N \to \infty,
\end{equation}
and 
\begin{equation*}
A(\zeta)+ C(\zeta) = o (B(\zeta)) \quad \mbox{ as } N \to \infty. 
\end{equation*}
\end{prop}
The proof of this proposition will be given in Section \ref{Prooftec2}. 

\medskip
{\it Proof of Theorem~\ref{mainth}.\ref{th:mainth:b}}.
First note that we can asymptotically neglect $\tfrac{R_j}{\mu_N r_{N-1}}$ with the same arguments as those used for $E_4(j)$ and $F_5(j)$, see Section~\ref{subsecsmallE4ref} and Section~\ref{subsection:ratchetcorrectedmistakes}.\\ 
(i) For $j=j_N = O(1/(m(1-\rho)))$ and any sequence $\xi=\xi_N$ converging to zero we have 
\begin{equation*}
    1-\exp\left( -(1\pm 2\xi) 2m(1-\rho)j/\rho\right) \sim 1-\exp\left( - 2m(1-\rho)j/\rho\right).
\end{equation*}
Hence Proposition~\ref{lem_tech2} together with the first bullet point of 
Lemma~\ref{lem_equiv_Pij}  gives 
\begin{eqnarray*}
    \E_j[T_0] \sim B(\zeta) \sim  \frac{1-\exp\left( - 2m(1-\rho)j/\rho\right) }{ 2m(1-\rho) }\sqrt{\frac{\pi}{mN}} \frac{2\rho}{1-\rho}\exp\left( 2m(1-\rho)^2 N H(1)\right) ,
\end{eqnarray*}
(ii) For $j\gg 1/(m(1-\rho))$ Proposition~\ref{lem_tech2} together with the second bullet point of Lemma~\ref{lem_equiv_Pij} gives 
\begin{eqnarray*}
    \E_j[T_0] \sim B(\zeta) \sim \frac{\rho}{m(1-\rho)^2} \sqrt{\frac{\pi}{mN}} \exp\left( 2m(1-\rho)^2 N H(1)\right).
\end{eqnarray*}
\\
(iii) To conclude 
\eqref{expclr} from (i) and (ii) it suffices to observe that
$\eta(m, \rho)= H(1)$, and that the assumption on $j$ in (ii) implies the convergence $1-\exp(-2m(1-\rho) j_N/\rho) \to 1$ as $N\to \infty$.
The claimed asymptotics \eqref{threecases}  is an immediate consequence of \eqref{expclr}.
\hfill\\
(iv) It remains to prove the claim on the expected number of excursions from $a:= \lfloor \mathfrak a\rfloor$, with $\mathfrak a = (1-\rho)N$ being the asymptotic center of attraction of $Y$. In view of \eqref{generalgreen1} this expected number equals
\begin{equation}
\label{asympV}(\lambda_a + \mu_a)G(a,a) = (\lambda_a + \mu_a)\frac{R_a}{\lambda_a r_a}.
\end{equation}
In order to estimate $r_a$ we observe that
\eqref{newsandwich}, when expressed in terms of the function $H$ (which was defined in \eqref{def_H}), gives the asymptotics
\begin{equation}\label{asra}
\frac 1{r_a} \sim \exp\left( 2m (1-\rho)^2 N H(1)\right) \quad \mbox{ as } N\to \infty.
\end{equation}
In addition, the second bullet point of Lemma~\ref{lem_equiv_Pij} gives 
\begin{equation}\label{asRa}
R_a \sim \frac{\rho}{2m(1-\rho)} \quad \mbox{ as } N\to \infty.
\end{equation}
Since $\lambda_a \sim \mu_a$ as $N\to \infty$, the combination of \eqref{asra} and \eqref{asRa} shows that \eqref{asympV} is asymptotically equivalent to \eqref{defvN}. \hfill$\Box$

\subsection{Proofs of Lemmas~\ref{lem:green}, \ref{lemma:betterpi} and \ref{lemma_tech3}}\label{sec:greenproof}\label{sec:prooflemmabetterpi}
\begin{proof}[Proof of Lemma~\ref{lem:green}]
We denote the time-discrete embedded process corresponding to $Y$ by $\mathcal Y$, and write $\mathcal G(m,n)$ for the expected number of visits at $n$ of  $\mathcal Y$ when starting in $m$. 
Let us start with an analysis of $\mathcal G(n,n)$. 
By standard arguments we have
\begin{equation}\label{tildeG}
\mathcal  G(n,n) = \frac 1 {\phi(n)},
\end{equation}
where $\phi(n)$ is the escape probability of $\mathcal Y$ from the state $n$, i.e.
\begin{equation}\label{escape}
\phi(n) =\frac {\mu_n}{ \mu_n+\lambda_n}(1-h^{(n)}(n-1)),
\end{equation}
where $h^{(n)}: \{0,1,\ldots, n\} \to [0,1]$ is  $\mathcal Y$-harmonic on $\{1,\ldots, n-1\}$
and satisfies the boundary conditions $h^{(n)}(0)=0$,  $h^{(n)}(n)=1$. Hence  
\begin{equation}\label{hnn}
h^{(n)}(\ell) = \frac {\sum_{j=0}^{\ell-1} r_j}{\sum_{k=0}^{n-1}r_k}, \quad \ell= 0,\ldots, n,
\end{equation}
with the oddsratio product $r_k$ as in \eqref{oddsrpr}. 
From~\eqref{hnn} we obtain
\begin{equation}\label{oneminush}
1-h^{(n)}(n-1) = \frac{r_{n-1}}{\sum_{k=0}^{n-1}r_k}.
\end{equation}
For $G(n,n)$, the expected time spent by ${ Y}$ in $n$ when starting in $n$, we thus obtain the relation
$$G(n,n) = \frac {\mathcal G(n,n)}{\lambda_n+\mu_n}  = \frac{1}{\phi(n)}\cdot  \frac{1}{\mu_n+\lambda_n}.$$
Combining this with \eqref{tildeG}, \eqref{escape} and \eqref{oneminush} we arrive at
\begin{equation}\label{Green1}
G(n,n) = \frac{1}{\mu_n} \cdot \sum_{k=0}^{n-1} \frac {r_k}{r_{n-1}} =  \frac{1}{\mu_n} \sum_{l=0}^{n-1} \prod_{k=l+1}^{n-1} \frac{\lambda_k}{\mu_k}.
\end{equation}
For $j > n$ we have
\begin{equation}\label{mbiggern}
  G(j,n) = G(n,n),
\end{equation}
while for $j<n$ 
\begin{equation*}
   G(j, n)=   h^n(j)G(n,n) = \frac{ \sum_{l=0}^{j-1}r_l }{\sum_{l=0}^{n}r_l} G(n,n). 
 \end{equation*}
Together with \eqref{Green1} this gives for $j < n$
 \begin{eqnarray}
    G(j, n)
           &=& \frac{1}{\mu_n} \sum_{k=0}^{n-1} \frac{r_k}{r_{n-1}} \cdot \frac{ \sum_{l=0}^{j-1}r_l }{\sum_{l=0}^{n-1}r_l}\nonumber\\
   &=&  \frac{1}{\mu_n} \sum_{l=0}^{j-1} \prod_{k=l+1}^{n-1} \frac{\lambda_k}{\mu_k}.\label{msmallern}
 \end{eqnarray}
It remains to observe that the three cases $j>n$, $j=n$, $j<n$ (which are covered by  \eqref{mbiggern},  \eqref{Green1}   \eqref{msmallern}) combine to  \eqref{eq:generalgreen}
\end{proof} 
\begin{proof}[Proof of Lemma~\ref{lemma:betterpi}]
  We have already seen 
\begin{eqnarray*}
\log r_j = \sum_{k=1}^j \log \left(\frac{\mu_k}{\lambda_k}\right)&=& j \log\left( \frac{1+2m}{1+2m/\rho} \right)+ \sum_{k=1}^j \log \left(\frac{1-k/(1+2m)N}{1-k/N}\right)\\ 
& =&j \log\left( \frac{1+2m}{1+2m/\rho} \right)+\sum_{k=1}^j \sum_{l=1}^{\infty} \left(1- \frac{1}{(1+2m)^l}\right)  \frac{1}{l}\left(\frac{k}{N}\right)^l.
\end{eqnarray*}
As $j \leq N-1$, we may apply Fubini's theorem to write 
$$ \sum_{k=1}^j \sum_{l=1}^{\infty} \left(1- \frac{1}{(1+2m)^l}\right)  \frac{1}{l}  \left(\frac{k}{N}\right)^l= \sum_{l=1}^{\infty}\left(1- \frac{1}{(1+2m)^l}\right)  \frac{1}{l}  \sum_{k=1}^j \left(\frac{k}{N}\right)^l. $$

Now sandwiching arguments yield
\begin{eqnarray*}
    \frac{j^{l+1}}{(l+1)N^l}&=& \int_0^j \left(\frac{x}{N}\right)^l\dif{x}\\&\leq&  \sum_{k=1}^j\left(\frac{k}{N}\right)^l \\&\leq&  \int_1^{j} \left(\frac{x}{N}\right)^l\dif{x} +\left(\frac{j}{N}\right)^l = \frac{1}{(l+1)N^l}(j^{l+1}-1)+\left(\frac{j}{N}\right)^l. 
\end{eqnarray*}
This means that if we introduce
$$\Delta_j:=   \sum_{l=1}^{\infty}\left(1- \frac{1}{(1+2m)^l}\right)  \frac{1}{l}  \sum_{k=1}^j \left(\frac{k}{N}\right)^l-\sum_{l=1}^{\infty}\left(1- \frac{1}{(1+2m)^l}\right)  \frac{1}{l(l+1)}  \frac{j^{l+1}}{N^l}, $$
we have
$$ 0 \leq \Delta_j \leq  \sum_{l=1}^{\infty}\left(1- \frac{1}{(1+2m)^l}\right)\frac{1}{l}\left(\frac{j}{N}\right)^l  $$
From the observation that
$$ 1- \frac{1}{(1+2m)^l} \leq 2ml, $$
we get
$$0\leq \Delta_j\leq 2m  \sum_{l=1}^{\infty}  \left(\frac{j}{N}\right)^l. $$
Now let $\xi \gg m$. Then
$j \leq (1-2\xi)N$ implies that for $N$ marge enough, $j+1 \leq (1-\xi)N$. Hence
$$ \Delta_j\leq 2m  \sum_{l=1}^{\infty}  \left(1-\xi\right)^l = 2m \frac{1-\xi}{\xi}=o(1).
 $$
 This entails that there exists $C$ such that for any $j \leq (1-2\xi)N$ and $N$ large enough,
\begin{align}
 0\leq \log r_j - j \log\left( \frac{1+2m}{1+2m/\rho} \right)- \sum_{l=1}^{\infty}\left(1- \frac{1}{(1+2m)^l}\right)  \frac{1}{l(l+1)}  \frac{j^{l+1}}{N^l}  \leq \mathrm{const}\cdot \frac{m}{\xi}.
\end{align}
\end{proof}
\begin{proof}[Proof of Lemma~\ref{lemma_tech3}] We set out from \eqref{newsandwich}. 
Notice that for any $x,y \geq 0$, $1-(1+x)^{-y}\leq xy$. Hence, for  $k \leq K \sqrt{N/m}$,
\begin{eqnarray*}
 \sum_{l=1}^{\infty} \left(1- \frac{1}{(1+2m)^l}\right)  \frac{k^{l+1}}{l(l+1)N^l}&\leq&
 2m\sum_{l=1}^{\infty} \frac{k^{l+1}}{(l+1)N^l}\leq
 m\sum_{l=1}^{\infty} \frac{k^{l+1}}{N^l}\\
 &=&   k^2 m/N + mN \sum_{l=3}^{\infty} \frac{k^{l}}{N^l}\\
 &=&   k^2 m/N + mN \frac{k^{3}}{N^3} \frac{1}{1-k/N}\\
 & \leq & k^2 m/N + \frac{K^3}{\sqrt{mN}-K}.
\end{eqnarray*}
Conversely we have
\begin{equation*}
 \sum_{l=1}^{\infty} \left(1- \frac{1}{(1+2m)^l}\right)  \frac{k^{l+1}}{l(l+1)N^l}\geq \left(1- \frac{1}{(1+2m)}\right)  \frac{k^{2}}{2N}
 \sim k^2 m/N.
\end{equation*}
Finally, for $k \leq K \sqrt{N/m}$ we also have
\begin{equation*}
\left|k \log\left( \frac{1+2m}{1+2m/\rho} \right)\right| =k \left|\log\left(1- \frac{2m(1-\rho)}{\rho+2m}\right)\right|
\sim  2mk (1-\rho) \leq 2K\sqrt{m(1-\rho)^2N}=o(1).
\end{equation*}
This concludes the proof of \eqref{piPismallj1}. The estimate \eqref{piPismallj2} then follows by approximating the sum $R_k := \sum_{l=0}^{k-1} r_l$ from  below by an integral from $0$ to $k-1$ and from above by an integral from $1$ to $k$. 
\end{proof}
\subsection{Proofs of Lemmas \ref{lemma_tech10}, \ref{Rrhighj} and \ref{lemma:antonnov} }\label{sec:proofstechlemmapolynomial}
\begin{proof}[Proof of Lemma~\ref{lemma_tech10}]
Noting that our choices of $K$ and $\zeta$ entail 
\begin{equation*}
    \frac{k^2 m}{(1+2m)N} \gg 4 (1-\rho) \sqrt{mN} \qquad \forall k\geq \zeta \sqrt{N/m}
\end{equation*}
as well as 
\begin{eqnarray*}
    4(1-\rho) K \sqrt{mN} &\rightarrow&0\\
    \frac{K^3}{\sqrt{mN}-1}&\rightarrow&0 .
\end{eqnarray*}
We can deduce
$$ \sum_{k = \zeta \sqrt{N/m}}^{K\sqrt{N/m}} \frac{R_k}{k r_k} \sim \int\limits_{y = \zeta \sqrt{N/m}}^{K\sqrt{N/m}} e^{- \frac{m}{N}y^2}\frac{\dif{y}}{y} \int\limits_{z=0}^y e^{ \frac{m}{N}z^2}\dif{z}=:I_N$$
from
\eqref{piPismallj1} and \eqref{piPismallj2}. 
We thus need to find an equivalent of $I_N$ for large $N$. Three successive changes of variables entail the equalities:
\begin{eqnarray*}
I_N&=& \int\limits_{y = \zeta \sqrt{N/m}}^{K\sqrt{N/m}}\frac{\dif{y}}{y} \int\limits_{z=0}^y e^{ -\frac{m}{N}(y^2-z^2)}\dif{z}\\&=& \int\limits_{y = \zeta \sqrt{N/m}}^{K\sqrt{N/m}}\dif{y} \int\limits_{\lambda=0}^1 e^{ -\frac{m}{N}y^2(1-\lambda^2)}d\lambda= \sqrt{\frac{N}{m}}\int\limits_{\lambda=0}^1 d\lambda\int\limits_{z = \zeta}^{K} e^{ -z^2(1-\lambda^2)}\dif{z}\\&=& \sqrt{\frac{N}{m}}\int\limits_{\lambda=0}^1 \frac{d\lambda}{\sqrt{1-\lambda^2}}\int\limits_{w = \zeta\sqrt{1-\lambda^2}}^{K\sqrt{1-\lambda^2}} e^{ -w^2}dw.
\end{eqnarray*}
Hence
$$ I_N \leq  \sqrt{\frac{N}{m}}\int\limits_{\lambda=0}^1 \frac{d\lambda}{\sqrt{1-\lambda^2}}\int\limits_{w =0}^{\infty} e^{ -w^2}dw=  \sqrt{\frac{N}{m}}\frac{\pi}{2}\frac{\sqrt{\pi}}{2}= \sqrt{\frac{N}{m}} \frac{\pi^{3/2}}{4}. $$
For the lower estimate we proceed as follows:
\begin{eqnarray*}
 \frac{\pi^{3/2}}{4}-I_N \sqrt{\frac{m}{N}}&= &\int\limits_{\lambda=0}^1 \frac{d\lambda}{\sqrt{1-\lambda^2}}\left(\int\limits_{w = 0}^{\zeta\sqrt{1-\lambda^2}} e^{ -w^2}dw + \int\limits_{w = K\sqrt{1-\lambda^2}}^\infty e^{ -w^2}dw\right)\\
 &\le&\int\limits_{\lambda=0}^1 \frac{d\lambda}{\sqrt{1-\lambda^2}}\left(\zeta + \int\limits_{w = K\sqrt{1-\lambda^2}}^\infty e^{ -w^2}dw\right)\\
 &\leq& \frac{\pi}{2}\zeta + \int\limits_{\lambda=0}^1 \frac{d\lambda}{\sqrt{1-\lambda^2}}\left(  \int\limits_{w = K\sqrt{1-\lambda^2}}^\infty e^{ -w^2}dw\right).
\end{eqnarray*}
We write the double integral on the r.h.s. as
\begin{equation}\label{eq:doubleinttobound}
    \int_0^1 \int_{z=K}^\infty e^{-z^2 (1-\lambda^2)}\dif{z}\dif{\lambda}= \int_{z=K}^\infty e^{-z^2} \int_0^1 e^{z^2 \lambda^2} \dif{\lambda}\dif{z}.
    \end{equation}
    By substituting $x=\lambda z$ the inner integral is equal to
    \begin{equation*}
        \frac{1}{z}\int_0^z  e^{x^2} \dif{x},
    \end{equation*}
    which by \cite[(1),(9)]{wolframDwason} is bounded from above by $\mathrm{const}\cdot e^{z^2}/z^2$, such that in total \eqref{eq:doubleinttobound} is bounded from above by
    \begin{equation*}
        \int_{K}^\infty e^{-z^2}\left(\mathrm{const}\cdot  \frac{e^{z^2}}{z^2} \right) \dif{z} \leq \mathrm{const}\cdot \frac{1}{K},
    \end{equation*}
    so in total
    \begin{equation*}
        \frac{\pi^{3/2}}{4}-I_N \sqrt{\frac{m}{N}} =  O(\zeta)+O\left(\frac 1K\right).
    \end{equation*}
Hence, as $\lambda_k \sim k/2$ for $k \ll N$, we have  proved that 
\begin{equation*}
\sum_{k = \zeta \sqrt{N/m}}^{K\sqrt{N/m}} \frac{R_k}{\lambda_k r_k} \sim \sum_{k = \zeta \sqrt{N/m}}^{K\sqrt{N/m}} 2 \frac{R_k}{k r_k} =  \sqrt{\frac{N}{m}} \left(\frac{\pi^{3/2}}{2}+O\left(\zeta\right)+O\left(\frac 1K\right)\right).  \end{equation*}
\end{proof} 
\begin{proof}[Proof of Lemma~\ref{Rrhighj}]
Lemma~\ref{lemma:betterpi} enables us to write
\begin{equation*}
  r_j \sim \exp\left( \frac{m}{N}j^2 f(j) \right)\qquad \mbox{ for }j\leq N - \xi N. 
\end{equation*}
So
\begin{eqnarray*}
  R_j &=& \sum_{l=0}^{j-1} r_l \sim \sum_{l=0}^{j-1} \exp\left( \frac{m}{N}l^2 f(l) \right)\\
&\leq& \mathrm{const} \cdot \int_0^j  \exp\left( \frac{m}{N}x^2 f(x) \right) \dif{x}
\end{eqnarray*}
and
\begin{eqnarray}
  \frac{R_j}{r_j}&\leq &\mathrm{const}  \exp\left( -\frac{m}{N}j^2 f(j) \right)  \cdot \int_0^j  \exp\left( \frac{m}{N}x^2 f(x) \right) \dif{x}\nonumber\\
                 &=&  \mathrm{const} \cdot \int_0^j  \exp\left( \frac{m}{N}\left(x^2 f(x)-j^2 f(j)\right) \right) \dif{x}.\label{eq:Rirj_int_tobound}
\end{eqnarray}
Since $f$ is non-decreasing this is bounded from above by
\begin{equation}\label{expRr}
 \mathrm{const} \exp\left( -\frac{m}{N} j^2f(j)  \right)  \cdot \int_0^j \exp\left( \frac{m}{N} x^2f(j)  \right) \dif{x}.
\end{equation}
By substituting $z= \sqrt{\tfrac{m}{N} f(j)}x$ the integral is equal to
\begin{equation*}
   \sqrt{\frac{N}{m f(j)}} \int_0^{ \sqrt{ \frac{m}{N}}jf(j) } e^{z^2} \dif{z}.
 \end{equation*}
 Since $f(x)\geq \frac12$ for $x \gg \sqrt{N/m}$, we obtain by applying \cite[(1),(9)]{wolframDwason}, that  this is bounded from above by 
\begin{equation*}
    \mathrm{const}\cdot \frac{ e^{ j^2 \cdot \frac{m}{N} }  }{j    }\cdot \frac{N}{m f(j)}.
\end{equation*}
So \eqref{expRr} - and hence also \eqref{eq:Rirj_int_tobound} - is asymptotically bounded by $\mathrm{const}\cdot \tfrac{N}{mj f(j)}$, which concludes the proof.
\end{proof}

\begin{proof}[Proof of Lemma~\ref{lemma:antonnov}]
 Note, that the mapping $k\mapsto r_k$ is decreasing on $\{0, \ldots, \mathfrak a\}$ and increasing on $\{\mathfrak a+1, \mathfrak a+2, \ldots\}$. Hence for $0\le k \le j$,
\begin{equation*}
\label{antonnov:cases}\frac {r_k}{r_j} \le \begin{cases} \frac {r_0}{r_{\mathfrak a}} \,\, \mbox{ if } k \le \mathfrak{a}\\ 1 \quad \mbox{ if } k > \mathfrak{a}.  \end{cases} \end{equation*}
From Lemma~\ref{lemma:betterpi} we get 
\begin{equation}\label{antonnov:rquot}
\frac{r_0}{r_{\mathfrak a}} \sim  e^{\varphi(0)-\varphi({\mathfrak a})} =\exp\left(-\int_0^{\mathfrak a} \varphi'(x)\, dx\right), 
\end{equation}
where for $x \ge 1$
\begin{eqnarray}
\varphi(x) &:=& \frac mN x^2 f(x) \nonumber\\
           &=&x\log\left(\frac{1+2m}{1+2m/\rho}\right) 
+\sum_{\ell=1}^\infty \left( 1 - \frac{1}{(1+2m)^\ell} \right) \frac{x^{\ell+1}}{\ell(\ell+1)N^\ell}.\nonumber
\end{eqnarray}
The derivative of $\varphi$ is
$$ \varphi'(x)= \log\left(\frac{1+2m}{1+2m/\rho}\right) +\sum_{\ell=1}^\infty \left( 1 - \frac{1}{(1+2m)^\ell} \right) \frac{x^\ell}{(\ell+1)N^\ell}.$$
We observe that for all sufficiently large $N$ and all $x < N$
\begin{equation*}\label{antonnov:phiprime}
\varphi'(x) \ge \log\left(\frac{1+2m}{1+2m/\rho}\right) \ge -4m(1-\rho).
\end{equation*}
This shows that the exponent in~\eqref{antonnov:rquot} is bounded from above by
$$4m(1-\rho) {\mathfrak a} = 4Nm(1-\rho)^2,$$
which converges to $0$ as $N\to \infty$, thus proving that
\begin{equation}\label{antonnov:r1ralimit}
\frac {r_1}{r_{\mathfrak a}} \to 1 \quad \mbox{ as } N\to \infty.
\end{equation}
Because of $1\ll Nm \ll (1-\rho)^{-2}$ we have
\begin{equation}\label{antonnov:Rjrj}
\frac{R_j}{r_j} = \sum_{k=0}^{j-1} \frac{r_k}{r_j} \le j \frac {r_0}{r_{\mathfrak a}}.
\end{equation}
Finally, by definition, 
\begin{equation}\label{antonnov:lambda}
\lambda_j \ge \frac j4 \quad \mbox {for all } j\le \frac N2.
\end{equation}
Combining~\eqref{antonnov:r1ralimit}, \eqref{antonnov:Rjrj} and~\eqref{antonnov:lambda} we arrive at~\eqref{antonnov:Rrbound}.
\end{proof}
\subsection{Proofs of Lemma~\ref{lem_equiv_Pij}  and Proposition \ref{lem_tech2}}\label{Prooftec2}\label{sec:caprjasympt}
\begin{proof}[Proof of Lemma~\ref{lem_equiv_Pij}]
  We start by collecting a few properties of the function $H$ defined in \eqref{def_H}. The first two derivatives of $H$ are
  \begin{eqnarray}
 H'(y)&=&  -\frac{1}{2m} \left[\frac{1}{1-\rho}\log \left( \frac{1+2m}{1+2m/\rho}  \right)+ \sum_{l=1}^{\infty} \left(1- \frac{1}{(1+2m)^l}\right) \frac{(1-\rho)^{l-1}y^{l}}{l}\right] \nonumber\\
 &=&  -\frac{1}{2m(1-\rho)} \left[\log \left( \frac{1+2m}{1+2m/\rho}  \right)+ \sum_{l=1}^{\infty} \left(1- \frac{1}{(1+2m)^l}\right) \frac{(1-\rho)^{l}y^{l}}{l}\right] \nonumber\\
 &=&  -\frac{1}{2m(1-\rho)}\log \left( \frac{1+2m}{1+2m/\rho} \frac{1- (1-\rho)y/(1+2m)}{1-(1-\rho)y} \right) \nonumber\\
 &=&  -\frac{1}{2m(1-\rho)}\log \left( \frac{1+2m- (1-\rho)y}{(1+2m/\rho)(1-(1-\rho)y)} \right) \nonumber \\
 &=&  -\frac{1}{2m(1-\rho)}\log \left( 1 + \frac{2m}{\rho}  \frac{(1-\rho)(y-1)}{(1+2m/\rho)(1-(1-\rho)y)} \right) \label{der_H}.
  \end{eqnarray}
  Since $\rho\geq \rho_0$ we have that for $y < \tfrac12$ and $N$ large enough
  \begin{equation*}
      \left\vert \frac{2m}{\rho}  \frac{(1-\rho)(y-1)}{(1+2m/\rho)(1-(1-\rho)y)}  \right\vert \geq \frac{2m}{\rho} \frac{ (1-\rho)\frac12 }{1+2m/\rho},
  \end{equation*}
  such that
  \begin{equation*}
      \left\vert \log\left(1+\frac{2m}{\rho}  \frac{(1-\rho)(y-1)}{(1+2m/\rho)(1-(1-\rho)y)} \right)  \right\vert  \geq \frac12 \frac{2m(1-\rho)}{\rho(1+2m/\rho)}, 
  \end{equation*}
  which gives
  \begin{equation}\label{eq:HderBound}
      H'(y) \geq \frac{1}{2m(1-\rho)} \frac12 \frac{2m(1-\rho)}{\rho(1+2m/\rho)} \geq \frac{1}{2\rho} \frac{1}{1+2m/\rho_0} \geq \frac14
  \end{equation}
  for $y\leq \tfrac12$ and $N$ large enough. 
We continue with the analysis of $H''$ and obtain
\begin{eqnarray*}
   H''(y)&=& -\frac{1}{2m(1-\rho)} \left[ -\frac{(1-\rho)}{1+2m- (1-\rho)y}+\frac{(1-\rho)}{1-(1-\rho)y} \right]\\
 &=& \frac{1}{2m} \left[ \frac{1}{1+2m- (1-\rho)y}-\frac{1}{1-(1-\rho)y} \right]\leq 0. 
\end{eqnarray*}
 Hence, $H(0)=0$, $H$ reaches its maximum at $y=1$, and then decreases, and  as $N\rightarrow \infty$
 $$ H(1)=-\frac{1}{2m} \left[\frac{1}{1-\rho}\log \left( \frac{1+2m}{1+2m/\rho}  \right)+ \sum_{l=1}^{\infty} \left(1- \frac{1}{(1+2m)^l}\right) \frac{(1-\rho)^{l-1}}{l(l+1)}\right] ,$$
 and 
 $$H''(1)=\frac{1}{2\rho m} \left[ \frac{1}{1+2m/\rho}-1 \right]\sim -\frac{1}{2\rho m}2m/\rho= - \frac{1}{\rho^2}, \quad \mbox{ as }N \to \infty.$$
Moreover, $H$ is non-negative on $[0,y_0]$ and negative on $(y_0,\infty)$, 
with $y_0$ satisfying
\begin{equation*}
y_0 \sim \frac{2}{\rho} \quad \mbox{ as }N \to \infty.
\end{equation*}
For later use we also  notice that from \eqref{der_H} we get for all $y\in\R \setminus \{1\}$
\begin{equation}
H'(y) \sim \frac{1-y}{\rho(1-(1-\rho)y)} \quad \mbox{ as }N\rightarrow\infty. \label{derHasym}    
\end{equation} 
We now focus on the {\bf second bullet point} of the lemma.
So $j$ is of the form
\begin{equation}\label{jframe} j=\frac{g(N)}{2 m(1-\rho)} \end{equation}
with $g(N)$  satisfying 
$$ g(N) \to \infty \quad \mbox{ and }\quad g(N) \leq (2/\rho-\xi) 2m (1-\rho)^2 N  . $$
In particular, this means
$$ \frac{1}{m(1-\rho)}\ll j \leq (2/\rho-\xi)(1-\rho)N\wedge N(1-\sqrt{m}). $$
Using \eqref{ln_pij} we obtain by a sandwiching argument
\begin{eqnarray*}
  R_j & \sim& \int_0^j \exp \left( x \log \left( \frac{1+2m}{1+2m/\rho}  \right)+ \sum_{l=1}^{\infty} \left(1- \frac{1}{(1+2m)^l}\right) \frac{1}{l(l+1)}\frac{x^{l+1}}{N^l} \right)\dif{x}\\
& =& j \int_0^1 \exp \left( j y\left[ \log \left( \frac{1+2m}{1+2m/\rho}  \right)+ \sum_{l=1}^{\infty} \left(1- \frac{1}{(1+2m)^l}\right) \frac{1}{l(l+1)}\frac{( jy)^{l}}{N^l} \right]\right)\dif{y}\\
& = &j \int_0^1 \exp \left( -2m(1-\rho)^2 N H\left(\frac{jy}{N(1-\rho)}\right) \right)\dif{y}\\
    & =& j \left(\int_0^{\eps} \exp \left( -2m(1-\rho)^2 N H\left(\frac{jy}{N(1-\rho)}\right) \right)\dif{y}\right.\\
  &&\left.+\int_{\eps}^1 \exp \left( -2m(1-\rho)^2 N H\left(\frac{jy}{N(1-\rho)}\right) \right)\dif{y}\right)
\end{eqnarray*}
where an adequate choice of $\varepsilon$ (independent of $N$) will be made below.  
By \eqref{jframe} and the above stated properties of the function $H$ we get that for all $y\in [\eps, 1]$,
\begin{equation}
    H\left(\frac{jy}{N(1-\rho)}\right) \geq H\left(\frac{g(N){\eps}}{2 m(1-\rho)^2N}\right)  \wedge H\left(2/\rho-\xi\right).\label{eq:Hbound} 
\end{equation}
Moreover, because 
$$H'(0)\sim \frac{1}{\rho}  \quad \mbox{and}\quad H'(2/\rho) \sim -\frac{1}{\rho}  \quad \mbox{as } N\to \infty,$$
combining \eqref{eq:HderBound} and \eqref{eq:Hbound} and setting $\varepsilon=\tfrac14$
we obtain for all $y\in [{\eps}, 1]$ and $N$ large enough,
\begin{align*}   H\left(\frac{jy}{N(1-\rho)}\right) &\geq \inf_{0\leq u \leq \eps} H'((2/\rho-\xi)u)\frac{g(N){\eps}}{2\rho m(1-\rho)^2N}  \wedge \inf_{0\leq u \leq y_0-2/\rho +\xi} |H'(u)|\xi  \\
&\geq \frac{g(N){\eps}}{8\rho m(1-\rho)^2N}  \wedge \frac{\xi}{4{\rho}}. 
\end{align*}
Hence
\begin{eqnarray*} &&\int_{{{\eps}}}^1 \exp \left( -2m(1-\rho)^2 N H\left(\frac{jy}{N(1-\rho)}\right) \right)\dif{y}\\&\leq& \exp \left( -2m(1-\rho)^2 N \left(\frac{g(N){{\eps}}}{8{\rho} m(1-\rho)^2N}  \wedge \frac{\xi}{4{\rho}}\right)\right) \\
 & = &\exp \left( -\frac{g(N){{\eps}}}{4{\rho}}\right) \vee \exp \left( -\frac12 m(1-\rho)^2 N \xi{/\rho}\right). \end{eqnarray*}
The equivalence $H'(0)\sim 1{/\rho}$ also entails that
\begin{eqnarray*}  &&j \int_0^{{{\eps}}} \exp \left( -2m(1-\rho)^2 N H\left(\frac{jy}{N(1-\rho)}\right) \right)\dif{y} \\
&\sim& j \int_0^{{{\eps}}} \exp \left( -2m(1-\rho)^2 N \frac{jy}{N{\rho}(1-\rho)} \right)\dif{y} \\ & =&j
 \int_0^{{{\eps}}} \exp \left( -2m(1-\rho)jy /{\rho}\right)\dif{y} \\
 & \sim &\frac{{\rho}}{2m(1-\rho)} = \frac{j\rho}{g(N)}. \end{eqnarray*}
 In total, 
as 
 $$ \exp \left( -\frac{g(N){{\eps}}}{4{\rho}}\right) = o \left( \frac{1}{g(N)} \right)$$
 and
 $$\exp \left( -\frac12 m(1-\rho)^2 N \xi{/\rho}\right) \leq \frac{2}{(Nm(1-\rho))^2} \leq \left(\frac{8}{\rho g(N)}\right)^2 =  o \left(\frac{1}{g(N)} \right) $$ this gives $R_j \sim {\rho}/(2m (1-\rho))$ and thus proves the second bullet point of the lemma. \\
 
We now turn to the {\bf third bullet point}. 
So $j$ is of the form $j=C(1-\rho)N$ with $C>2/\rho$ and $j\leq N(1-\sqrt{m}) $. (Note that $C(1-\rho)N \leq N$ and $ C>2/\rho$ imply $\rho > 2/3$.) 
Using  \eqref{newsandwich} and sandwiching arguments yields
\begin{equation*}
R_j  \sim  (1-\rho)N \int_0^C \exp \left( -2m(1-\rho)^2N H(y) \right)\dif{y}.
\end{equation*}
$H(y)$ is non-negative for $y\le y_0$ with $y_0 \sim 2/\rho$, and negative for $y > y_0$. Moreover, $H''<0$. Consequently we get
\begin{equation*}
R_j  \sim  (1-\rho)N \int_{y_0}^C \exp \left( -2m(1-\rho)^2N H(y) \right)\dif{y}.
\end{equation*}
By \eqref{derHasym} there exists $\mathfrak{x}\geq 0$ such that 
$$H'(y) \in [-\mathfrak{x} \xi - (C-1)/(\rho(1-C(1-\rho))),\mathfrak{x} \xi - (C-1)/(\rho(1-C(1-\rho))) ], \quad \forall y \in [C-\xi, C+\xi].$$
This entails
\begin{eqnarray*}
  &&\int_{C-\xi }^C e^{ -2m(1-\rho)^2N ( (C-1)/(\rho(1-C(1-\rho)))+\mathfrak{x} \xi )(C-y)}\dif{y}\\ & \leq& \int_{C-\xi }^C e^{-2m(1-\rho)^2N (H(y)-H(C))}\dif{y}
\\ &  \leq &\int_{C-\xi }^C e^{-2m(1-\rho)^2N ( (C-1)/(\rho(1-C(1-\rho)))-\mathfrak{x} \xi )(C-y)}\dif{y}.
\end{eqnarray*}
Moreover, 
\begin{eqnarray*}
 \int_{y_0}^{C-\xi } \exp \left( -2m(1-\rho)^2N H(y) \right)\dif{y} &\leq& C\exp \left( -2m(1-\rho)^2N H(C-\xi )\right)\\
 & = &o \left(\exp \left( -2m(1-\rho)^2N H(C)\right)\right).  \end{eqnarray*}
We deduce that
\begin{eqnarray*}
    R_j  &\sim&  (1-\rho)N \frac{\rho(1-C(1-\rho))\exp \left( -2m(1-\rho)^2N H(C)\right)}{2m(C-1)(1-\rho)^2N}\\
    &=&\frac{ \rho(1-C(1-\rho)) }{2m(C-1)(1-\rho)}\exp \left( -2m(1-\rho)^2N H(C)\right),
\end{eqnarray*}
which proves the third bullet point of the lemma.\\

Finally, we focus on the {\bf first bullet point}. 
Let   $j \leq \xi  (1-\rho)N$ with $\xi \leq \tfrac12$. From the observation that
$$ 1- \frac{1}{(1+2m)^l} \leq 2ml  $$
we get
\begin{eqnarray*}
  \sum_{l=1}^{\infty} \left(1- \frac{1}{(1+2m)^l}\right) \frac{1}{l(l+1)}\frac{j^{l}}{N^l} 
 & \leq& 2m \sum_{l=1}^{\infty} \frac{1}{l+1}\frac{j^{l}}{N^l} \\
 & \leq & m \sum_{l=1}^{\infty} \frac{j^{l}}{N^l}\\
 & \leq &m \frac{\xi (1-\rho)}{1- \xi (1-\rho)} \leq \frac{m}{\rho}\xi (1-\rho).
\end{eqnarray*}
As 
$$ \log \left(\frac{1+2m/\rho}{1+2m}\right)=
\log \left(1+ \frac{2m}{\rho}\frac{1-\rho}{1+2m}\right)\sim \frac{2m}{\rho}(1-\rho)$$
we deduce that
$$\sum_{l=1}^{\infty} \left(1- \frac{1}{(1+2m)^l}\right) \frac{1}{l(l+1)}\frac{j^{l}}{N^l} \leq \xi  \log \left(\frac{1+2m/\rho}{1+2m}\right).  $$
Hence for $j\leq \xi (1-\rho)N \leq (1-\rho)N$,
\begin{eqnarray*}
  -(1+\xi )\frac{2m}{\rho}(1-\rho)j& \leq& j \log\left( \frac{1+2m}{1+2m/\rho} \right)+ \sum_{l=1}^{\infty} \left(1- \frac{1}{(1+2m)^l}\right) \frac{j^{l+1}}{l(l+1)N^l} \\
  &\leq& -(1-\xi )\frac{2m}{\rho}(1-\rho)j.
\end{eqnarray*}
Consequently,
\begin{eqnarray*}
  \frac{1- e^{-(1+2\xi )(2m/\rho)(1-\rho)j} }{(1+2\xi )(2m/\rho)(1-\rho)}&= &\int_0^j e^{-(1+2\xi )(2m/\rho)(1-\rho)x}\dif{x}  \\
 &\leq R_j&
 \leq \int_0^j e^{-(1-2\xi )(2m/\rho)(1-\rho)x}\dif{x} =
\frac{1- e^{-(1-2\xi )(2m/\rho)(1-\rho)j} }{(1+2\xi )(2m/\rho)(1-\rho)}. \end{eqnarray*}
This ends the proof of Lemma~\ref{lem_equiv_Pij}.
\end{proof}
\begin{proof}[Proof of Proposition \ref{lem_tech2}]
Let us first study the asymptotics of $B(\zeta)$.
Thanks to the second bullet point of Lemma~\ref{lem_equiv_Pij} we know that for any $ (1-\zeta)(1-\rho)N\leq k \leq (1+\zeta)(1-\rho)N$ one has $R_k \sim \rho/(2m(1-\rho))$. Hence
\begin{equation*}
   B(\zeta) \sim \left( R_j \wedge \frac{\rho}{2m(1-\rho)} \right)\sum_{k=(1-\zeta)(1-\rho)N+1}^{(1+\zeta)(1-\rho)N} \frac{1}{\lambda_k r_k}.
\end{equation*}
Moreover, for $k\in \left[ (1-\zeta)(1-\rho)N, (1+\zeta)(1-\rho)N\right]$ we have $ \lambda_k \sim \rho k/2$.
Hence
\begin{eqnarray*}
  \frac{2}{\rho (1+2\zeta)(1-\rho)N} \sum_{k=(1-\zeta)(1-\rho)N+1}^{(1+\zeta)(1-\rho)N} \frac{1}{r_k} &\leq& \sum_{k=(1-\zeta)(1-\rho)N+1}^{(1+\zeta)(1-\rho)N} \frac{1}{\lambda_k r_k} \\&\leq& \frac{2}{\rho (1-2\zeta)(1-\rho)N}\sum_{k=(1-\zeta)(1-\rho)N+1}^{(1+\zeta)(1-\rho)N} \frac{1}{r_k}.
\end{eqnarray*}
We are left with the study of
$$ \sum_{k=(1-\zeta)(1-\rho)N+1}^{(1+\zeta)(1-\rho)N} \frac{1}{r_k} \sim N(1-\rho) \int_{1-\zeta}^{1+\zeta}  \exp \left( 2m(1-\rho)^2N H(y) \right)\dif{y}, $$
where the equivalence is a consequence of \eqref{newsandwich}. (Note that the function $H$, see \eqref{def_H}, appears in \eqref{newsandwich}.)
Since the function $H$ reaches its maximum at $1$ and $H''(1) \sim {-1/\rho^2}$, and since $\zeta$ satisfies 
$$ \zeta \sqrt{mN}(1-\rho) \to \infty \quad \mbox{ as } N\rightarrow\infty,$$ an application of the Laplace method yields
\begin{eqnarray*}  \int_{1-\zeta}^{1+\zeta}  \exp \left( 2m(1-\rho)^2N H(y) \right)\dif{y}&\sim& \sqrt{\frac{2\pi\rho^2}{2m(1-\rho)^2 N}}\exp \left( 2m(1-\rho)^2N H(1) \right)\\
& =& \sqrt{\frac{\pi}{m N}}\frac{\rho}{1-\rho}\exp   \left( 2m(1-\rho)^2N H(1) \right). \end{eqnarray*}
Hence
\begin{equation*}
   B(\zeta) \sim \left( R_j \wedge \frac{\rho}{2m(1-\rho)} \right) \sqrt{\frac{\pi}{m N}}\frac{2}{1-\rho}\exp   \left( 2m(1-\rho)^2N H(1) \right).
\end{equation*}
This completes the proof of \eqref{simBeps}.\\

To bound $A(\zeta)$, it is enough to notice that for any $k \leq (1-\zeta)(1-\rho)N$, 
$$ \lambda_k \geq \frac{k}{2}(1- (1-\zeta)(1-\rho)), \quad r_k \geq  \exp \left(- 2m(1-\rho)^2 N H(1-\zeta/2)\right),  \quad R_{k \wedge j} \leq \frac{\rho}{2m(1-\rho)}. $$
The first inequality is a direct consequence of the definition of $\lambda_k$ in \eqref{bestupproof}, the second one stems from equality \eqref{ln_pij}, and the last one is a consequence of Lemma \ref{lem_equiv_Pij}. Altogether this yields that
\begin{eqnarray*}
 A(\zeta)&\leq& \frac{(1-\zeta)(1-\rho)\rho N}{2m(1-\rho)}\frac{2\exp \left( 2m(1-\rho)^2 N H(1-\zeta/2)\right)}{(1-\zeta)(1-\rho)N(1- (1-\zeta)(1-\rho))} \\
 &=& o \left(B(\zeta) \right).
\end{eqnarray*}
The term $C(\zeta)$ is more delicate to bound and we have to decompose it into several terms. This decomposition depends on the value of $\rho$:

Let us begin with the simplest case, that is $\rho\leq  2/3$. In this case $(2/\rho)(1-\rho)\geq 1$ and thus  $(2/\rho)(1-\rho)N\geq N$. Recall that the positive root $y_0$ of $H$ satisfies $y_0 \sim 2/\rho$.
We may decompose $C(\zeta)$ as follows:
\begin{eqnarray*}
C(\zeta) &=& \sum_{k=(1+\zeta)(1-\rho)N+1}^{(y_0(1-\rho)\wedge 1)N(1-\sqrt{m})} \frac{R_{k\wedge j}}{\lambda_k r_k} +\sum_{k=(y_0(1-\rho)\wedge 1)N(1-\sqrt{m})+1}^{N} \frac{R_{k\wedge j}}{\lambda_k r_k}  \\
&:= &C_\alpha(\zeta)+C_\beta(\zeta).
\end{eqnarray*}
Using that $H$ is non-negative and decreasing on $[1,y_0]$ and $y_0 \sim 2/\rho$ we obtain  from equality \eqref{ln_pij} that for any $(1+\zeta)(1-\rho)N+1 \leq k \leq (y_0(1-\rho)\wedge 1)N(1-\sqrt{m})$,
$$\lambda_k r_k \geq k m \exp \left(- 2m(1-\rho)^2 N H(1+\zeta/2)\right)\quad \text{and} \quad R_{k \wedge j} \leq \frac{\rho}{2m (1-\rho)}.$$
Hence
\begin{eqnarray*}
 C_\alpha(\zeta)&\leq& \frac{\rho N}{2m(1-\rho)}\frac{1}{ (1-\rho)N m} \exp \left( 2m(1-\rho)^2 N H(1+\zeta/2)\right)
= o \left( B(\zeta) \right).
\end{eqnarray*}
To bound $C_\beta(\zeta)$ we apply \eqref{ext_from_N} with $\xi \sim \sqrt{m}$ satisfying
$$ (y_0(1-\rho)\wedge 1)N(1-\sqrt{m})=N(1-\xi) .$$
This shows that $C(\zeta) = o(B(\zeta))$ in the case $\rho\leq 2/3$.

Let us now consider the case $\rho>2/3$. We decompose $C(\zeta)$ into three terms as follows:
\begin{eqnarray*}
C(\zeta) &=& \sum_{k=(1+\zeta)(1-\rho)N+1}^{(2/\rho-\zeta)(1-\rho)N} \frac{R_{k\wedge j}}{\lambda_k r_k} +\sum_{k=(2/\rho-\zeta)(1-\rho)N+1}^{(2/\rho+\zeta)(1-\rho)N} \frac{R_{k\wedge j}}{\lambda_k r_k} +\sum_{k=(2/\rho+\zeta)(1-\rho)N+1}^N \frac{R_{k\wedge j}}{\lambda_k r_k} \\
&:= &C_1(\zeta)+C_2(\zeta)+C_3(\zeta).
\end{eqnarray*}
$C_1(\zeta)$ may be bounded with similar arguments as for $A(\zeta)$: for any $(1+\zeta)(1-\rho)N \leq k \leq (2/\rho-\zeta)(1-\rho)N$,
$$ \lambda_k r_k \geq k m \exp \left(- 2m(1-\rho)^2 N H(1+\zeta/2)\right) \quad \text{and} \quad R_{k \wedge j} \leq \frac{\rho}{2m (1-\rho)}.$$
This entails
\begin{eqnarray*}
 C_1(\zeta)&\leq& \frac{(1-2\zeta)(1-\rho)N}{2m\rho(1-\rho)}\frac{1}{(1-\rho)Nm} \exp \left( 2m(1-\rho)^2 N H(1+\zeta/2)\right)
= o \left( B(\zeta) \right).
\end{eqnarray*}

Now, recalling the third bullet point of Lemma~\ref{lem_equiv_Pij},  that $R_.$ is increasing and that $r_k$ is increasing with $k$ when $k$ is larger than $(1+\zeta)(1-\rho)N$, we get for 
$ (2/\rho-\zeta)(1-\rho)N\leq k \leq (2/\rho+\zeta)(1-\rho)N$,
$$ \lambda_k r_k \geq km \exp \left( - 2m(1-\rho)^2N H(2/\rho-2\zeta) \right)$$ 
and
$$ R_k \leq R_{(2/\rho+\zeta)(1-\rho)N} \sim 
\rho(1-(2/\rho+\zeta)(1-\rho))\frac{\exp \left( -2m(1-\rho)^2N H(2+\zeta)\right)}{2m(2/\rho+\zeta-1)(1-\rho)}.$$
Using that $H'(2/\rho) \sim -1/\rho$, we deduce that there exists a constant $\mathfrak{K}$ such that
\begin{eqnarray*}
C_2(\zeta)&\leq& \frac{\mathfrak{K}\zeta(1-\rho)N}{(1-\rho)^2Nm^2} \exp \left( 2m(1-\rho)^2 N (H(2-2\zeta)-H(2+2\zeta))\right)\\
& \leq& \frac{\mathfrak{K}\zeta}{(1-\rho)m^2}\exp \left( \mathfrak{K} m(1-\rho)^2 N \zeta\right) = o \left(B(\zeta) \right).
\end{eqnarray*}

Notice that for  $k \geq (2/\rho+\zeta)(1-\rho)N-1$, $r_k = \sup \{ r_l, \ l \leq k\}.$
Hence, for any $j$, 
\begin{eqnarray*}
C_3(\zeta)& =& \sum_{k=(2/\rho+\zeta)(1-\rho)N+1}^N \frac{R_{k\wedge j}}{k q_k r_{k-1}}\\
& \leq& \frac{1}{m} \sum_{k=(2/\rho+\zeta)(1-\rho)N+1}^N \frac{R_{k}}{kr_{k-1}}\\
& = &\frac{1}{m} \sum_{k=(2/\rho+\zeta)(1-\rho)N+1}^N \frac{r_0+...+r_{k-2}+r_{k-1}}{kr_{k-1}} \leq \frac{N}{m} = o \left(B(\zeta) \right).
\end{eqnarray*}
This shows that $C(\zeta) = o(B(\zeta))$ in the case $\rho>2/3$, and thus concludes the proof of Proposition~\ref{lem_tech2}.
\end{proof}


\paragraph{\bf Acknowledgment} We are grateful to an anonymous reviewer for a careful reading of a previous version and for suggestions that led to an improvement of the presentation.
We thank the {\it Allianz f\"ur Hochleistungsrechnen Rheinland-Pfalz} for granting us access to the High Performance Computing \textsc{Elwetritsch}, on which the simulations underlying our figures have been performed. 
On a more personal note, one of the authors (A.W.) remembers with pleasure the inspiring collaboration with Alison Etheridge and Peter Pfaffelhuber which led to the paper \cite{EPW09}.  Together with \cite{GSW23}, \cite{EPW09}  constitutes the main root of the results and insights achieved in the present paper. \\
\paragraph{\bf Funding}This work was partially funded by the Chair "Modélisation Mathématique
et Biodiversité" of VEOLIA-Ecole Polytechnique-MNHN-F.X. AGC was supported by PAPIIT‐UNAM (grant IN101722) and by CONACYT Ciencia Básica (grant A1‐S‐14615).

\bibliography{IGSW} 

\newcommand{\etalchar}[1]{$^{#1}$}
\begin{thebibliography}{GCSW23}
\expandafter\ifx\csname url\endcsname\relax
  \def\url#1{\texttt{#1}}\fi
\expandafter\ifx\csname doi\endcsname\relax
  \def\doi#1{\burlalt{doi:#1}{http://dx.doi.org/#1}}\fi
\expandafter\ifx\csname urlprefix\endcsname\relax\def\urlprefix{URL }\fi
\expandafter\ifx\csname href\endcsname\relax
  \def\href#1#2{#2}\fi
\expandafter\ifx\csname burlalt\endcsname\relax
  \def\burlalt#1#2{\href{#2}{#1}}\fi

\bibitem[AP13]{audiffren2013muller}
J.~Audiffren and E.~Pardoux.
\newblock Muller’s ratchet clicks in finite time.
\newblock {\em Stochastic Processes and their Applications}, 123(6):2370--2397,
  2013.
\newblock \doi{10.1016/j.spa.2013.02.008}.

\bibitem[BFM18]{EvComp1}
T.~B{\"a}ck, D.~B. Fogel, and Z.~Michalewicz.
\newblock {\em Evolutionary computation 1: Basic algorithms and operators}.
\newblock CRC press, 2018.

\bibitem[BS22]{brautigam2022diffusion}
C.~Bräutigam and M.~Smerlak.
\newblock Diffusion approximations in population genetics and the rate of
  {M}uller’s ratchet.
\newblock {\em Journal of Theoretical Biology}, 550:111236, 2022.
\newblock \doi{10.1016/j.jtbi.2022.111236}.

\bibitem[BT96]{BT961}
T.~Blickle and L.~Thiele.
\newblock A comparison of selection schemes used in evolutionary algorithms.
\newblock {\em Evolutionary Computation}, 4(4):361--394, 1996.
\newblock \doi{10.1162/evco.1996.4.4.361}.

\bibitem[CCM16]{chazottes2016sharp}
J.~R. Chazottes, P.~Collet, and S.~M{\'e}l{\'e}ard.
\newblock Sharp asymptotics for the quasi-stationary distribution of
  birth-and-death processes.
\newblock {\em Probability Theory and Related Fields}, 164(1):285--332, 2016.
\newblock \doi{10.1007/s00440-014-0612-6}.

\bibitem[DSS05]{doering}
C.~R. Doering, K.~V. Sargsyan, and L.~M. Sander.
\newblock Extinction times for birth-death processes: exact results, continuum
  asymptotics, and the failure of the {F}okker-{P}lanck approximation.
\newblock {\em Multiscale Model. Simul.}, 3(2):283--299, 2005.
\newblock \doi{10.1137/030602800}.

\bibitem[EPW09]{EPW09}
A.~M. Etheridge, P.~Pfaffelhuber, and A.~Wakolbinger.
\newblock {\em How often does the ratchet click? Facts, heuristics,
  asymptotics}, page 365–390.
\newblock London Mathematical Society Lecture Note Series. Cambridge University
  Press, 2009.
\newblock \doi{10.1017/CBO9781139107020.016}.
\newblock \burlalt{arXiv:0709.2775}{http://arxiv.org/abs/0709.2775}.

\bibitem[Fel74]{felsenstein1974evolutionary}
J.~Felsenstein.
\newblock {The evolutionary advantage of recombination}.
\newblock {\em Genetics}, 78(2):737--756, 10 1974.
\newblock \doi{10.1093/genetics/78.2.737}.

\bibitem[GC00]{Gordo2000TheDO}
I.~Gordo and B.~Charlesworth.
\newblock {The Degeneration of Asexual Haploid Populations and the Speed of
  Muller's Ratchet}.
\newblock {\em Genetics}, 154(3):1379--1387, 03 2000.
\newblock \doi{10.1093/genetics/154.3.1379}.

\bibitem[GCSW23]{GSW23}
A.~Gonz\'{a}lez~Casanova, C.~Smadi, and A.~Wakolbinger.
\newblock Quasi-equilibria and click times for a variant of {M}uller's ratchet.
\newblock {\em Electron. J. Probab.}, 28:Paper No. 159, 37, 2023.
\newblock \doi{10.1214/23-ejp1055}.

\bibitem[Hai78]{H78}
J.~Haigh.
\newblock The accumulation of deleterious genes in a population—{M}uller's
  ratchet.
\newblock {\em Theoretical Population Biology}, 14(2):251--267, 1978.
\newblock \doi{10.1016/0040-5809(78)90027-8}.

\bibitem[IGSW24]{IGSW}
J.~Igelbrink, A.~{González Casanova}, C.~Smadi, and A.~Wakolbinger.
\newblock Muller’s ratchet in a near-critical regime: Tournament versus
  fitness proportional selection (published version of arxiv:2306.00471v2
  [q-bio.pe]).
\newblock {\em Theoretical Population Biology}, 158:121--138, 2024.
\newblock \doi{10.1016/j.tpb.2024.06.001}.

\bibitem[Lam05]{L}
A.~Lambert.
\newblock The branching process with logistic growth.
\newblock {\em The Annals of Applied Probability}, 15(2):1506--1535, 2005.
\newblock \doi{10.1214/105051605000000098}.

\bibitem[ME13]{ME13}
J.~J. Metzger and S.~Eule.
\newblock Distribution of the fittest individuals and the rate of muller's
  ratchet in a model with overlapping generations.
\newblock {\em PLOS Computational Biology}, 9(11):1--10, 2013.
\newblock \doi{10.1371/journal.pcbi.1003303}.

\bibitem[MPV20]{MPV20}
M.~Mariani, E.~Pardoux, and A.~Velleret.
\newblock Metastability between the clicks of the {M}uller ratchet, 2020,
  \burlalt{arXiv:2007.14715}{http://arxiv.org/abs/2007.14715}.

\bibitem[Mul64]{M64}
H.~Muller.
\newblock The relation of recombination to mutational advance.
\newblock {\em Mutation Research/Fundamental and Molecular Mechanisms of
  Mutagenesis}, 1(1):2--9, 1964.
\newblock \doi{10.1016/0027-5107(64)90047-8}.

\bibitem[NS12]{neher2012fluctuations}
R.~A. Neher and B.~I. Shraiman.
\newblock Fluctuations of fitness distributions and the rate of muller’s
  ratchet.
\newblock {\em Genetics}, 191(4):1283--1293, 2012.
\newblock \doi{10.1534/genetics.112.141325}.

\bibitem[PBB{\etalchar{+}}15]{Pai2015}
T.~Paixão, G.~Badkobeh, N.~Barton, D.~Çörüş, D.-C. Dang, T.~Friedrich,
  P.~K. Lehre, D.~Sudholt, A.~M. Sutton, and B.~Trubenová.
\newblock Toward a unifying framework for evolutionary processes.
\newblock {\em Journal of Theoretical Biology}, 383:28--43, 2015.
\newblock \doi{0.1016/j.jtbi.2015.07.011}.

\bibitem[PSW12]{pfaffelhuber2012muller}
P.~Pfaffelhuber, P.~R. Staab, and A.~Wakolbinger.
\newblock {Muller’s ratchet with compensatory mutations}.
\newblock {\em The Annals of Applied Probability}, 22(5):2108 -- 2132, 2012.
\newblock \doi{10.1214/11-AAP836}.

\bibitem[SaSha13]{sagitov2013extinction}
S.~Sagitov and A.~Shaimerdenova.
\newblock Extinction times for a birth--death process with weak competition.
\newblock {\em Lith. Math. J.}, 53(2):220--234, 2013.
\newblock \doi{10.1007/s10986-013-9204-x}.

\bibitem[SCS93]{Stephan1993TheAO}
W.~Stephan, L.~Chao, and J.~G. Smale.
\newblock The advance of muller's ratchet in a haploid asexual population:
  approximate solutions based on diffusion theory.
\newblock {\em Genetics Research}, 61(3):225–231, 1993.
\newblock \doi{10.1017/S0016672300031384}.

\bibitem[Wei]{wolframDwason}
E.~W. Weisstein.
\newblock Dawson's integral.
\newblock \urlprefix\url{https://mathworld.wolfram.com/DawsonsIntegral.html}.
\newblock Visited on February 21, 2024.

\bibitem[WL10]{waxman2010stochastic}
D.~Waxman and L.~Loewe.
\newblock A stochastic model for a single click of muller's ratchet.
\newblock {\em Journal of Theoretical Biology}, 264(4):1120--1132, 2010.
\newblock \doi{10.1016/j.jtbi.2010.03.014}.

\end{thebibliography}
\bibliographystyle{habbrv}
\end{document}